\documentclass{vki}

\usepackage{lmodern,pdftexcmds,amsmath}
\usepackage{placeins}

\usepackage{amsfonts,amsthm,bm} 

\DeclareMathAlphabet{\mathsfbi}{OT1}{\sfdefault}{bx}{sl}
\DeclareMathVersion{sfletters}
\SetSymbolFont{letters}{sfletters}{OML}{ntxsfmi}{b}{it}

\makeatletter
\newcommand{\mathbfsbilow}[1]{%
	\text{\mathversion{sfletters}$\m@th#1$}%
}

\DeclareRobustCommand{\tensor}[1]{%
	\begingroup
	\ifcat\noexpand #1\relax
	\edef\greek@test{\detokenize{#1}}%
	\edef\greek@test{\expandafter\@cdr\greek@test\@nil}%
	\edef\greek@test{\expandafter\@car\greek@test\@nil}%
	\edef\x{\the\lccode\expandafter`\greek@test}%
	\edef\y{\number\expandafter`\greek@test}%
	\ifnum\x=\y\relax
	\mathbfsbilow{#1}%
	\else
	\mathsfbi{#1}%
	\fi
	\else
	\mathsfbi{#1}%
	\fi
	\endgroup
}
\makeatother

\usepackage[most]{tcolorbox}
\usepackage{booktabs}
\usepackage{amsmath}
\usepackage{wrapfig}
\usepackage[linesnumbered,ruled,vlined]{algorithm2e}

\usepackage{amssymb}
\usepackage{mathtools}
\tcbset{enhanced,colback=cyan!2!white,colframe=cyan!90!white,coltitle=blue!20!black,fonttitle=\bfseries}

\usepackage{listings}
\usepackage{hyperref}
\hypersetup{
	colorlinks=true,
	linkcolor=blue,
	filecolor=magenta,      
	citecolor=blue,
	urlcolor=cyan,
}

\begin{document}


\pagenumbering{roman}
\title{Statistical Treatment, \\Fourier and Modal Decomposition}
\author{Miguel Alfonso Mendez
\thanks{mendez@vki.ac.be.  }\\
von Karman Institute for Fluid Dynamics }
\date{4 November 2021}
\maketitle

These are the lecture notes for the lecture ``Statistical Treatment, Fourier and Modal Decompositions'', given at the VKI Lecture series ``Fundamentals and Recent Advances in Particle Image Velocimetry and Lagrangian Particle Tracking'' (full program \href{https://www.vki.ac.be/index.php/events-ls/events/eventdetail/520/-/online-on-site-lecture-series-fundamentals-and-recent-advances-in-particle-image-velocimetry-and-lagrangian-particle-tracking}{here}). The course was held at the von Karman Institute for fluid dynamics from 15 November to 18 November 2021.

This lecture provides a guided tour through the processing of data acquired via image velocimetry. Far from being an exhaustive account of the field, which would require an entire course on its own, the scope is to provide a hands-on tutorial. This begins with basic statistical treatment, briefly reviews frequency and modal analysis, and conclude with more advanced research topics such as multiscale modal decompositions and nonlinear dimensionality reduction. The material covered should hopefully propel newcomers into the subject while remaining of interest to experienced practitioners. 

All the codes related to this lecture are available at \href{https://github.com/mendezVKI/Lecture_10_VKI_PIV_PTV_LS}{this github repository.}\\
\\\vspace{20mm}\\
How to cite this work
\bigskip
\bigskip
\bigskip
\begin{centering}
	\begin{lstlisting}
@InProceedings{Mendez2020,
 author = {Miguel A. Mendez},
 title = {Statistical Treatment, Fourier and Modal Decompositions},
 booktitle = {Fundamentals and Recent Advances in Particle
 Image Velocimetry and Lagrangian Particle Tracking, VKI Lecture Series},
 editor = {Discetti, S. and Mendez, M. A},
 year = {2021},
 publisher = {von Karman Institute},
 ISBN = {978-2-87516-181-9},
}
	\end{lstlisting}
\end{centering}

\pagenumbering{arabic}
\setcounter{page}{1}
\clearpage{\pagestyle{empty}} 

\tableofcontents
\clearpage{\pagestyle{empty}}

\section{Lecture Format and Organization}\label{sec1}

Modern image velocimetry produces data with spatio-temporal resolutions that were inconceivable a couple of decades ago \citep{Raffel2018,Adrian2005}. This lecture is about processing that data and reviewes a broad range of techniques. 

With the intent of being accessible while providing some hands-on experience, the lecture is cast a sequence of exercises in {Python}. Thus, these notes should be read while keeping a {Python} terminal at hand, ready to run the provided scripts. {Python} is nowadays the lingua franca of data processing and ``soft computing''; it is an extremely powerful language, it is free, and it is easy to use. While the reader might not be an experienced programmer, it is hoped that the presented scripts are simple enough to be intelligible. Running and interacting with these is essential to the correct delivery of the presented material. Readers interest in learning Python for scientific computing are referred to \cite{SveinLinge2019,Langtangen2016,Johansson2018,PaulJ.Deitel2019}.

All the presented tools will be applied to a specific time-resolved PIV dataset. The details on this dataset and the instructions to download it are provided in Section \ref{sec2}. The provided exercises guide the reader to the implementation of the most important data processing operations in statistical analysis, frequency analysis and modal decompositions. The theoretical background is kept to a minimum, and the reader is urged to consult the provided references to learn more. 

The first part, in Section \ref{sec2}, covers some essential statistical treatment, primarily focusing on time series analysis. There we treat the dataset as a set of independent time series corresponding to the velocity signal in various positions. These time series are, of course, not independent. However, the methods presented in this section do not explore nor exploit their spatial dependency. The tools introduced in this section are also the oldest: these were the only ones covered in data processing courses before image velocimetry made space and time resolution easily accessible. The second part, in Section \ref{sec3}, moves from a ``local processing'' to a ``global processing'', i.e. including both space and time analysis. This takes us to modal analysis, and the section reviews the notion of flow \emph{modes} and \emph{modal decompositions}. These are introduced first for the familiar Discrete Fourier Transform and then for the two most popular \emph{data-driven} decompositions: the Proper Orthogonal Decomposition (POD) and the Dynamic Mode Decomposition (DMD). We shall explore ``pro'' and ``cons'' of each with the hope that the reader quickly realizes that the distinction between  ``pro'' and ``cons'' is meaningless: these are simply different tools that serve different purposes and answer different questions.

The last part, in Section \ref{sec5}, covers two advanced research topics. The first topic is an extension of the previous methods while keeping their linear framework: a recent decomposition that breaks the dichotomy POD/DMD and proposes a bridge between the two. This is the multi-scale Proper Orthogonal Decomposition (mPOD), extensively described in \cite{Mendez2019,MendezLS2}. The second topic also extends the previous methods but brings the treatment to a nonlinear framework, with kernels functions defining nonlinear similarity metrics. This section touches briefly on nonlinear dimensionality reduction and manifold learning. These are well-known tools in machine learning \citep{Bishop2006} but are now gaining popularity also in fluid mechanics, offering new ways of looking at the dynamics of fluid flows. Section closes \ref{sec6} with some final remarks.



\section{Collecting the Dataset}\label{sec2}

The selected test case is the flow past a cylinder of $d=5$ mm diameter and $L=20cm$ length in transient conditions, with varying free stream velocity. The experiments were carried out in the L10 low-speed wind tunnel of the von Karman Institute, instrumented with a TR-PIV system from Dantec Dynamics. The details of the experiment are described in \cite{Mendez2020}. The dataset can be downloaded by the provided script Get\_Cylinder\_DATA.py, and particularly the following lines:

\begin{centering}
	\begin{lstlisting}[language=Python,linewidth=17cm,xleftmargin=.05\textwidth,xrightmargin=.05\textwidth,backgroundcolor=\color{yellow!10}]
import urllib.request
print('Downloading Data PIV for Chapter 10...')
url = 'https://osf.io/47ftd/download'
urllib.request.urlretrieve(url, 'Ex_5_TR_PIV_Cylinder.zip')
print('Download Completed! I prepare data Folder')
# Unzip the file 
from zipfile import ZipFile
String='Ex_5_TR_PIV_Cylinder.zip'
zf = ZipFile(String,'r'); 
zf.extractall('./DATA_CYLINDER'); zf.close()
	\end{lstlisting}
\end{centering}

The dataset consists of $n_t=13200$ velocity fields, sampled at $f_s=3kHz$ over a grid of $71\times30$ points. The spatial resolution is approximately $\Delta x=0.85 mm$. A plot of an exemplary time step, e.g. time step $10$, can be printed as follows:

\begin{centering}
	\begin{lstlisting}[language=Python,breaklines,linewidth=17cm,xleftmargin=.05\textwidth,xrightmargin=.05\textwidth,backgroundcolor=\color{yellow!10}]
import os
import matplotlib.pyplot as plt
from Functions_Chapter_10 import Plot_Field_TEXT_Cylinder
FOLDER='DATA_CYLINDER'
Name=FOLDER+os.sep+'Res%05d'%10+'.dat' # Check it out: print(Name)
Name_Mesh=FOLDER+os.sep+'MESH.dat'
n_s,Xg,Yg,Vxg,Vyg,X_S,Y_S=Plot_Field_TEXT_Cylinder(Name,Name_Mesh) 
Name='Snapshot_Cylinder.png'
plt.savefig(Name, dpi=200) 
	\end{lstlisting}
\end{centering}

\begin{figure}[htbp]
	\centering
	\includegraphics[keepaspectratio=true,width=0.75 \columnwidth]
	{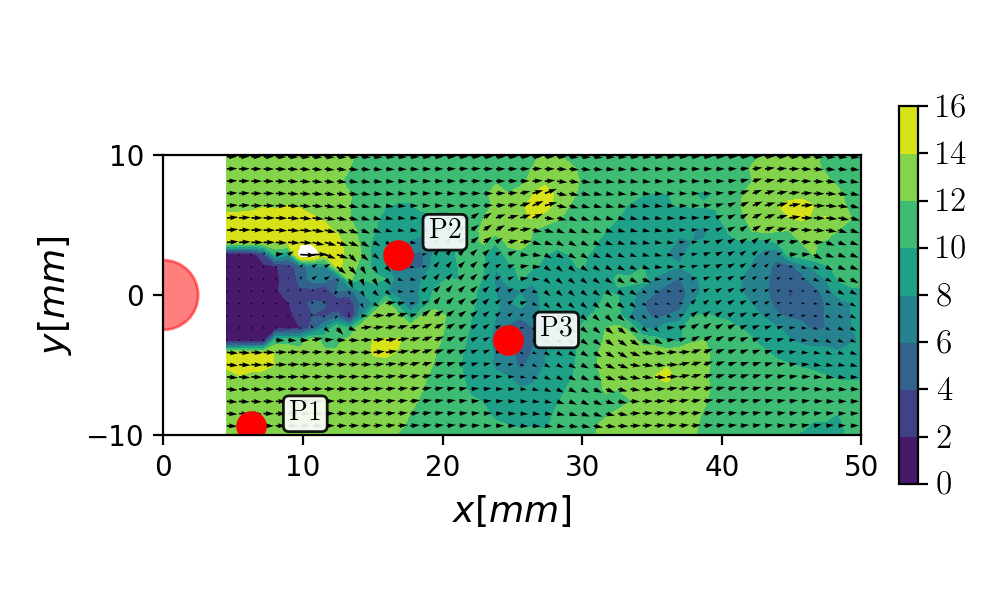}\\
	\vspace{-0.8cm}
	\caption
	{Example snapshot for the dataset analyzed in this lecture. This is the flow past a cylinder in transient condition from \cite{Mendez2020}. In the following section, the time series obtained at three virtual probes P1, P2, P3 will be analyzed. The location of these probes is shown in the figure.}\label{fig1}
\end{figure}

The figure produced from this script is shown in Figure \ref{fig1}. Three probes, denoted as P1, P2 and P3, are placed to extract time series data for the following exercises. The location of these probes is also shown in the plot.

The following code extracts the time series of the velocity in the three locations:

\begin{centering}
	\begin{lstlisting}[language=Python,linewidth=17cm,xleftmargin=.05\textwidth,xrightmargin=.05\textwidth,backgroundcolor=\color{yellow!10},breaklines]
#%% Get Time Series
import numpy
import numpy as np
# Define the location of the probes in the grid
i1,j1=2,4; xp1=Xg[i1,j1]; yp1=Yg[i1,j1]
i2,j2=14,18; xp2=Xg[i2,j2]; yp2=Yg[i2,j2]
i3,j3=23,11; xp3=Xg[i3,j3]; yp3=Yg[i3,j3]
# Prepare the time axis
n_t=13200; Fs=3000; dt=1/Fs 
t=np.linspace(0,dt*(n_t-1),n_t) # prepare the time axis# 
# Initialize Probes P1,P2,P3
P1=np.zeros(n_t);P2=np.zeros(n_t);P3=np.zeros(n_t);
# number of velocity points 
#(from Plot_Field_TEXT_Cylinder)
nxny=int(n_s/2) 
		
for k in range(0,n_t):
 # Name of the file to read
 Name=FOLDER+os.sep+'Res%05d'%(k+1)+'.dat' 
 # Read data from a file
 # Here we have the two colums
 DATA = np.genfromtxt(Name,usecols=np.arange(0,2),max_rows=nxny+1) 
 Dat=DATA[1:,:] # Remove the first raw with the header
 # Get U component and reshape as grid
 V_X=Dat[:,0]; V_X_g=np.reshape(V_X,(n_y,n_x))
 # Get V component and reshape as grid
 V_Y=Dat[:,1]; V_Y_g=np.reshape(V_Y,(n_y,n_x))
 # Get velocity magnitude
 V_MAGN=np.sqrt(V_X_g**2+V_Y_g**2)
 # Sample the three probes
 P1[k]=V_MAGN[i1,j1];P2[k]=V_MAGN[i2,j2];P3[k]=V_MAGN[i3,j3];
 print('Loading Step '+str(k+1)+'/'+str(n_t)) 
 \end{lstlisting}
\end{centering}

Figure \ref{fig2} shows the signals in P1 (on the left) and P3 (on the right), together with their histogram and their power spectral densities. The meaning of these quantities will be evident by the end of the lecture. For the moment, it suffices to notice that this test case is characterized by a large scale variation of the free-stream velocity. 

This is evident from the velocity evolution of both probes, but it is clearer in P1 because this probe is sufficiently far from the cylinder wake to capture the evolution of the free-stream flow. In the first $1.5s$, the free stream velocity is at approximately $U_\infty\approx12\mbox{m/s}$. During this time, the flow is stationary (more about this in the next session).

Between $t=1.5\mbox{s}$ and $t=2.5\mbox{s}$ the velocity drops down to $U_\infty\approx8\mbox{m/s}$, reaching a second stationary phase that last until the end of the acquisition. The variation of the flow velocity is sufficiently low to let the vortex shedding adapt and hence preserve an approximately constant Strouhal number of $St=f d/ U_{\infty}\approx 0.19$, with $d=5mm$ the diameter of the cylinder. Consequently, the vortex shedding varies from $f\approx 459 Hz$ to $f\approx 303 Hz$. The two peaks in the PSD in probes P3 clearly shows both frequencies, although a frequency analysis does not allow to identify \emph{when} these occur. This will also be briefly addressed later. The same peaks are much weaker but still present in P1.

In the histogram of both probes, two distributions are visible. These corresponds to the first ($t_k<1$) and the second ($t_k>2.5$) stationary conditions. The next section is dedicated to statistical analysis time series like these.

\begin{center}%
	
	\begin{minipage}{.49\linewidth}
		\hspace{3.7cm}Probe 1\\
		\includegraphics[width=1\linewidth]{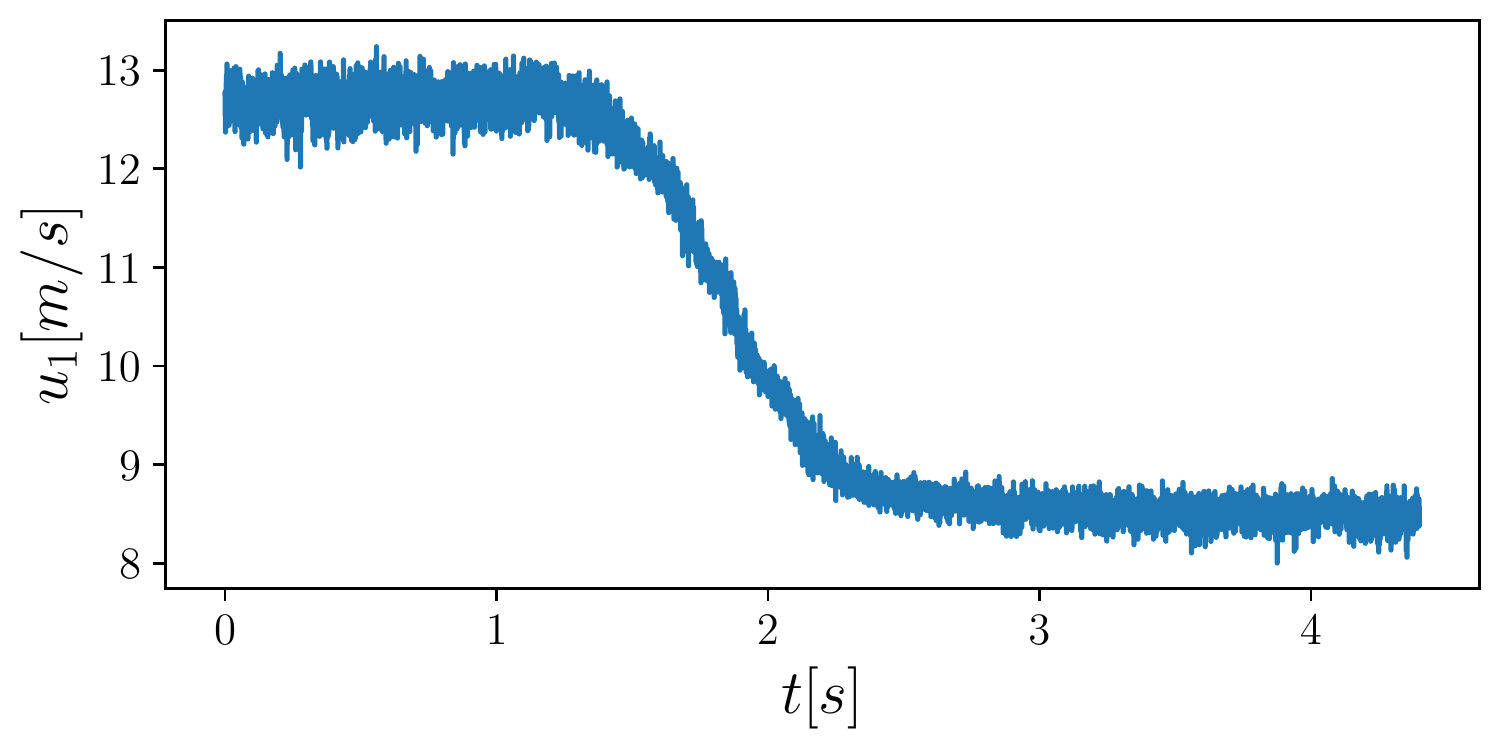}\\
		\includegraphics[width=1\linewidth]{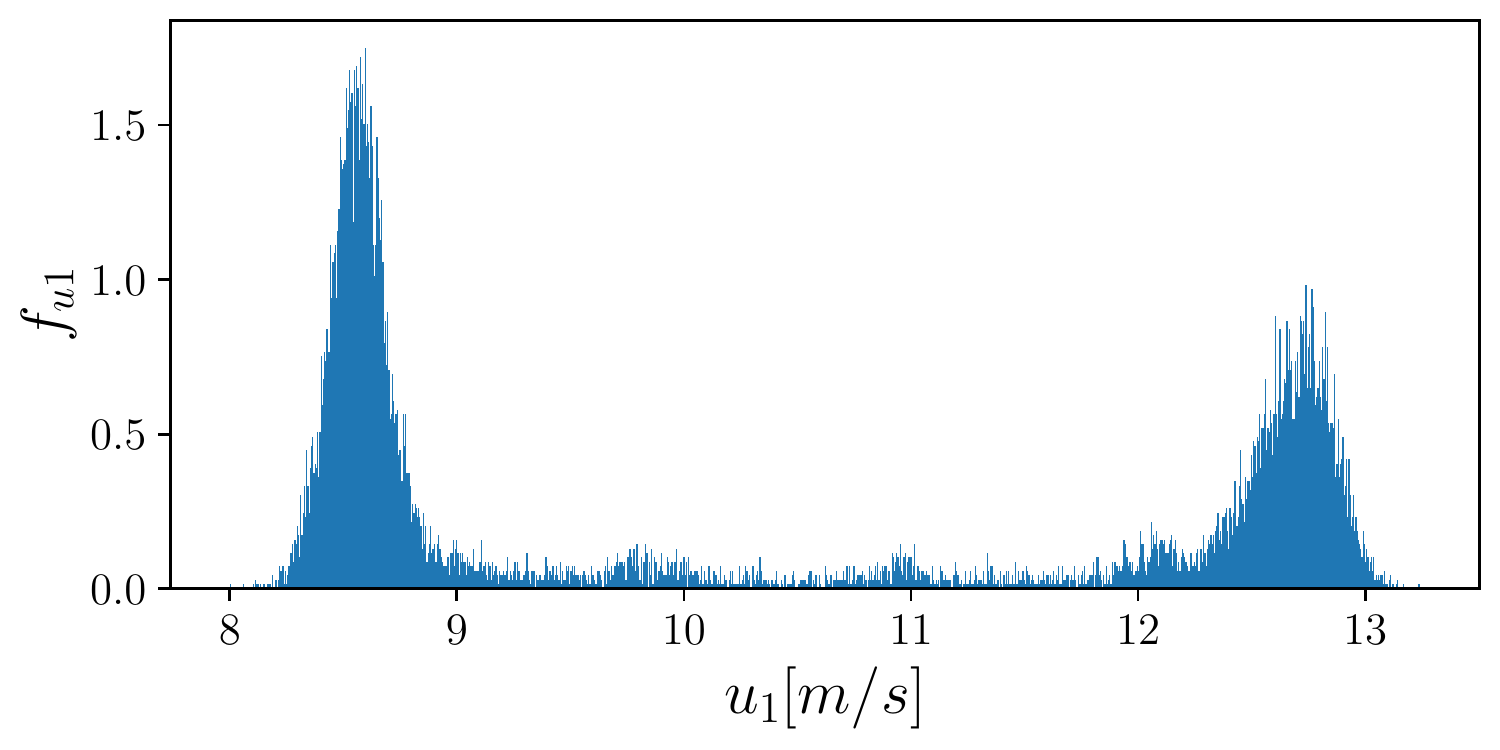}\\
		\includegraphics[width=1\linewidth]{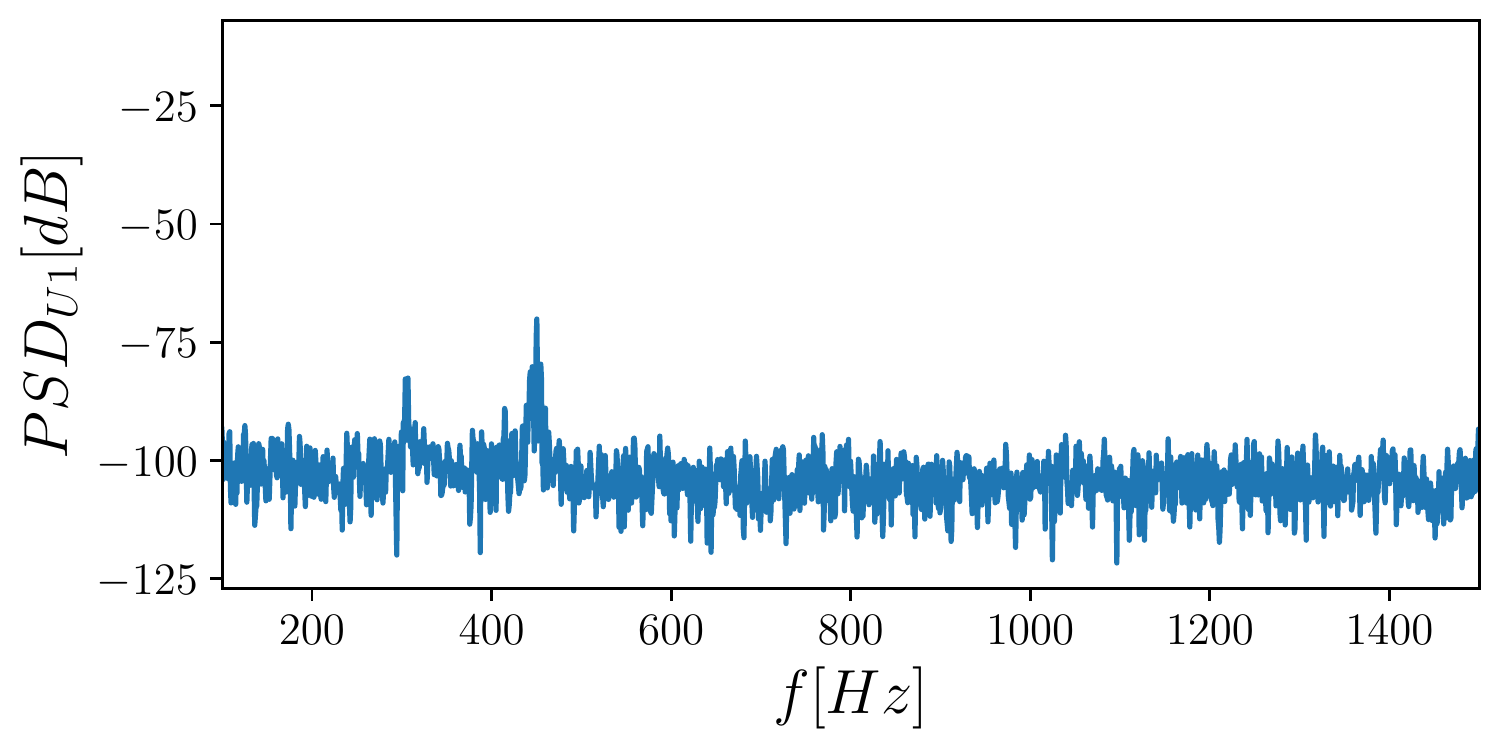}
	\end{minipage}
	\begin{minipage}{.49\linewidth}
		\hspace{3.7cm} Probe 3\\
		\includegraphics[width=1\linewidth]{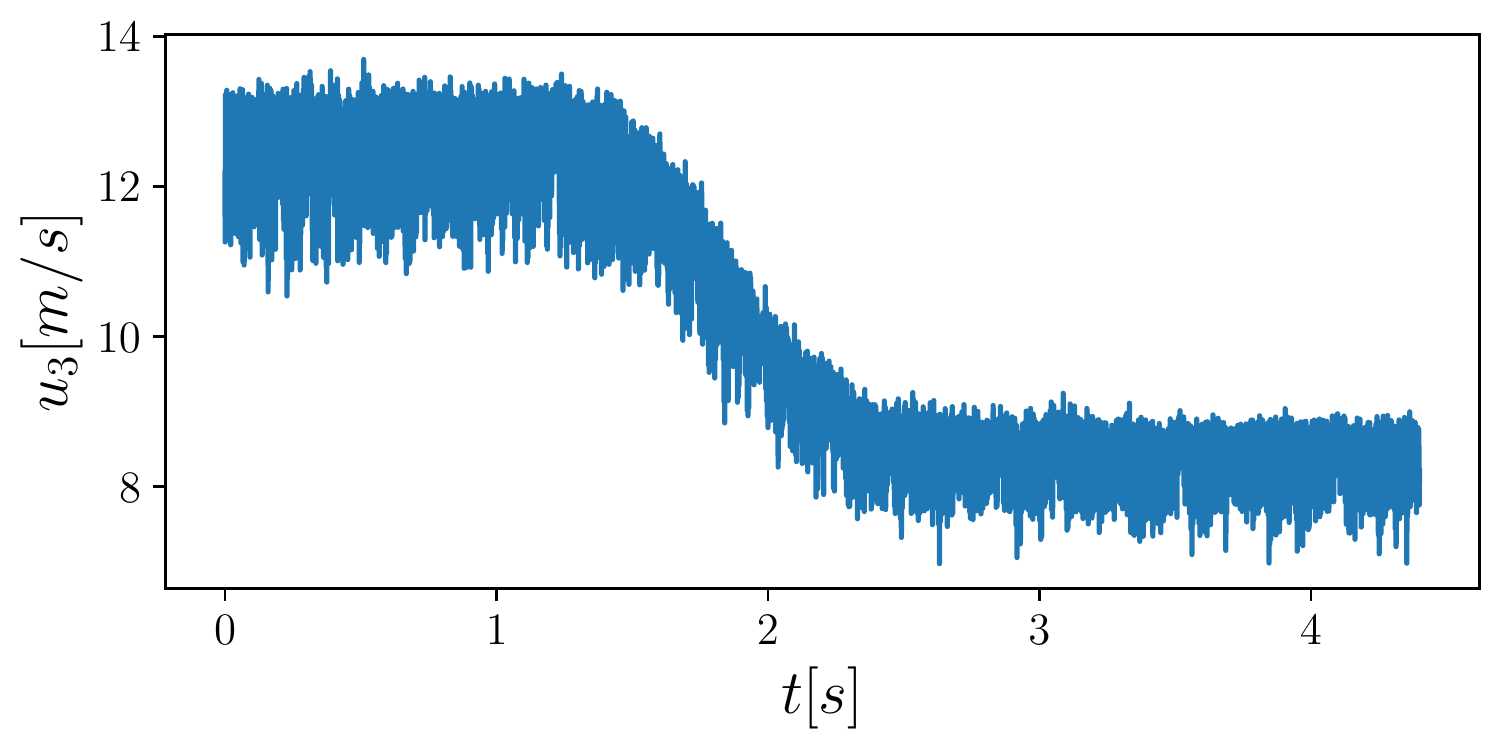}\\
		\includegraphics[width=1\linewidth]{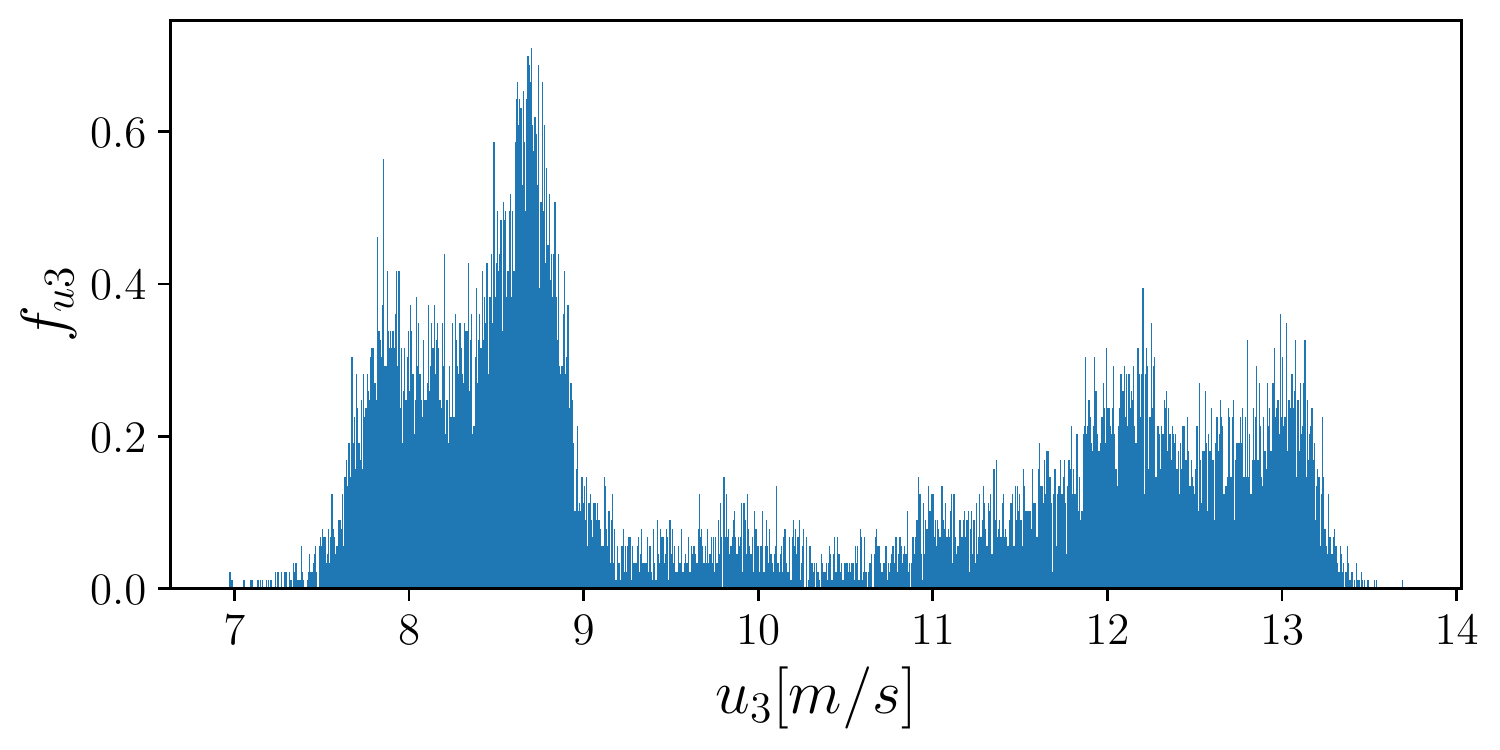}\\
		\includegraphics[width=1\linewidth]{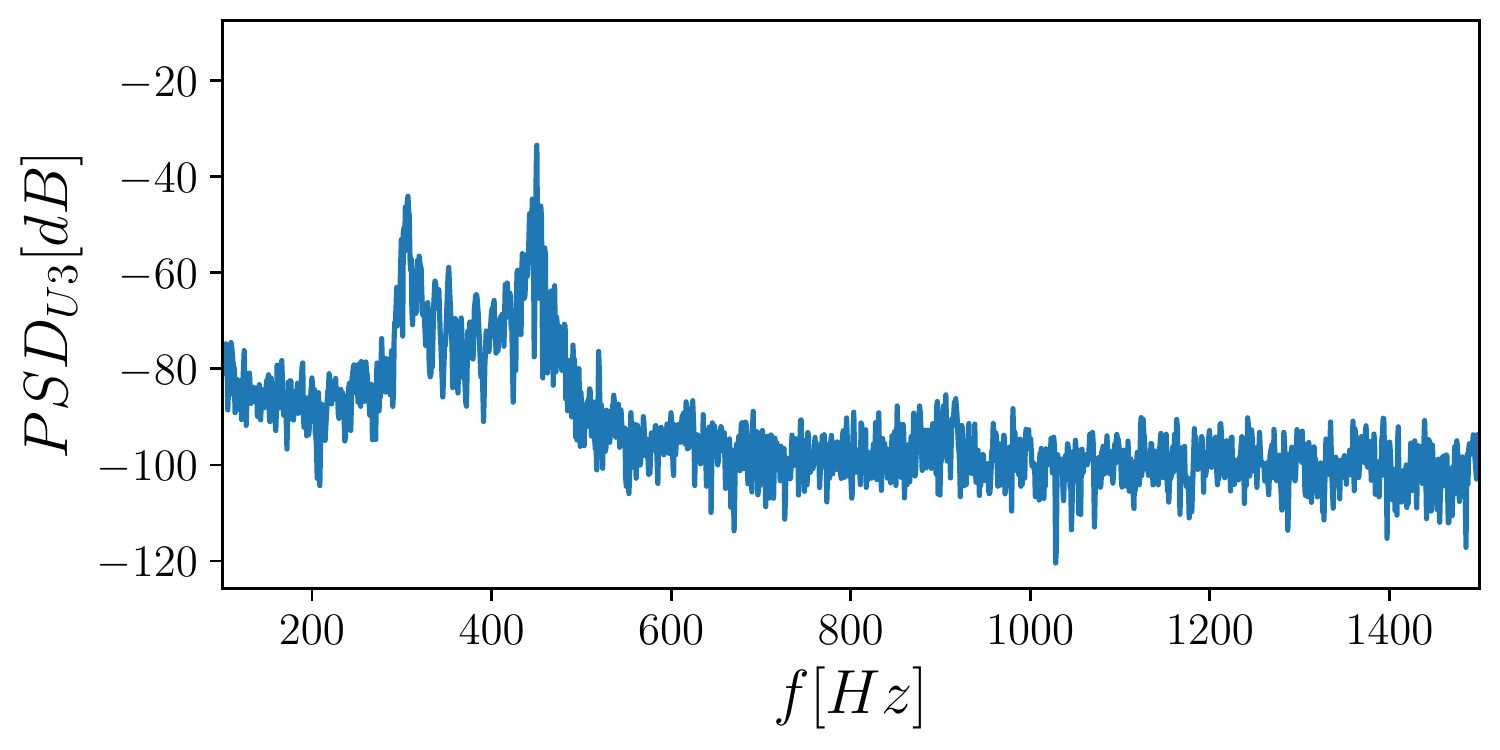}
	\end{minipage}\\
	\vspace{0.1cm}
	\captionof{figure}{	Velocity time series in probes P1 (on the left) and P3 (on the right) for the TR-PIV dataset analyzed in this lecture. See Figure \ref{fig1} for the probe location. }	
	\label{fig2}
\end{center}


\section{Statistics}\label{sec3}

This section presents some basic concepts and tools to study time series. We introduce the notion of random processes and random variables in the subsection \ref{sec2p1}, along with the relevant first and second-order statistics. As a first exercise, we study the ensemble statistics of an Ornstein Uhlenbeck process before recalling the notion of stationarity in the strong and weak sense.

The subsection \ref{sec2p2} introduce the notion of ergodicity and the time statistics for finite duration signals. These are linked, in \ref{sec2p3}, to fundamental statistical quantities characterizing turbulent flows, further discussed in the second exercise.

Finally, \ref{sec2p4} completes the introduction by reviewing the spectral characterization of stochastic signals. The review thus includes the notion of power spectral density and coherency explored in the third exercise.

The material covered falls within well-known textbooks in statistical inference and signal processing. Excellent references with a focus on deterministic signals are \cite{Oppenheim1996a,Ljung1994,Hsu2013,Hayes2011}, while \cite{DFT_Smith,FIR_Smith,Spectral_Smith} provides a comprehensive treatment of Fourier analysis and filters.
A gentle introduction to stochastic signals is provided by \cite{Kay1993,Williamson1999,Oppenheim2015} while \cite{Wang2009} provides an excellent review of orthogonal transforms.

\subsection{First and Second Order Statistics}\label{sec2p1}

Each of the $n_x=71\times n_y=30$ grid points in the TR-PIV measurements generates a time series of $n_t=13200$ equally spaced samples in time. We will use the notation $u_n(t_k)$ to refer to the time series generated at the $n-th$ spatial location, with $n\in [0,n_x\, n_y]$ and $t_k=k\Delta t$ the discrete time axis. Denoting as $\mathbf{x}=(x_i,y_j)$ the Cartesian grid from which the PIV data is available, with $i\in[0,n_y-1]$ and $j\in[0,n_x-1]$, we can see $n$ as a matrix linear index scanning all the points of the grid. Accordingly, we use $\mathbf{x}_n$ to denote the location of the probe.

Unless otherwise stated, the time series $u_n$ under analysis is the velocity magnitude at that location\footnote{With a little overload of the notation, we use $u_n$ to refer to the time series and $u$ to denote the first vector component.} $\mathbf{x}_n$: $$u_n(t_k)=\sqrt{u(\mathbf{x}_n,t_k)^2+v(\mathbf{x}_n,t_k)^2}\,.$$

In a statistical treatment of signals, we admit that it is not possible (or simply not convenient) to describe signals with a deterministic model: we will not be able (nor we want to attempt) to predict the \emph{exact} value $u_n(t_k)$ observed at a certain time. Instead, we will be content with the ability to attribute a \emph{probability} to $u_n(t_k)$.

More concretely, we treat a time series as the realization of a \emph{random process}, and we treat the value that the time series takes at a specific time as a \emph{random variable}. A random process can be seen as a family of jointly distributed random variables indexed by the time $t_k$.
Let $U_1, U_2,\dots U_{n_t}$ denote the set of random variables associated to the value of $u_n(t_1),u_n(t_2), \dots u_n(t_{n_t})$. A full probabilistic characterization requires the joint probability density function of $n_t$ random variables:

\begin{equation}
	\label{pdfS}
	f_{U_1, U_2,\dots U_{n_t}}(u_n(t_1),u_n(t_2), \dots u_n(t_{n_t}))\,\,.
\end{equation}

This is a $n_t$-order description, and it is clearly impractical in most cases of interest. It is also not particularly interesting because we are usually not interested in computing the probability that a specific realization of $n_t$ samples occurs.

On the other hand, we are always interested in a first order description. In other words, we seek to predict the probability of observing a certain $u_n(t_k)$ at a certain time $t_k$. This is a 'one-point statistics', governed by the univariate probability density function (pdf) $f_{U}(u_n(t_k))$ associated to the random variable $U_n$. This pdf allows to assign the probability that $u_m\leq u_n(t_k)<u_M$:

\begin{equation}
	\label{eq_uni}
	P\{ u_m\leq u_n(t_k)<u_M\}=\int^{u_M}_{u_m} f_{U_k}(U_k)dU_k\,.
\end{equation} 

Given this distribution, the mean (expectation) of the time series at $t_k$ is:

\begin{equation}
	\label{mu_x}
	\mu_{U_n}(t_k)=\mathbb{E}_{\sim E}[u_n(t_k)]=\langle u_n(t_k)\rangle_E =\int^{\infty}_{-\infty} U_k f_{U_k}(U_k) dU_k\,,
\end{equation} having introduced the expectation operator $\mathbb{E}$ and the short hand notation with brackets $\langle \rangle$. Similarly, the variance var$(u_n(t_k))=\sigma^2_{U_k}(t_k)$ of the time series at time $t_k$ is:

\begin{equation}
	\label{sigma_x}
	\sigma^2_{U_k}(t_k)=\mathbb{E}_{\sim E}[(u_n(t_k)-\mu_{U_n}(t_k))^2]= \int^{\infty}_{-\infty} (U_k-\mu_{U_n}(t_k))^2 f_{U_k}(U_k) dU_k\,,
\end{equation} where $\sigma_{U_k}(t_k)=\sqrt{\mbox{var}(u_n(t_k))}$ is the standard deviation, also referred to as the root mean square of $u_n(t_k)$.

In a second order description of the time series, we might be interested in the statistical links between the value that the time series takes at time $t_k$ and the values it takes at time $t_j$. The expectation of the product $u_n(t_k) u_n(t_j)$ is the all important \emph{autocorrelation function}, whose definition requires the (bivariate) joint probability function $f_{U_k, U_j}(U_k, U_j)$:

\begin{equation}
	\label{Auto_corr}
	R_{UU}(t_k,t_j)=\langle u_n(t_k) u_n(t_j)\rangle_E=\int^{\infty}_{-\infty}\int^{\infty}_{-\infty}U_k\, U_j\,f_{U_k, U_j}(U_k, U_j) dU_k dU_j\,.
\end{equation}

Similarly, one can define the \emph{autocovariance function} as 

\begin{equation}
	\label{Auto_cov}
	C_{UU}(t_k,t_j)=\langle [u_n(t_k)-\mu_{U_n}(t_k)][ u_n(t_j)-\mu_{U_n}(t_j)]\rangle_E=R_{UU}(t_k,t_j)-\mu_{U_n}(t_k) \mu_{U_n}(t_j)\,.
\end{equation}

It is easy to note that $C_{UU}(t_k,t_k)=\sigma^2(t_k)$. 

Finally, we might be interested in the statistical link between two different time series; for examples the ones generated at two different locations. Denoting these as $u_n(t_k)$ and $u_l(t_j)$ we recall the celebrated \emph{cross-correlation} (the engine of PIV analysis!), which requires the (bivariate) joint probability function $f_{U_k, W_j}({U_k, W_j})$ where $U_k$ is the random variable associated to $u_n(t_k)$ and $W_j$ is the one associated to $u_l(t_j)$. We have: 

\begin{equation}
	\label{Cross_Co}
	R_{UW}(t_k,t_j)=\langle u_n(t_k) u_l(t_j)\rangle_E=\int^{\infty}_{-\infty}\int^{\infty}_{-\infty}U_k W_j\,f_{U_k, W_j}({U_k, W_j}) dU_k dW_j\,\,
\end{equation} and the cross-covariance is simply

\begin{equation}
	\label{Cross_cov}
	C_{UW}(t_k,t_j)=\langle [u_n(t_k)-\mu_{U_n}(t_k)][ u_l(t_j)-\mu_{U_l}(t_j)]\rangle_E=R_{UW}(t_k,t_j)-\mu_{U_n}(t_k) \mu_{U_l}(t_j)\,.
\end{equation}

All the above quantities are difficult to evaluate because none of the invoked distributions is usually known. To compute these important statistics, one must then also estimate the underlying distributions, and the only way to do this is to sample many instances of the time series. The set of instances we will be working with is called \emph{ensamble} and the expectation operator should now be generalized to the notion of \emph{ensamble averaging}. Let us review this concept in some detail for the mean in \eqref{mu_x}; we leave the extensions of the others statistics as an exercise. Let us assume that a total of $n_r$ realizations has been collected (each being of length $n_t$ and indexed by $t_k$) and let $u_n(t_k;r)$ be the $r-th$ realization at time $t_k$. 

Adopting a frequentist approach to probabilities, we can compute the probability that the value $u_n(t_k)$ is taken at $t_k$ as $f_V(u_n(t_k),t_k)=n_u/n_r$, where $n_u$ is the number of times such value has been observed at time $t_k$ within the full ensemble\footnote{A trivial example might help understanding. Compute the mean of the random variable $\mathbf{x}=[2,2,4,3,3]$. Using the a frequentist approach, we note that this variable only takes three values ($2,4,3$) with probabilities $p_j=[2/5,1/5,2/5]$. Then the expectation is $$\mu_X=\sum^{3}_{j=1}x_j\,p_j =\frac{2}{5}\,2+\frac{1}{5}\,4+\frac{2}{5} 3\,.$$ This is no different than the simple arithmetic average in \eqref{mu_discrete}.}.

Then, replacing the integral with a summation leads to 

\begin{equation}
	\label{mu_discrete}
	\mu_{Un}(t_k)=\langle u(t_k)\rangle_E(t_k)=\sum_{r=0}^{n_r-1} u_n(t_k;r) f_{U_k}(u_n(t_k))=\frac{1}{n_r} \sum^{n_r}_{j=0}u_n(t_k; r)\,.
\end{equation} 

For reasons that will soon become obvious, it is convenient to construct a snapshot matrix $\bm{U}_n\in\mathbb{R}^{n_t\times n_r}$ with every column collecting a realization of length $n_t$ of the stochastic process in the probe located at $n$, and every row containing the $n_r$ realizations at time $t_k$. Every realization, at the moment, must be thought of as a different experiment. With this arrangement of the data, the ensemble average of the series is the average across the rows of $\bm{U}_n$. This is equivalent to multiplying the matrix by a vector of length $n_t$ containing all $1/n_r$ terms. Similarly, the ensemble variance at all times can be computed along all rows of  $\bm{U}_n$. The computation of both quantities is shown in the first exercise.

Moving to second order statistics, the computation of the autocorrelation function requires averaging all products $u_n(t_k)u_n(t_j)$. The ensamble autocorrelation in the matrix formalism becomes\footnote{A note of warning: in statistics, the correct normalization would be $1/(n_r-1)$ and not $1/n_r$. At sufficiently large $n_r$ this distinction is not relevant; the reader is referred to \cite{Kay1993} for more.}

\begin{equation}
	\label{Auto_corr_E}
	R_{UU}(t_k,t_j)=\langle u_n(t_k),u_n(t_j)\rangle_E=\frac{1}{n_r }\bm{U}^k_n\,(\bm{U}^j_n)^T\,,
\end{equation} where $\bm{U}^k_n\in\mathbb{R}^{1\times n_r}$ is the \emph{row vector} collecting all the realizations $u_n(t_k;r)$ at time $t_k$, the superscript $^T$ denotes transposition. This is essentially an inner product. The autocovariance is simply $C_{UU}(t_k,t_j)=R_{UU}(t_k,t_j)-\mu_U(t_k)\mu_U(t_j)$, and the reader should notice that $C_{UU}(t_k,t_k)=\sigma^2_U(t_k)$. 

The cross correlation between two variables is analogous to \eqref{Auto_corr_E}: denoting as $\bm{U}_n\in\mathbb{R}^{n_t\times n_r}$ and $\bm{W}_n\in\mathbb{R}^{n_t\times n_r}$ the snapshot matrices collecting the sets of $n_r$ ensambles for the two random processes, the cross correlation reads

\begin{equation}
	\label{cross_corr}
	R_{UW}(t_k,t_j)=\langle u_n(t_k),w_n(t_j) \rangle_E=\frac{1}{n_r }\bm{U}^k_n\,(\bm{W}^j_n)^T\,.
\end{equation} 

Finally, it is worth noticing that in many fields, the definitions of $C_{UU}$, $C_{UW}$ are normalized so that their value is in the range $[-1,1]$. The normalized quantities are distinguished with a hat and read:

\begin{equation}
	\label{norm_cov}
	\hat{C}_{UU}(t_j,t_k)=C_{UU}(t_j,t_k)/\sigma^2_U(t_k) \quad \quad \hat{C}_{UW}(t_j,t_k)=C_{UW}(t_j,t_k)/(\sigma_U(t_j)\sigma_W(t_k))\,.
\end{equation}

As a note of warning, the reader should note that in a vast portion of the literature in experimental fluid mechanics, statistics are computed on the `fluctuating' component of the velocity fields, following a classic Reynolds decomposition \citep{Pope2000} (see also Section \ref{sec2p4}). Therefore, the notion of autocovariance and autocorrelation are used interchangeably.

It is now worth pausing and practice with an exercise.

\medskip

\begin{tcolorbox}[breakable, opacityframe=.1, title=Exercise 1: The statistics of an Ornstein-Uhlenbeck Process]
	
	Consider the Ornstein-Uhlenbeck process governed by the following stochastic differential equation (see \cite{Oksendal1998,Borodin2002,Pope2000})
	
	\begin{equation}
		\label{eqXX}
		d X_t= \kappa (\theta-X_t) dt+\sigma d W_t
	\end{equation}
	
	where the subscript $t$ here refers to a time step, $\kappa>0$ is called rate of reversion and controls how quickly the process reaches its (stationary) long-term behaviour, characterized by mean $\theta$ and random fluctuation with standard deviation $\sigma$. The second term $W_t$ is a stochastic term, and it is here taken as a normal Gaussian with zero mean and unitary standard deviation, here written as $\mathcal{N}(0,1)$. The first term is often referred to as \emph{drift}, the second as \emph{diffusion}.
	
\hspace{2mm} Consider a case with initial condition $X_0=0$, $\kappa=1.2$, $\theta=2$ and $\sigma=1$. Consider $n_t=1001$ samples, with a sampling of $\Delta t=0.01$ s. This corresponds to a physical observation time of $T=10 s$. Plot four realizations to explore the process. 
	
\hspace{2mm} The questions are two. (1) compute the ensemble mean, the ensemble standard deviation and the ensemble autocovariance as a function of time using $n_r=100$. (2) Study the convergence of these statistical quantities at two time steps (say at $k=10$ and $k=700$) as a function of the number of samples in the ensemble.

	\medskip
	\textbf{Solution}.  Let us begin by creating a function that produces a sample of the process, taking as input the four process parameters $\kappa,\theta,\sigma$ and vector of times $t_k$. Using a simple loop, the following function does the job (see Python exercise 1):

	\begin{centering}
		\begin{lstlisting}[language=Python,linewidth=15.5cm,xleftmargin=.05\textwidth,xrightmargin=.05\textwidth,backgroundcolor=\color{yellow!10}]
import numpy as np
# Function definition
def U_O_Process(kappa,theta,sigma,t):
 n_T=len(t) 
 # Initialize the output
 y=np.zeros(n_T)
 # Define Drift and Diffusion functions in the process
 drift= lambda y,t: kappa*(theta-y)
 diff= lambda y,t: sigma
 noise=np.random.normal(loc=0,scale=1,size=n_T)*np.sqrt(dt)
 # Solve Stochastic Difference Equation
 for i in range(1,n_T):
   y[i]=y[i-1]+drift(y[i-1],i*dt)*dt+diff(y[i-1],i*dt)*noise[i]
 return y
		\end{lstlisting}
	\end{centering}
	
\hspace{2mm}The following script creates $n_r=500$ realizations and store them in a matrix $\bm{U}_n$ of size $n_r\times n_r$. Then, it plots four randomly chosen samples, namely the numbers $r=1,10,22,55$. The four realizations are shown in Figure \ref{a)}, with the plot axis being customized (see the provided codes).

	\begin{centering}
		\begin{lstlisting}[language=Python,linewidth=15.5cm,xleftmargin=.05\textwidth,xrightmargin=.05\textwidth,backgroundcolor=\color{yellow!10}]
import matplotlib.pyplot as plt
# Initial and final time
t_0=0; t_end=10
# Number of Samples
n_t=1001
# Process Parameters
kappa=1.2; theta=3; sigma=0.5
# Create the time scale
t=np.linspace(t_0,t_end,n_t); dt=t[2]-t[1]
# Collect 500 sample and store in U_N
n_r=500; U_N=np.zeros((n_t,n_r))
for l in range(n_r):
  U_N[:,l]=U_O_Process(kappa,theta,sigma,t)
  # Plot the results 
  plt.figure(1)
  plt.plot(t,U_N[:,1])
  plt.plot(t,U_N[:,10])
  plt.plot(t,U_N[:,22])
  plt.plot(t,U_N[:,55])
		\end{lstlisting}
	\end{centering}
	
\hspace{2mm} The reader should play with the code long enough to realize that this process is characterized by a transient time of the order of $0.3s$ within which all signals move from zero to a stationary conditions (a precise definition of stationary signals is yet to come). Having arranged the data into `snapshot matrices', the computation of the time average and the temporal standard deviation can be done in one line each:

	\begin{centering}
		\begin{lstlisting}[language=Python,linewidth=15.5cm,xleftmargin=.05\textwidth,xrightmargin=.05\textwidth,backgroundcolor=\color{yellow!10}]
U_Mean=np.mean(U_N,axis=1) # Ensamble Mean
U_STD=np.std(U_N,axis=1) # Ensamble STD
		\end{lstlisting}
	\end{centering}
	
\hspace{2mm}These are essentially `row-wise' statistics.
	
	\hspace{2mm}Figure \ref{b)} shows the time average together with the range $\mu_U(t_k)\pm \sigma_U(t_k)$. It appears that the standard deviation grows gently until it reaches a constant value after about $t_k>4s$.

\hspace{2mm}Finally, we analyze the ensamble autocorrelation of this random process. The ensamble cross-correlation between the time steps $t_j$ and $t_k$ for the ensambles $\mathbf{U}_n$ and $\mathbf{W}_n$ of two random variables (see \eqref{cross_corr}) can be combuted with the following function:
	
	\begin{centering}
		\begin{lstlisting}[language=Python,linewidth=15.5cm,xleftmargin=.05\textwidth,xrightmargin=.05\textwidth,backgroundcolor=\color{yellow!10}]
def Norm_Autocov_0(U_N,W_N,k,j):
 n_r,n_t=np.shape(U_N)
 # Select all realizations at time t_k for U
 U_N_k=np.expand_dims(U_N[k,:], axis=0);
 # Select all realizations at time t_kj for W
 W_N_k=np.expand_dims(W_N[j,:],axis=0)
 # Note (These are row vectors)
 # Compute the average products      
 PR=U_N_k.T.dot(W_N_k)
 R_UW=np.mean(PR)/(np.std(U_N_k)*np.std(W_N_k))   
 return R_UW
		\end{lstlisting}
	\end{centering}

	\begin{center}%
		\setcounter{subfigure}{0}%
		
		\begin{minipage}{.49\linewidth}
			\includegraphics[width=\linewidth]{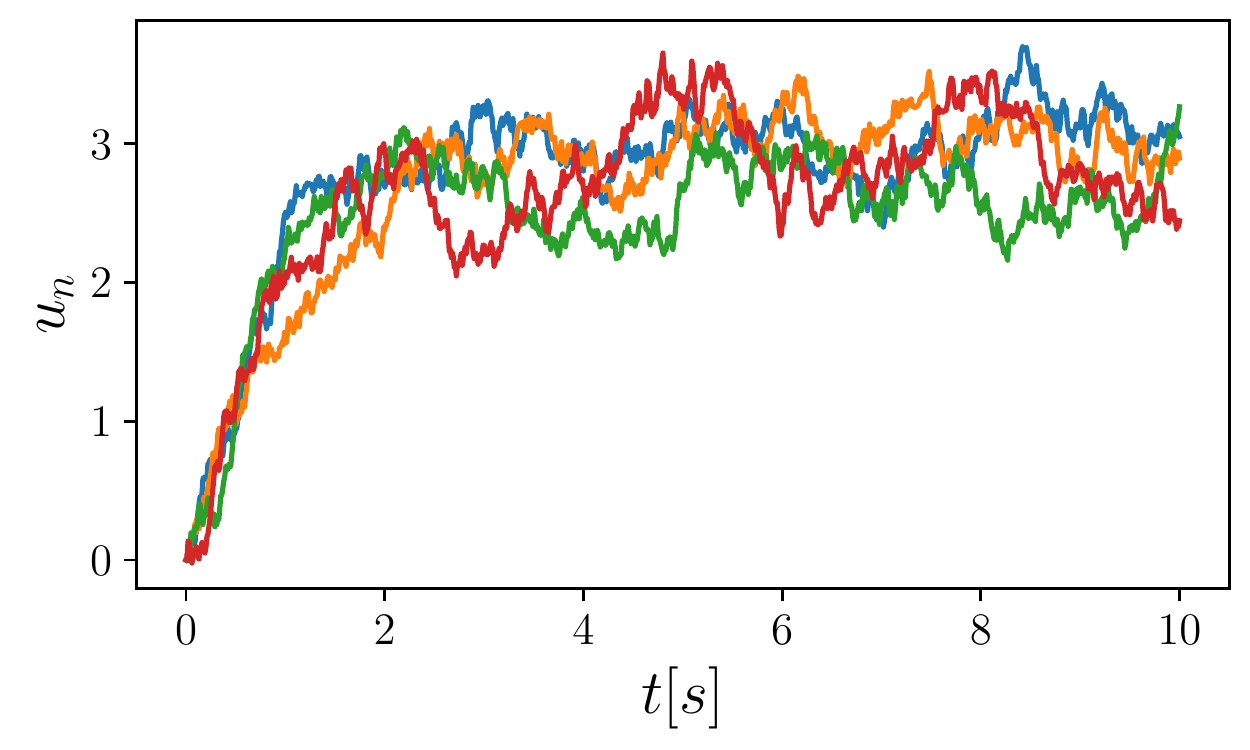}
			\label{a)}%
			\vspace{-3mm}
			\center{a)}
		\end{minipage}
		\begin{minipage}{.49\linewidth}
			\includegraphics[width=\linewidth]{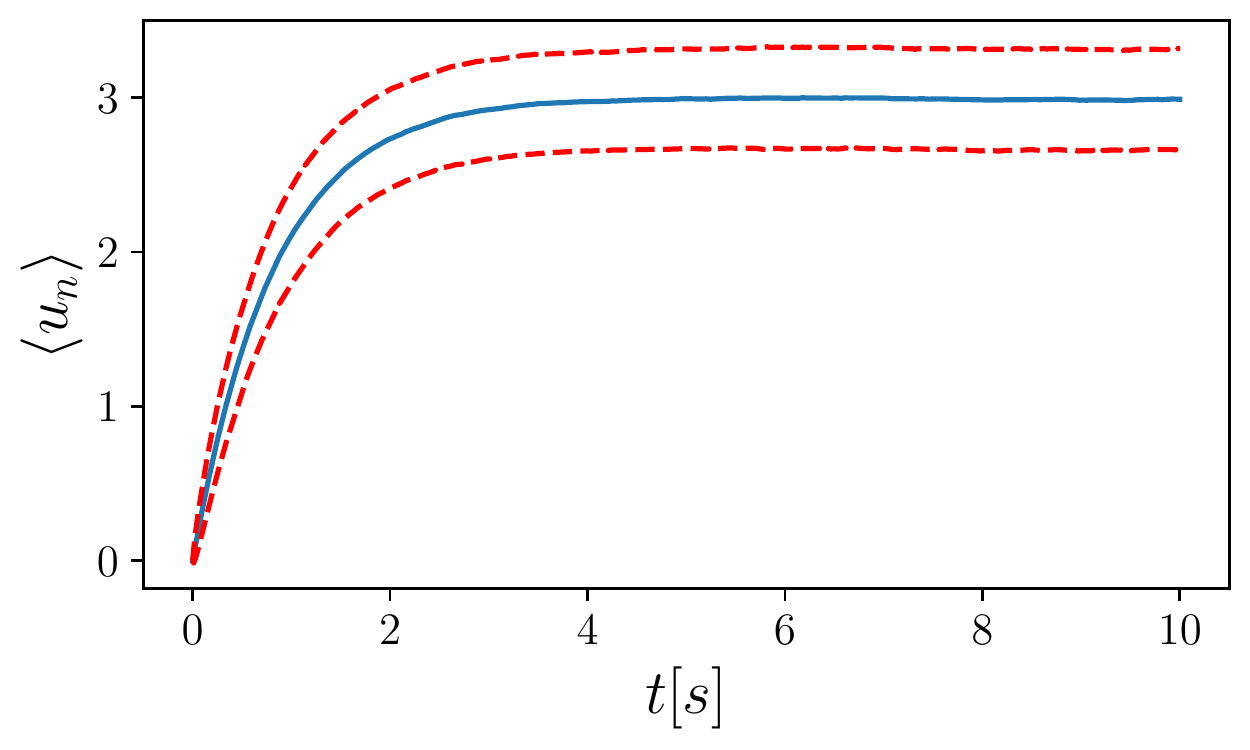}%
			\label{b)}%
			\vspace{-3mm}
			\center{b)}
		\end{minipage}	
		\vspace{-0.05cm}
		\captionof{figure}{	Fig (a): Four randomly chosen samples of the Ornstein- Uhlenbeck process; Fig (b): Ensamble average of the process (in blue), along with the $\mu+\sigma$ and $\mu-\sigma$ curves.}	
		\label{figEx1}
	\end{center}
	
	Having added a normalization. Then, the following script computes the autocorrelation of 100 randomly pairs, separated by a lag of $50$ samples:
	
	\begin{centering}
		\begin{lstlisting}[language=Python,linewidth=15.5cm,xleftmargin=.05\textwidth,xrightmargin=.05\textwidth,backgroundcolor=\color{yellow!10}]
# Define lag (in number of samples)
lag=50 
# Study the autocorrelation between two points at equal lags
N_S=100; R_UW=np.zeros(N_S)
# Select a 100 random points i (larger than 500)
J=np.random.randint(500,800,N_S); K=J+50
for n in range(N_S):
  R_UW[n]=Norm_Autocov_0(U_N,U_N,J[n],K[n])
		\end{lstlisting}
	\end{centering}
	
\hspace{2mm}The resulting set of autocorrelations has a mean of $0.545$ and a standard deviation of $0.0059$: in other words, it does not matter what the exact pair $j,k$ is, as long as they differ by the same lag (in this case, 50). This is because here we are sampling in time intervals from 500 to 800, and here the process has reached its stationary condition. The reader is encouraged to repeat the exercise at an earlier interval. 
	
\hspace{2mm}Finally, we analyze the convergence of the statistics as a function of the number of realizations. The following script computes the mean and the standard deviation at $k=10$ and $k=700$ for a number of realizations that goes from $1$ to $1000$. 
	
	\begin{centering}
		\begin{lstlisting}[language=Python,linewidth=15.5cm,xleftmargin=.05\textwidth,xrightmargin=.05\textwidth,backgroundcolor=\color{yellow!10}]
n_R=np.round(np.logspace(0.1,3,num=41))
# Prepare the outputs at k=100
mu_10=np.zeros(len(n_R))
sigma_10=np.zeros(len(n_R))
# Prepare the outputs at k=700
mu_700=np.zeros(len(n_R))
sigma_700=np.zeros(len(n_R))
# Loop over all n_R's.
for n in range(len(n_R)):
 # show progress
 print('Computing n='+str(n)+' of '+str(len(n_R)))
 n_r=int(n_R[n]) # Define the number of ensambles
 U_N=np.zeros((n_t,n_r)) # Initialize the ensamble set
 for l in range(n_r): # Fill the Ensamble Matrix
   U_N[:,l]=U_O_Process(kappa,theta,sigma,t)
   # Compute the mean and the std's
   mu_10[n]=np.mean(U_N[10,:]) # Ensamble Mean
   sigma_10[n]=np.std(U_N[10,:]) # Ensamble STD
   mu_700[n]=np.mean(U_N[700,:]) # Ensamble Mean
   sigma_700[n]=np.std(U_N[700,:]) # Ensamble STD		
		\end{lstlisting}
	\end{centering}
	
\hspace{2mm}	By analyzing the vector n\_R in the script, the reader should note that some of the entries with a small number of samples are taken multiple times to show the variability in the prediction. The results are shown in Figure \ref{figEx1b}.
	
\hspace{2mm}	It can be shown that the convergence of both the mean and the standard deviation is $\propto \sqrt{n_r}$, but the convergence of the mean is $\propto \sigma_U$ while the convergence of the standard deviation is $\propto \sigma^2_U$ (see also \cite{AndreaLS}): this explains why a larger number of samples is needed to converge the second-order statistics, and why the convergence at $k=100$ is slower. The reader is encouraged to explore the provided scripts further, to observe this result in action at different points.

	\begin{center}%
		\setcounter{subfigure}{0}%
		
		\begin{minipage}{.49\linewidth}
			\includegraphics[width=\linewidth]{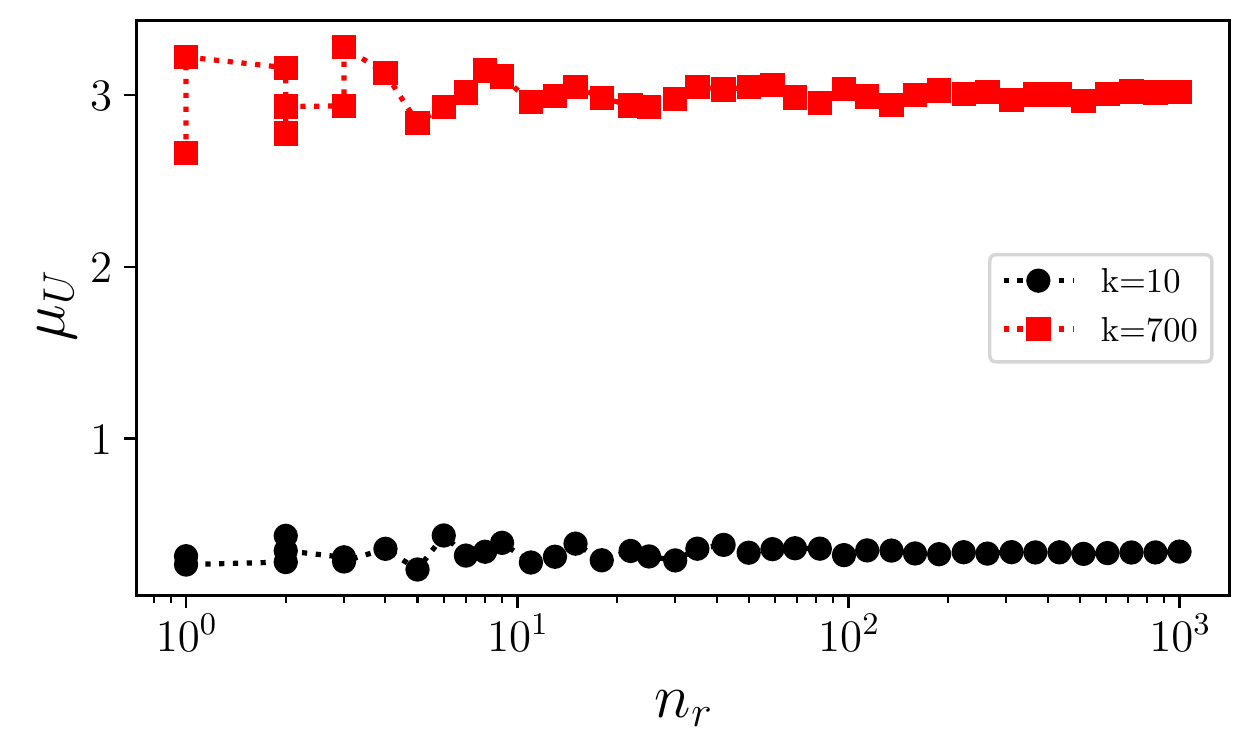}
			\label{2a)}%
			\vspace{-3mm}
			\center{a)}
		\end{minipage}
		\begin{minipage}{.49\linewidth}
			\includegraphics[width=\linewidth]{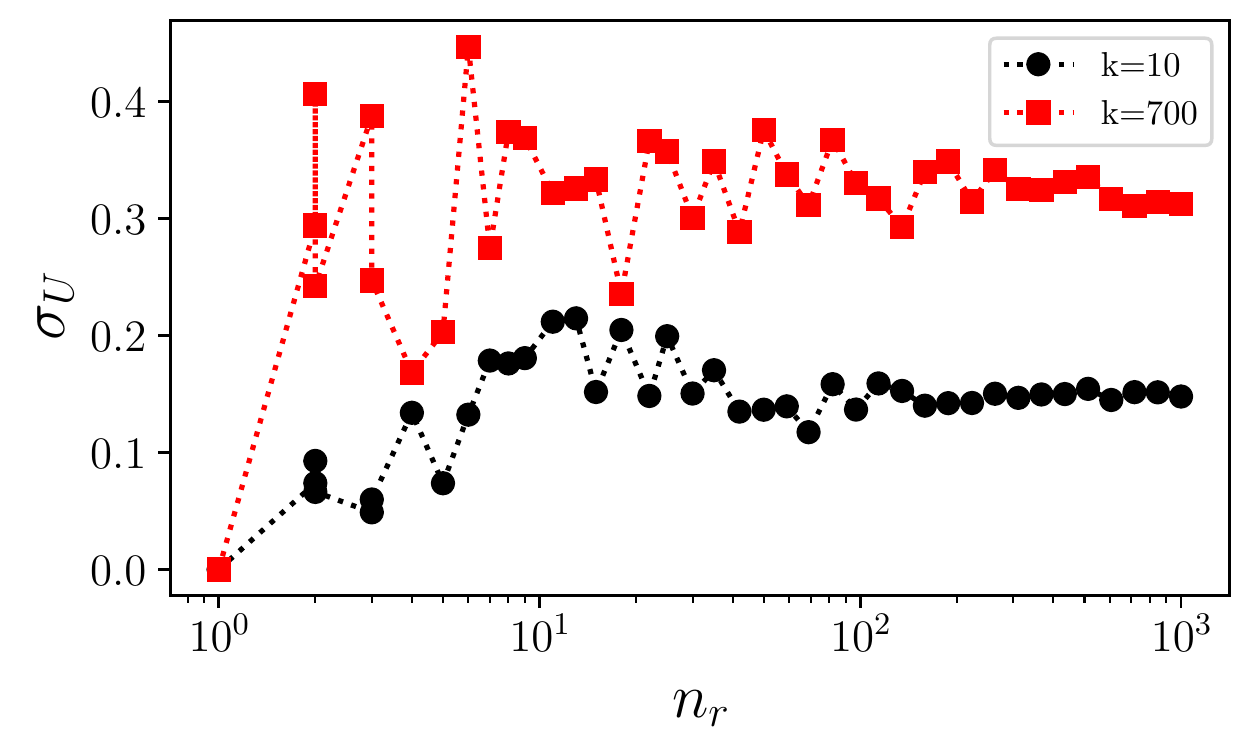}%
			\label{2b)}%
			\vspace{-3mm}
			\center{b)}
		\end{minipage}	
		\vspace{-0.05cm}
		\captionof{figure}{	Fig (a): Four randomly chosen samples of the Ornstein- Uhlenbeck process; Fig (b): Ensamble average of the process (in blue), along with the $\mu+\sigma$ and $\mu-\sigma$ curves.}	
		\label{figEx1b}
	\end{center}

\end{tcolorbox}

We close this section with the definition of stationarity. A process is \emph{stationary in a strict sense} (SSS) if \emph{all} its statistics are invariant to a shift of the origin. In other words, we could add any $t_0$ to the times $t_1, t_2, \dots t_{n_t}$ in \eqref{pdfS} and still have the same distribution. A less restrictive concept is that of a process which is \emph{stationary in a weak sense} (WSS). In this case, only the first and second order statistics are time invariant. As a result, the ensamble mean is constant, i.e. $\mu_{Un}(t_k)=\mathbb{E}[u_n(t_k)]=\mu_U \in\mathbb{R}$, and the autocorrelation/autocovariance are only function of the time delay between the two points considered: $R_{UU}(t_k,t_k+\tau)=R_{UU}(\tau)$ and $C_{UU}(t_k,t_k+\tau)=C_{UU}(\tau)$ $\forall t_k$.

\subsection{Ergodicity and Finite Durations}\label{sec2p2}

Ensemble averaging requires repeating experiments hundreds of times. This is not feasible. However, because stationary signals' statistics are time-invariant, it is reasonable to wonder whether a sufficiently long experiment can be statistically equivalent to many short ones.

This brings us to the notion of ergodicity. This is a sophisticated concept (see also \cite{Lumley1,Oppenheim2015,Bruun1995}), but the essential idea can be explained as follows: a process is ergodic if ensemble statistics can be replaced by temporal statistics on any particular realization\footnote{ It is worth noticing that a process might be ergodic for a given statistic but not with others (e.g. a process might be ergodic in the mean but not in the autocorrelation). We will not dig deeper (see \cite{Kay1993,Oppenheim2015} for more).}. We avoid introducing the statistics in the continuous domain and move straight to the statistics for a discrete sequence like the ones encountered in practice. The time average is defined as 

\begin{equation}
	\tilde{\mu}_U=\langle u_n(t_k)\rangle_T=\lim_{n_t \to \infty}\frac{1}{n_t}\sum^{n_t-1}_{k=0} u_n(t_k)\,,
\end{equation}

while the temporal autocorrelation is 

\begin{equation}
	\label{auto_corr_T}
	\tilde{R}_{UU}(\tau_l)=\langle u_n(t_k),u_n(t_k+\tau_l)\rangle_T=\lim_{n_t \to \infty}\frac{1}{2 n_t-1}\sum^{n_t-1}_{k=-n_t+1}u_n(t_k)u_n(t_k+\tau_l)\,.
\end{equation}

The expectation is now taken in time (i.e. $\mathbb{E}_{\sim T}\{\}$). A subscript $T$ is used to distinguish from the expectation in the ensemble set, and a tilde is used to distinguish the temporal statistics from the ensemble ones. While the definition of the time average is rather obvious, the one of autocorrelation deserves a deeper discussion.

Both quantities converge at sufficiently large $n_t$ for a stationary signal while none of them is meaningful for non-stationary signals. We come back to the convergence problem with an exercise at the end of this section. For the moment, the main concern in the reader's mind should be a practical one: even assuming that $n_t$ is large enough (hence the limit can be dropped), eq \eqref{auto_corr_T} cannot be computed for finite duration signals unless additional information is provided.

\begin{figure}[htbp]
	\centering
	\includegraphics[keepaspectratio=true,width=0.7 \columnwidth]
	{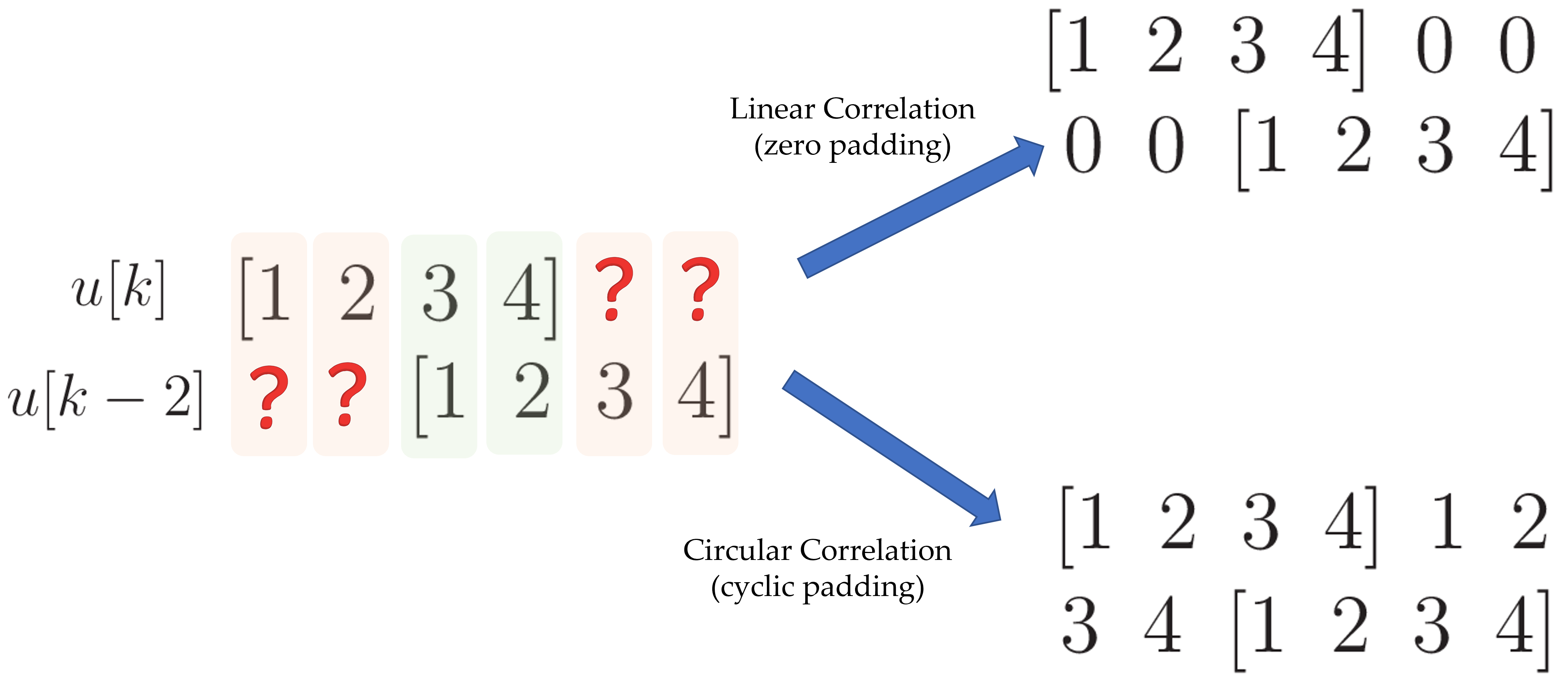}\\
	\vspace{-0.08cm}
	\caption
	{Sketch of the difference between linear and cyclic autocorrelation. The figure shows how these two definitions compute the autocorrelation entry $\tilde{R}_{UU}[-2]$ of the vector $u[k]=[1,2,3,4]$. The linear correlation pads with zero on both sides while the cyclic correlation assumes periodicity with period $n_t$.}\label{fig5}
\end{figure}

Let us illustrate this with an example: compute the autocorrelation of the time series $u(t_k)=[1,2,3,4]$. Consider a Python-like notation, hence $u(t_0)=0$ and $u(t_3)=4$. Considering $t_k=k\Delta t$, we could use an index notation $u[k]$ and write $u[0]=0$ or $u[3]=4$ regardless of the $\Delta t$. Because this signal has $n_t=4$ entries, we have a total of $2n_t-1=7$ possible lags $l$. The lags in \eqref{auto_corr_T} take the form $\tau_l=l \Delta t$, and the autocorrelation will thus be a vector of length $7$ which can also be indexed as $\tilde{R}_{UU}(\tau_l)=\tilde{R}_{UU}[l]$ with $l\in[-3,3]$. Figure \ref{fig5} illustrate the problem of computing $\tilde{R}_{UU}[-2]$: this is the correlation between the signal $u[k]$ and its copy shifted by $2$ entries to the right $u[k-2]$.

While it is clear that the summation will contain $3\times 1+4\times 2$, it is unclear how to deal with the entries that do not have a match in the other signal because of the finite size of the signal. On a practical level, the two classic solutions are zero padding and cyclic padding. 

The zero-padding leads to the \emph{linear correlation}: the vectors are padded by zeros in all the entries lacking information. The cyclic padding leads to the \emph{cyclic correlation}: the vectors are assumed periodic with period $n_t$. We distinguish these operators with and $L$ or a $C$, and the results for this example is:

\begin{equation}
	\tilde{R}_{UUL}[-2]=11/7 \quad 	\tilde{R}_{UUC}[-2]=22/7\,.
\end{equation}

Note that $R_{UUC}$ is periodic and for both operators we have $\tilde{R}_{UU}[l]=\tilde{R}_{UU}[-l]$. This is why one often plots only\footnote{The reader should check the understanding by using \eqref{auto_corr_T} (ignoring the limit) to compute the linear and the circular autocorrelation of $u[k]=[1,2,3,4]$. Focusing on the `positive shifts', one has $\tilde{R}_{UUL}=[30,20,11,4]/7$ and $\tilde{R}_{UUC}=[30,24,22,24]/7$.} the last $n_t$ vectors of the autocorrelation, corresponding to the lags from $l=0$ to $l=n_t-1$. The same considerations are valid for the computation of the autocovariance, the cross correlations and the cross covariance, and the reader should be able to derive their formulae.

When the time series are `short', $\tilde{R}_{UUL}$ and $\tilde{R}_{UUC}$ are very different, and it is essential to clarify which is being used. `Short' here means that relevant lags (giving normalized cross-correlation of $\sim 1$) are present within a time scale that is comparable to the duration of the signal. This is the case of PIV interrogation (see \cite{Theo}), in which time is replaced by space, and the discrete-time indices are the shifts in pixels.

The linear (acyclic) operators are unbiased estimators \citep{Kay1993} and provide better statistical convergence to their ensemble counterparts. On the other hand, the cyclic operators are computationally more interesting because the periodic assumption enables a link with the cyclic convolution, which can be computed in the frequency domain using the Fast Fourier Transform (FFT). Briefly, it is easy to note that the cyclic convolution between two vectors $x[k]$ and $y[k]$ differs from the cyclic correlation by the flipping of the second vector (see \cite{Hayes2011}). In the frequency domain, this flipping corresponds to the conjugation of the associated Fourier transform.

To be more specific, let $X(f_n)$ be the Discrete Fourier Transform (DFT) of $x[n]$, i.e.:

\begin{equation}
	\label{Fourier_Pairs1}
	X(f_n)=\mathcal{F}\{x[k]\}=\frac{1}{n_t}\sum^{n_t}_{k=0} x[k] e^{-2 \pi f_n k\Delta t}  \leftrightarrow	x[k]=\mathcal{F}^{-1}\{X[k]\}=\sum^{n_t}_{k=0} X(f_n) e^{2 \pi f_n k\Delta t} 
\end{equation}  where $f_n=n \Delta f $, with $n=0,n_t-1$ and $\Delta f=f_s/n_t$ the frequency resolution, with $fs=1/\Delta t$ the sampling frequency. The cyclic cross-correlation between two signals $x[k]$, $y[k]$, having DFT $X(f_n)$, $Y(f_n)$ can be computed as 

\begin{equation}
	\label{cross_co_dis}
	R_{XY}=\mathcal{F}^{-1}\biggl(\mathcal{F}\{x[k]\}\overline{\mathcal{F}\{y[k]\}}\biggr)=\mathcal{F}^{-1}\bigl(X(f_n) \overline{Y}(f_n)\bigr)\,,	
\end{equation} with the overbar denoting complex conjugation. A Python (and normalized) implementation of the cross covariance is thus provided by following function:

\begin{centering}
	\begin{lstlisting}[language=Python,linewidth=15.5cm,xleftmargin=.05\textwidth,xrightmargin=.05\textwidth,backgroundcolor=\color{yellow!10}]
def R_UW_C(u,v):
  RUU=np.fft.ifft(np.abs(np.fft.fft(u)*np.fft.fft(v))).real 
  c=(RUU/len(u)-(np.mean(u)*np.mean(v)))/(np.std(u)*np.std(v))
  return c[:len(u)//2]		
	\end{lstlisting}
\end{centering}

The auto-covariance is thus simply $R\_UW\_C(u,u)$ while moving from covariances to correlations is only a matter of removing or keeping the time average.

It is possible to compute the linear operators using a circular extension if an appropriate zero padding is used before the cyclic extension (see \cite{DFT_Smith}). This is what software packages like Scipy and Numpy do when computing the 'FFT-based' cross-correlation. The linear (and normalized) cross-covariance using the \textit{Scipy}'s function \emph{correlate} is implemented in the following function:

\begin{centering}
	\begin{lstlisting}[language=Python,linewidth=15.5cm,xleftmargin=.05\textwidth,xrightmargin=.05\textwidth,backgroundcolor=\color{yellow!10}]
from scipy import signal
def R_UW_L(u,v):
 # Call to the scipy function correlate:
 RUU = signal.correlate(u-np.mean(u), v-np.mean(v),\
 mode='same',method='auto')/len(u)/(np.std(u)*np.std(v))
 return RUU[RUU.size//2:]		
	\end{lstlisting}
\end{centering}

The entry `method', set by default to `auto', uses the fastest option between direct and FFT-based correlation, depending on the size of the array. The reader will need these functions for the next exercise. But first, we might need to recall some more definitions.

\subsection{A note on Turbulence}\label{sec2p3}

The statistical treatment of turbulence is a vast subject, far beyond the scope of this lecture. The reader is referred to \cite{Pope2000,Tennekes1972,Davidson2004,Mcdonough04introductorylectures} for an comprehensive introduction on the topic and to \cite{Saarenrinne2000,Lavoie2007,Segalini2014,Scharnowski2018,Ayegba2020,Wang2021} for a discussion on the impact of measurement resolution on the main turbulence variables. This section recalls the definitions that are required to solve the provided exercises.

In the classic Reynolds decomposition of stationary turbulent flows, the velocity field is decomposed into the sum of an average and a fluctuating part. For the provided 2D dataset, we write these as:

\begin{equation}
	\label{Reynolds}
	\mathbf{u}(\mathbf{x}_i,t_k)=\langle \mathbf{u} \rangle (\mathbf{x}_i)+\mathbf{u}'(\mathbf{x}_i,t_k) \,,
\end{equation} where $\mathbf{x}_i$ denotes a Cartesian grid, $\langle \mathbf{u}\rangle=(\langle u\rangle,\langle v\rangle)^T$ is the mean-field and $\mathbf{u}'=(u',v')^T$ is the fluctuating field. The averaging in this formulation is an ensemble averaging. However, under the assumption of stationarity and ergodicity, this can be safely replaced by a more convenient time averaging.

The intensity of the fluctuation part is measured in terms of standard deviation or root mean square velocity, with components $u_{rms}=\sigma_U(\mathbf{x}_i)$, $v_{rms}=\sigma_V(\mathbf{x}_i)$. A second-order description is provided by the time correlation matrix of the fluctuating components\footnote{Or, equivalently, the covariance matrix of the velocity components.}, which is a 2D flow reads:

\begin{equation}
	\label{Reynolds_STRESS}
	\langle u'_i,u'_j \rangle= \begin{pmatrix}
		\langle u'^2 \rangle & \langle u' v' \rangle \\
		\langle v' u' \rangle & \langle v'^2 \rangle
	\end{pmatrix}=\begin{pmatrix}
		u^2_{rms} & \tilde{R}_{U,V}(0) \\
		\tilde{R}_{V,U}(0) & v^2_{rms}
	\end{pmatrix}\,.
\end{equation}  This matrix varies at every location and in a statistical treatment of turbulence is linked to the \emph{Reynolds stress tensor} $\tau_{i,j}=-\rho \langle u'_i,u'_j\rangle $, namely the additional stresses used to model the non-linear diffusion effect produced by turbulence on the mean flow. The trace of this matrix is linked to the \emph{turbulence kinetic energy} $\kappa=1/2 (u^2_{rms}+v^2_{rms})$, while the ratios $TI_u=u_{rms}/\langle u\rangle$ and $TI_v=v_{rms}/\langle v\rangle$ define the turbulence intensity vector $TI=(TI_u,TI_v)$.

The Reynolds stress tensors provides information on the degree of anisotropy of turbulence: in \emph{isotropic} turbulence, one has $\langle u,v\rangle=0$ and $\langle u^2\rangle=\langle v^2\rangle$; hence the matrix in \eqref{Reynolds_STRESS} is diagonal. The eigenvalues of this matrix can be used to study the principal direction of turbulence momentum exchange \citep{Emory2014VisualizingTA,Simonsen2005} using graphical representation such as the popular Lumley's triangle \citep{Lumley1977}. This is essentially a principal component decomposition of the velocity components in a point.

Similarly, the covariance between gradients of the fluctuating components is linked to the \emph{turbulent dissipation rates} $\varepsilon$. In its complete 3D form, this is defined as $\varepsilon=\nu\langle  \partial_{x_i}u_j(\partial_{x_j}u_i+\partial_{x_i}u_j)\rangle$, with $\nu$ the kinematic viscosity, having used the indices $i,j=1,2,3$ to index the velocity components and axes and having implied summation. The full expression counts 12 terms and it is of difficult evaluation in a 2D measurement because of the missing information in the direction normal to the measurement plane (see \cite{Saarenrinne2000,Wang2021} for an interesting discussion). This quantity measures the rate at which turbulence energy is converted into thermal energy by viscous dissipation and is essential in the modeling of turbulent flows. Because of the strong simplifications required by its computation, and because of the presence of gradients (which need extra care!), the exercise does not ask to compute this quantity\footnote{but the curious reader should proceed nevertheless!}.

Finally, an important statistical quantity is the integral time scale $\Theta$. For the fluctuating component $u$, having normalized temporal autocorrelation $\tilde{R}_{UU}/\sigma_U$, this is defined (see also \cite{Oliveira2007}) as 

\begin{equation}
	\Theta=\int^{\infty}_0{ \tilde{R}}_{UU}(\tau)/\sigma_U d\tau
	\,.
\end{equation}

This integral converges (and the definition makes sense) only if the autocorrelation function tends to zero sufficiently fast. When this occurs, the integral time scale measures the time after which the random process becomes uncorrelated with itself or, in a more pictorial interpretation, the time within which the variable `remembers' its history.

The reader is now ready for the second exercise.

\begin{tcolorbox}[breakable, opacityframe=.1, title=Exercise 2: The statistics of a turbulent Flow]
	
	Study the statistics of the provided dataset, focusing on $t_k\leq 1$. (1) Compute the mean-field and the turbulence intensity field. (2) Analyze the convergence of these quantities in probe P1 and probe P2. (3) Compute the time autocorrelations in P1 and P2. If meaningful, compute the integral time scale. Test if the use of linear or cyclic operators makes any difference. Finally, (4) compute the Reynolds stress tensor at both locations.

	\medskip
	\textbf{Solution}.  We begin with a minor modification to code listing 1, where the velocity magnitude was sampled in the provided probes. We now store the full snapshots into a snapshot matrix $\mathbf{U}$ containing all fields of $u$ and a snapshot matrix $\mathbf{V}$ containing all fields of $v$. This matrix will be important for the second part of the lecture. We also extract both velocity components in the probes P1,P2,P3. Here's how the loop should be modified (see Exercise\_2.py):
	
	\begin{centering}
		\begin{lstlisting}[language=Python,linewidth=15.5cm,xleftmargin=.05\textwidth,xrightmargin=.05\textwidth,backgroundcolor=\color{yellow!10}]
# Initialize the probes
P1_U=np.zeros(n_t,dtype="float32") 
P1_V=np.zeros(n_t,dtype="float32") 
P2_U=np.zeros(n_t,dtype="float32") 
P2_V=np.zeros(n_t,dtype="float32") 
P3_U=np.zeros(n_t,dtype="float32") 
P3_V=np.zeros(n_t,dtype="float32") 
# Prepare the snapshot matrices U and V
D_U=np.zeros((n_s//2,n_t),dtype="float32") 
D_V=np.zeros((n_s//2,n_t),dtype="float32") 
for k in range(0,n_t):
 # Name of the file to read
 Name=FOLDER+os.sep+'Res%05d'%(k+1)+'.dat' 
 # Read data from a file
 # Here we have the two colums
 DATA = np.genfromtxt(Name,usecols=np.arange(0,2),\ 
 max_rows=nxny+1) 
 Dat=DATA[1:,:] # Remove the first raw with the header
 # Get U component and reshape as grid
 V_X=Dat[:,0]; V_X_g=np.reshape(V_X,(n_y,n_x))
 # Get V component and reshape as grid
 V_Y=Dat[:,1]; V_Y_g=np.reshape(V_Y,(n_y,n_x))
 # Assign the vector like snapshots to snapshot matrices
 D_U[:,k]=V_X; D_V[:,k]=V_Y
 # Sample the U components
 P1_U[k]=V_X_g[i1,j1];P2_U[k]=V_X_g[i2,j2];P3_U[k]=V_X_g[i3,j3];
 # Sample the V Components
 P1_V[k]=V_Y_g[i1,j1];P2_V[k]=V_Y_g[i2,j2];P3_V[k]=V_Y_g[i3,j3];
 print('Loading Step '+str(k+1)+'/'+str(n_t)) 
		\end{lstlisting}
	\end{centering}

\hspace{2mm}	Both the snapshot matrices and the velocity components sampled in probes are saved in a \emph{npz} file for the exercises in the second part. Moreover, to limit memory usage, we work with variables in float32 rather than float64; this is accurate enough for our purposes and let the exercise comply with the limited memory of the author's laptop.
	
\hspace{2mm}	Using the snapshot matrices, the mean flow and the root mean square velocity components can be computed in one line each: 
	
	\begin{centering}
		\begin{lstlisting}[language=Python,linewidth=15.5cm,xleftmargin=.05\textwidth,xrightmargin=.05\textwidth,backgroundcolor=\color{yellow!10}]
# U Component
mu_U=np.mean(D_U,axis=1) # Ensamble Mean
sigma_U=np.std(D_U,axis=1) # Ensamble STD
# V component
mu_V=np.mean(D_V,axis=1) # Ensamble Mean
sigma_V=np.std(D_V,axis=1) # Ensamble STD 
		\end{lstlisting}
	\end{centering}	
	
\hspace{2mm}	Plotting these fields is now only a matter of reshaping the results back onto a matrix and using the provided functions (see the python code). The mean flow and the turbulence intensity $TI=\sqrt{TI_u^2+TI_v^2}$ are shown in Figure \ref{Ex2}. Note that the latter is computed by normalizing the root mean square component with respect to the free stream velocity and not the local mean velocity. Most of the turbulence is localized in the wake region, where the fluctuation amplitude reaches up to $50\%$ of the free stream velocity. This answers question (1).

	\begin{center}%
		\setcounter{subfigure}{0}%
		
		\begin{minipage}{.49\linewidth}
			\includegraphics[width=\linewidth]{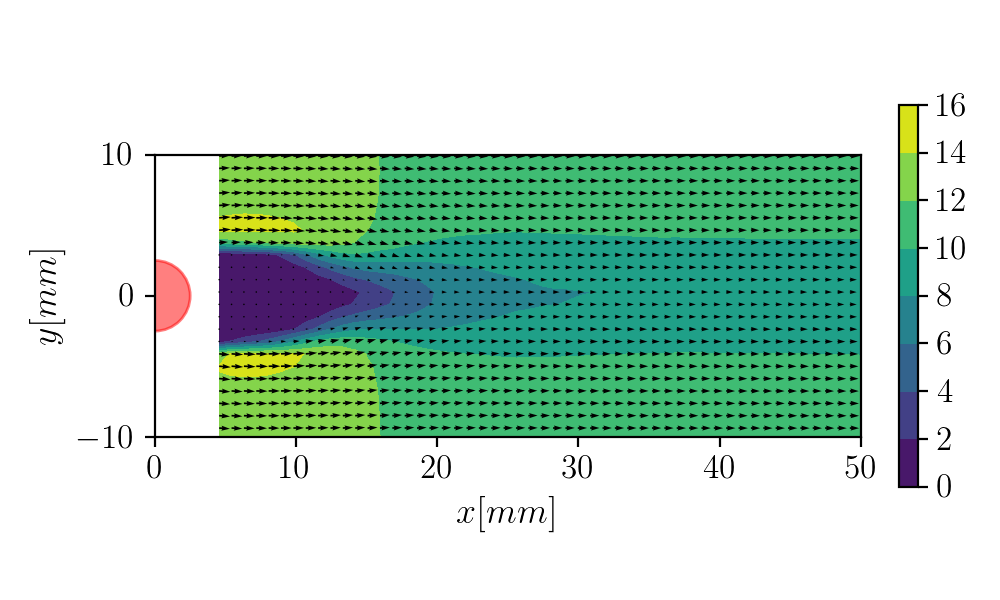}
			\label{Ex2a)}%
			\vspace{-3mm}
			\center{a)}
		\end{minipage}
		\begin{minipage}{.49\linewidth}
			\includegraphics[width=\linewidth]{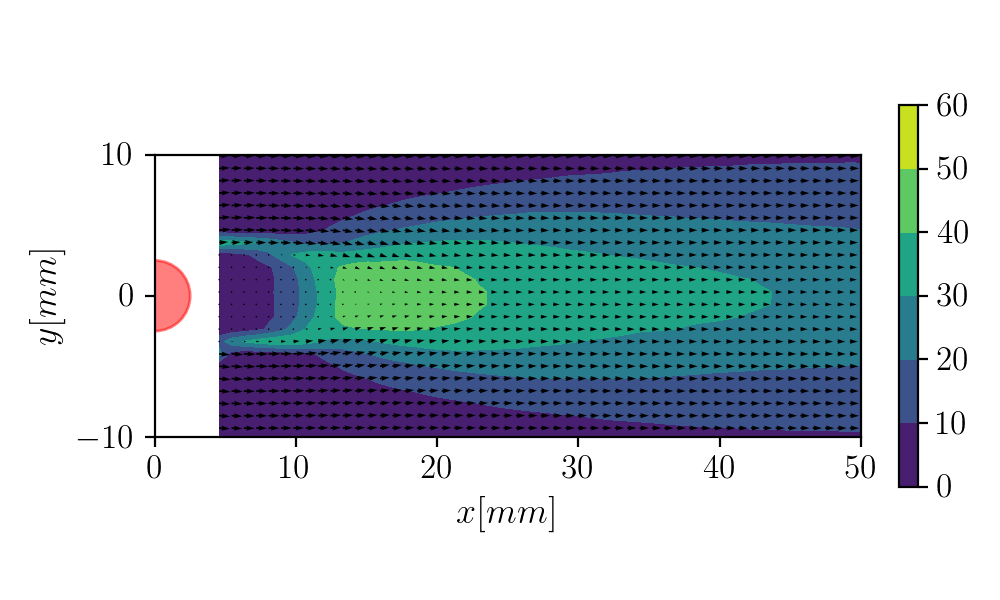}%
			\label{Ex2b)}%
			\vspace{-3mm}
			\center{b)}
		\end{minipage}	
		\vspace{-0.05cm}
		\captionof{figure}{	Fig (a): Mean flow of the provided dataset in the first $1s$ (3000 snapshots) (b): Turbulence intensity, normalizing the root mean square field by the free stream.}	
		\label{Ex2}
	\end{center}
	
\hspace{2mm}	The analysis of the convergence of the statistics at the probe locations P1 and P3 is similar to what was performed in the previous exercise. This is left as an exercise for the reader. The results are shown in Figure \ref{Ex22}. The convergence in P1 is particularly fast, as the root mean square velocity is low and the mean velocity is large. The same is not true in P3, where convergence is barely achieved. The reader should analyze these statistics alongside the histogram in Figure \ref{fig2}: the $t_k\leq 1s$ of acquisition is associated with the distribution centred at approximately $12$ m/s. 
	
\hspace{2mm}	The reader is referred to \cite{Bruun1995} (chapter 12) for a discussion on the convergence of the mean and standard deviation as a function of the turbulence intensity.

	\begin{center}%
		\setcounter{subfigure}{0}%
		\begin{minipage}{.49\linewidth}
			\includegraphics[width=\linewidth]{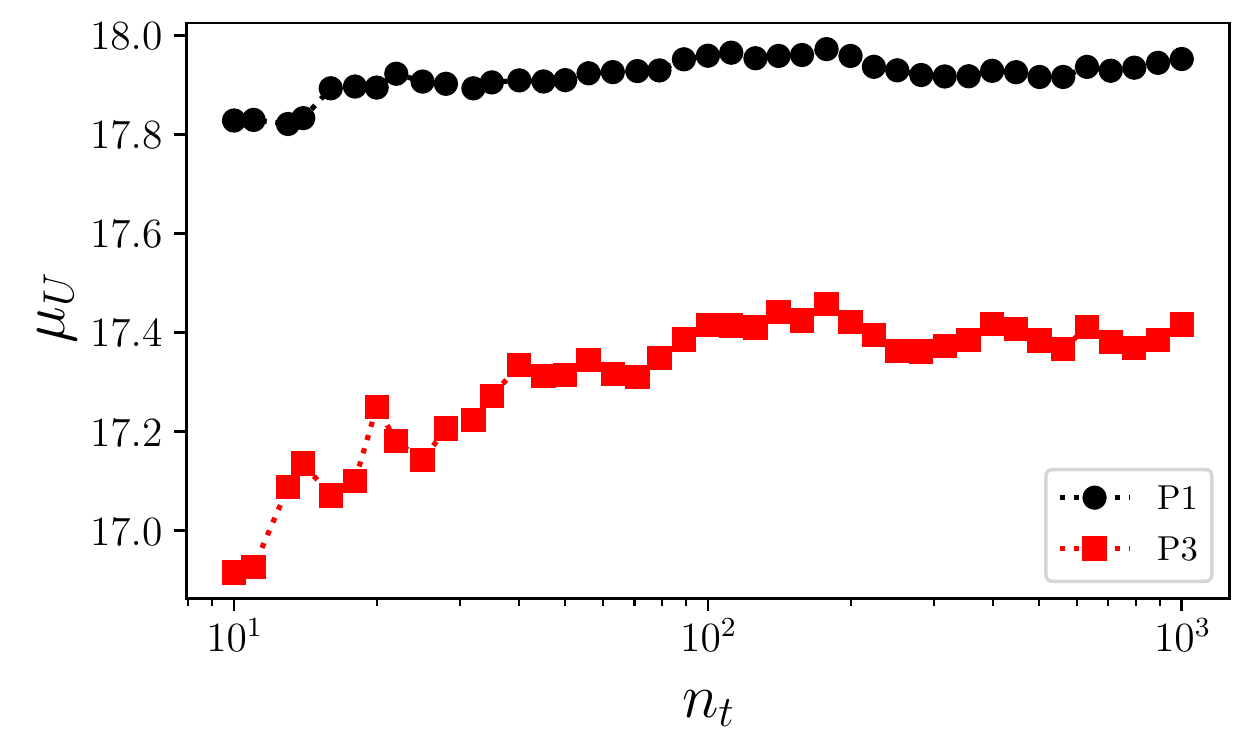}
			\label{Ex2aa)}%
			\vspace{-4mm}
			\center{a)}
		\end{minipage}
		\begin{minipage}{.49\linewidth}
			\includegraphics[width=\linewidth]{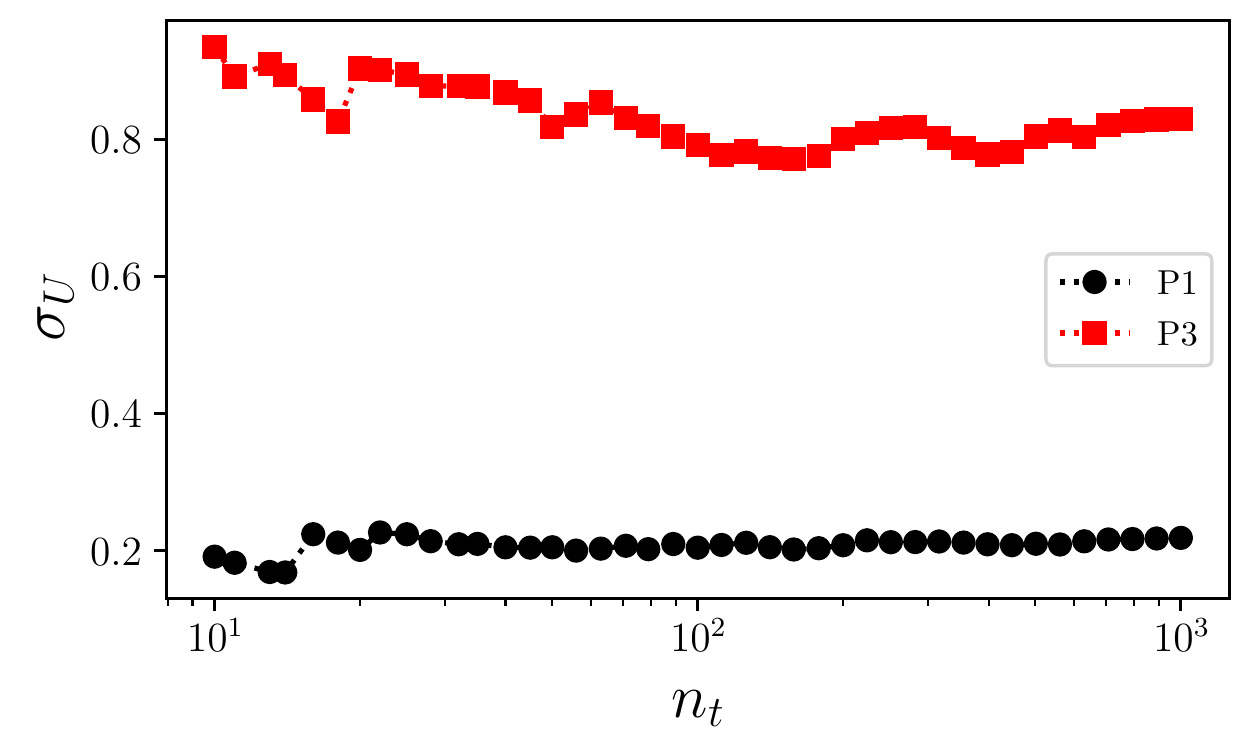}%
			\label{Ex2bb)}%
			\vspace{-4mm}
			\center{b)}
		\end{minipage}	
		\vspace{-0.05cm}
		\captionof{figure}{	Fig (a): Mean flow of the provided dataset in the first $1s$ (3000 snapshots) (b): Turbulence intensity, normalizing the root mean square field by the free stream.}	
		\label{Ex22}
	\end{center}
	
\hspace{2mm}	For the scope of this lecture, we can declare point (2) answered, although the reader is encouraged to proceed with further analysis\footnote{is the sampling rate `optimal' for statistical analysis? To have fast convergence of the statistics, one should aim to have every realization `independent' from one another. To this, one should have a time separation between samples that is sufficiently larger than the integral time scale. The optimal separation is $\Delta t=2 \Theta$ \citep{Bruun1995}). }.

\hspace{2mm}	The time autocorrelations in P1 and P2 can be computed with the functions provided in the previous sections. Because the time series under analysis are rather long (3000 samples), the difference between circular and linear autocorrelation is not appreciable for lags $\tau_l<0.1s$; the reader is nevertheless encouraged to explore the difference at larger lags. The results for P1 and P2, using circular autocorrelation, are shown in Figure \ref{fig7} for the first $0.01s$. In both probes, oscillations of the flow are visible, although these are much stronger in P3 than in P1. As a result, none of the autocorrelations operators converge to zero, meaning that the signals are not purely `stochastics' and a deterministic component could (and should!) be extracted \emph{before} an integral time scale is to be computed\footnote{the reader might want to come back to this point after the next section has introduced modal analysis: one could remove the `coherent' (i.e. deterministic) part and run statistical analysis in remaining (`unresolved') portion. This is the essence of more sophisticated versions of the Reynolds decomposition (see for example \cite{Baj2015}). }. Interestingly, the oscillations appear in phase, as if the entire flow field is locked at a given frequency.
	
	\begin{center}%
		\setcounter{subfigure}{0}%
		\begin{minipage}{1\linewidth}
			\centering
			\includegraphics[width=0.6\linewidth]{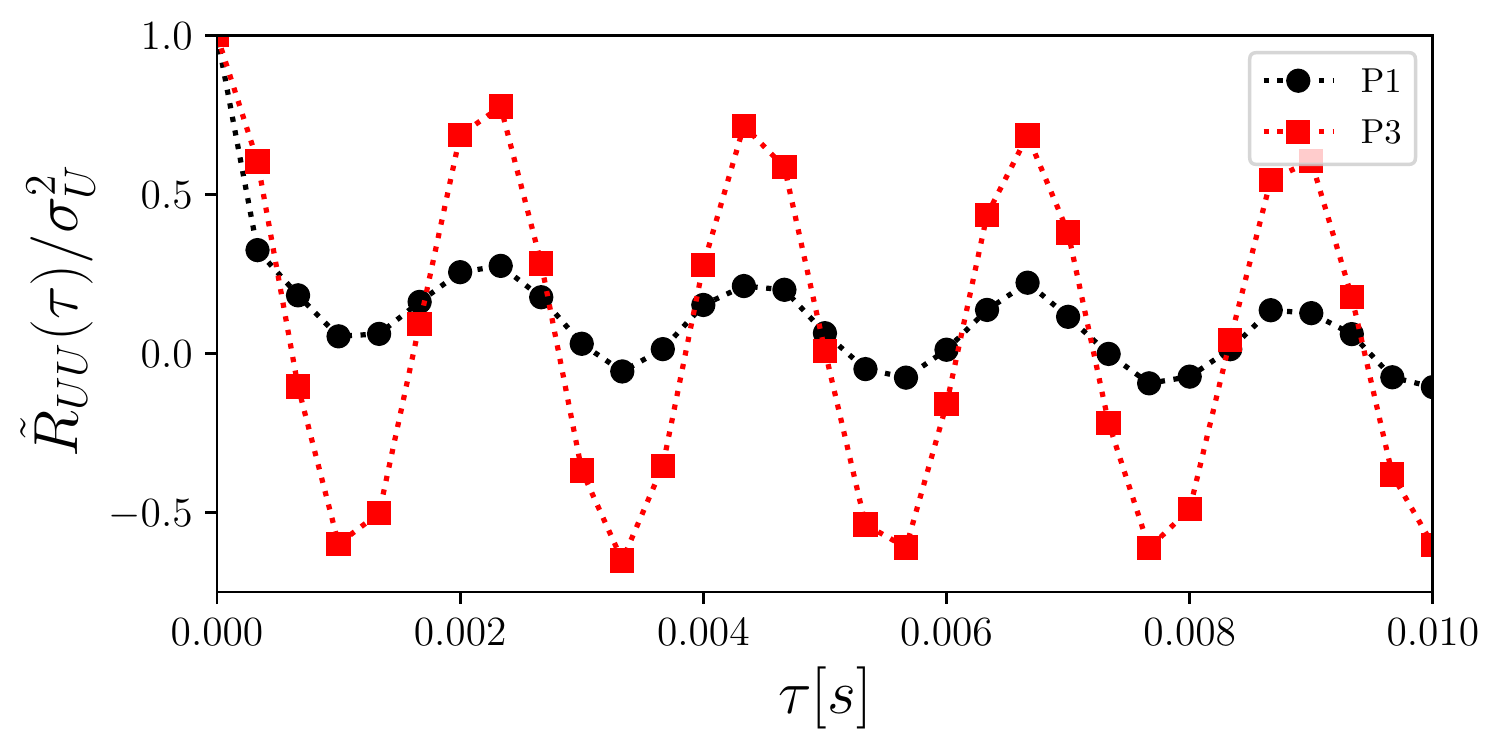}
			\label{fig7}%
			\captionof{figure}{	Fig (a): Mean flow of the provided dataset in the first $1s$ (3000 snapshots) (b): Turbulence intensity, normalizing the root mean square field by the free stream.}
		\end{minipage}
		
	\end{center}
	
\hspace{2mm}	Finally, we close with the Reynolds stress tensor. This is essentially a covariance matrix. Arranging the velocity information in matrices of dimention $3000\times 2$, this is simply:

	\begin{centering}
		\begin{lstlisting}[language=Python,linewidth=15.5cm,xleftmargin=.05\textwidth,xrightmargin=.05\textwidth,backgroundcolor=\color{yellow!10}]
u1_f=P1_U-np.mean(P1_U);  v1_f=P1_V-np.mean(P1_V)
U_f_1=np.vstack((u1_f, v1_f)).T
Re_1=1/len(P1_U)*U_f_1.T.dot(U_f_1)    
		\end{lstlisting}
	\end{centering}	
	
\hspace{2mm}	The results in points P1 and P2 are:
	
	\begin{equation}
		\label{Re_R}
		\langle u_i,u_j \rangle(P1)= \begin{pmatrix}
			0.0254 & -0.0022 \\
			-0.0022 & 0.0169
		\end{pmatrix} \quad \langle u_i,u_j  \rangle(P3)= \begin{pmatrix}
			0.3337 & -0.0068 \\
			-0.0068 & 0.2793
		\end{pmatrix}
	\end{equation}

	\begin{center}%
		\setcounter{subfigure}{0}%
		\begin{minipage}{.49\linewidth}
			\centering
			\includegraphics[width=0.9\linewidth]{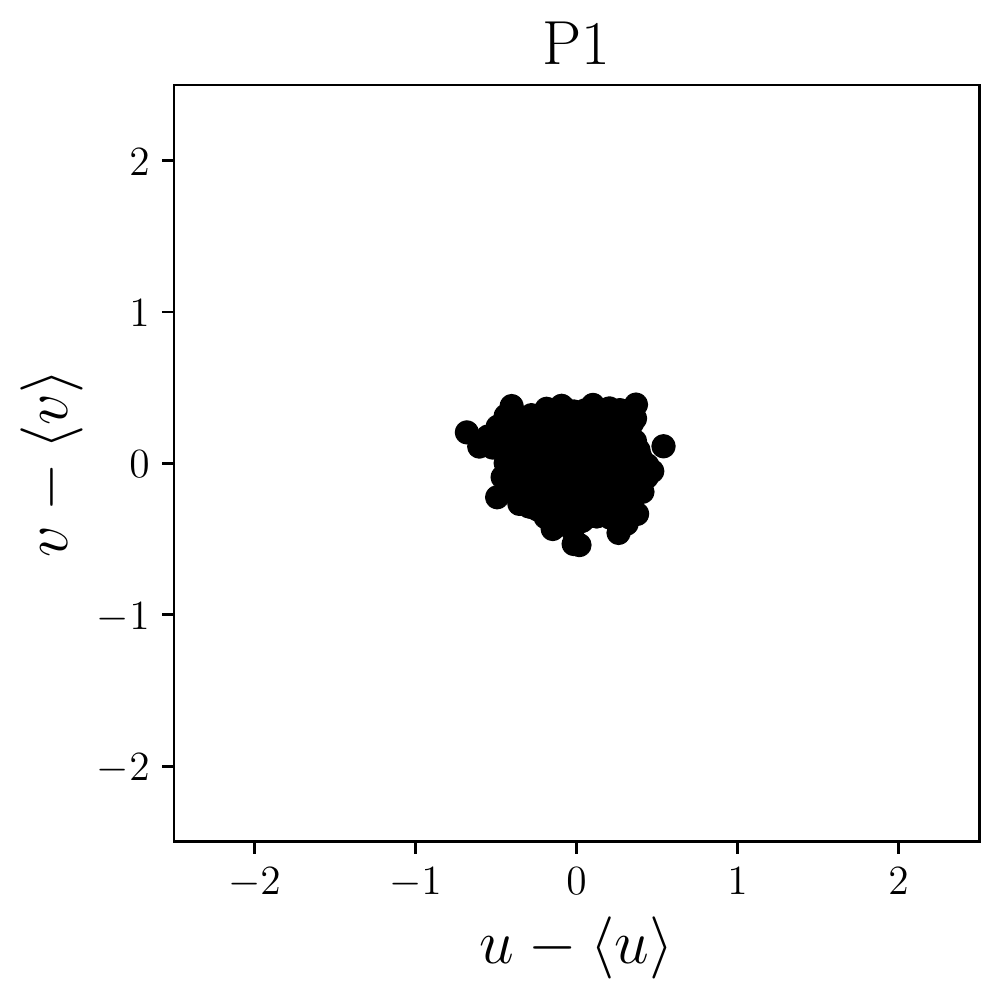}
			\label{Ex2aaa)}%
			\vspace{-3mm}
			\center{\hspace{5mm} a)}
		\end{minipage}
		\begin{minipage}{.49\linewidth}
			\centering
			\includegraphics[width=0.9\linewidth]{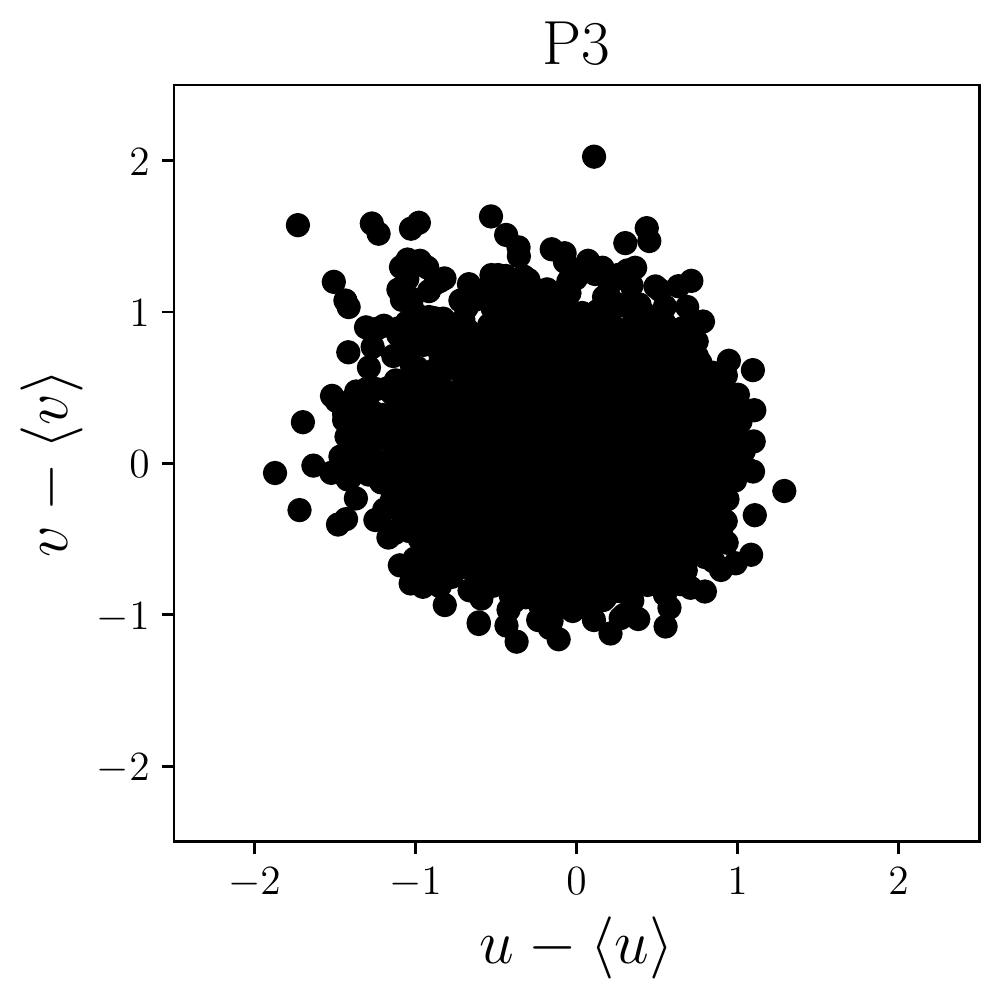}%
			\label{Ex2bbb)}%
			\vspace{-3mm}
			\center{\hspace{5mm} b)}
		\end{minipage}	
		\vspace{-0.05cm}
		\captionof{figure}{Scatter plot of the fluctuating components $v-\langle v\rangle$ and $u-\langle u\rangle$ at the location of probe P1 (Fig a) and probe P2 (Fig b).}	
		\label{Ex3Last}
	\end{center}

\hspace{2mm}	The scatter plot of the fluctuation components are shown in Figure \ref{Ex3Last}. It is clear that the turbulent kinetic energy and the turbulent intensity is much larger in P3 than P1, and hence the distribution is broader in the first. In both points, however, the scatter plot shows a rather uncorrelated distribution of fluctuation, resulting in a nearly diagonal covariance: the flow is \emph{almost} isotropic. `Almost', because $u_{rms}$ is slightly larger than $v_{rms}$ in both points.

\end{tcolorbox}

\subsection{Power Spectral Density and Coherence}\label{sec2p4}

We now move to the frequency representation of stochastic signals. This is slightly more involved than the frequency representation of deterministic signals. In a continuous domain, a stochastic signal is seldomly square-integrable, and it thus seldomly admits an ordinary Fourier transform. In the signal processing terminology, the definition of an appropriate frequency domain requires shifting the treatment from the notion of energy to the notion of power\footnote{Given an infinite duration stochastic signal $x[k]$, indexed by the integers $k$, the energy $\mathcal{E}$ and the power $\mathcal{P}$ are defined as follows:
	
	\begin{equation}
		\label{E_P}
		\mathcal{E}\{x[k]\}=\lim_{n_t\to \infty}\sum^{n_t}_{k=0} |x[k]|^2 \quad \quad \mathcal{P}\{x[k]\}=\lim_{n_t\to \infty}\frac{1}{2n_t-1}\sum^{n_t}_{k=0} |x[k]|^2\,.
\end{equation}}: stochastic signals have generally infinite energy but finite power \citep{Oppenheim2015,Hayes2011}. 

Therfore, we shall not focus on the Fourier transform of a signal but at the Fourier transform of its autocorrelation (which is square integrable). The following transform exist for all signals of interest:

\begin{equation}
	\label{R00}
	\mathbb{E}\{u^2_n(t_k)\}=\tilde{R}_{UU}(0)=\frac{1}{2\pi} \int^{\infty}_{-\infty} S_{UU}(\omega) d\omega\,,
\end{equation} where $\omega=f/2\pi$ is the pulsation and $f$ is the frequency and $S_{UU}$ is the continuous Fourier transform of the autocorrelation function and is known as \emph{power spectral density}. Note that this is real and even in $\omega$ because the time autocorrelation $R_{UU}$ is symmetric ($R_{UU}(\tau)=R_{UU}(-\tau)$). The Fourier pairs of interest are thus 

\begin{equation}
	\label{Fourier_PAIRS}
	S_{UU}(\omega)=\int^{\infty}_{-\infty} \tilde{R}_{UU}(\tau) e^{-\mathrm{j}\omega \tau} d\tau \quad \mbox{and} \quad 	\tilde{R}_{UU}(\tau)=\frac{1}{2\pi}\int^{\infty}_{-\infty} S_{UU}(\omega) e^{\mathrm{j}\omega \tau} d\omega\,.
\end{equation}

These are also known as Wiener-Khinchin relations. The notion of power spectral density is important because it allows to give a spectral representation to stochastic signals and thus to generalize the theory of linear time-invariant (LTI) systems (see \cite{MendezLS}). This theory brings powerful and simple tools for system identification, filtering and forecasting, and we show some of them in practice in the next exercise (see \cite{Oppenheim2015,FIR_Smith} for more). 

Concretely, we are interested in the notion of coherency in relation to the frequency content of signals. The question motivating this section is simple: is there any chance to predict what is happening in one probe (e.g. P2), at least within a certain range of frequencies, given the signal we have in another probe (e.g. P3)?

We first need some definitions. Let us assume that a stochastic signal $x[k]\in \mathbb{R}^{n_t}$ is the input of a linear and deterministic system which responds with a second stochastic signal $y[k] \in \mathbb{R}^{n_t}$. Uncorrelated noise might be added to the output of this system and we only see the resulting $y_n[k]=y[k]+N[k]$. If the power of the noise is too large, we might not be able to recover any reasonable estimate of $y[k]$ and we will say that there is a poor level of \emph{coherency} between $x[k]$ and $y_n[k]$. Conversely, we might be able to identify an approximation of the underlying linear linking input and output. This link is here to be analyzed frequency by frequency.

Considering finite duration signals, the convergence problems of the Fourier representation are less stringent, and we can waive some of the formalism required for the continuous world: under the assumption of circular extension of the signals\footnote{In this section we will only consider cyclic padding. The zero-padding requires some little extra care which is not essential for this lecture (see \cite{DFT_Smith}).)}, every digital signal has a Discrete Fourier Transform (DFT). Let $X(f_n)$ and $Y(f_n)$ denote the DFT of the input $x[k]$ and the output $y[k]$ (see \eqref{Fourier_Pairs1} for the definitions).

If a linear time invariant system links $y[k]$ to $x[k]$, the output can be computed via convolution of the input with the system's impulse response. From the convolution theorem, we know that in frequency domain this is a multiplication with the transfer function of the system $H(f_n)$, i.e. the DFT of the impulse response. We thus have:

\begin{equation}
	\label{Definitions}
	y[k]=\sum^{n_t-1}_{m=0} x[k]\,h[k-m] \,\,\longleftrightarrow	\,\, Y(f_n)= H(f_n)\, X(f_n)\,.
\end{equation}  

We now introduce the discrete equivalent of \eqref{Fourier_PAIRS}. It is possible to show (see  \cite{Oppenheim1996a} and eq. \eqref{cross_co_dis}) that the power spectral density of the input is

\begin{equation}
	\label{DefinitionsII}
	S_{XX}(f_n)=\sum^{n_t-1}_{k=0} R_{XX}[k] e^{-2 \pi f_n k\Delta t} =\mathcal{F}\{\mathcal{F}^{-1}\{X(f_n) \overline{X}(f_n)\}\}= X(f_n) \overline{X}(f_n)\,.
\end{equation} 

Similarly, the power spectral density of the output is $S_{YY}(f_n)=Y (f_n) \overline{Y}(f_n)$ and we can additionally define the \emph{cross-spectral density} as $S_{YX}=X(f_n) \overline{Y}(f_n)$.

We can now craft a spectral function which measures how well the spectrum of the output correlates with the spectrum of the input. This is the \emph{coherence function}:

\begin{equation}
	\label{DefinitionsIV}
	C_{YX}(f_n)=\frac{|S_{YX}(f_n)|^2}{S_{XX}(f_n) S_{YY}(f_n)}\,\,.
\end{equation}

At those frequencies for which $Y(f_n)=H(f_n) X(f_n)$ (i.e. for which a LTI system could model the input-output relation), one has 

$$S_{YX}(f_n)=Y(f_n)\overline{X}(f_n)=H(f_n) X(f_n)\overline{X}(f_n)\rightarrow S_{YX}(f_n)=H(f_n)S_{XX}(f_n)\,.$$ 

Moreover, noticing that $S_{YY}(f_n)=|H(f_n)|^2 S_{XX}(f_n)$, we recover $C_{YX}(f_n)=1$. At frequencies for which no LTI system can link the two spectra, the \emph{coherence} is much close to zero. The function is undefined at $f_n$'s for which either $S_{XX}$ or $S_{YY}$ is zero.

To conclude this section, it is worth noticing that working with one single spectrum for both input and outputs makes little sense: an ensemble of time series leads to an ensemble of spectra. The classic approach to evaluating a stochastic signal's frequency representation involves averaging, which could be either in the ensemble domain or in the time domain. Under the assumption of ergodicity, the second is usually preferred, and the result is the well known Welch's method \citep{Welch1967} or \emph{periodogram method}. The computation is performed by dividing the signal into successive (and overlapping) blocks, computing the DFT of these, and then averaging the results. In practice, a smoothing window $w[k]$ multiplies the signal in the time domain to cope with the problems arising by the (usually violated) assumption of periodicity. For $S_{XX}(f_n)$, for instance, we have:

\begin{equation}
	S_{XX}(f_n)=\frac{1}{n_L}\sum^{n_L-1}_{m=1} |XW_m(f_n)|^2 \, \mbox{with} \,  XW_m(f_n)=\mathcal{F}\{x[k] \,w[k]\}
\end{equation} 

Here $XW_m(f_n)$ is the DFT of the windowed block $x_m[k]w_[k]$, with $x_m[k]=x[k+m n_W]$, $w_m[k]$ the window function designed to taper gracefully to zero at both endpoints of the block, and $k=0,1,\dots n_w-1$ with $n_L$ denoting the number of blocks. Because the multiplication in the time domain is a convolution in the frequency domain, the method is essentially a smoothing operation on the signal's spectrum; this is why the method is sometimes also called `smoothed spectrum' and implemented in the frequency domain. 

The \textit{Python}'s \emph{scipy} package offers robust functions to compute both the power spectral densities and the spectral coherence. We use both in the next exercise.

\begin{tcolorbox}[breakable, opacityframe=.1, title=Exercise 3: Power Spectral Densities and Coherency]
	
	The power spectral density of the signal in the probes P1 and P3 was shown in Figure \ref{fig2}. As in the previous exercise, let us consider only the first $t_k\leq 1$ s of acquisition. First, practice with the notion of coherency by testing it on a synthetic signal $y$. To construct this signal, filter the signal in P2 with a moving average filter of order three, i.e., impulse response $h=[1,1,1]/3$, and add noise to the result. Next, show the original and synthetic signal's power spectral densities and study the coherence function between the two signals. (1) What do you get and why? Finally, (2) study the coherency between the signals in P2 and P3.

	\medskip
	\textbf{Solution}.  We assign the signal to the input variable $x$ and perform the filtering plus noise superposition in the following lines:
	
	\begin{centering}
		\begin{lstlisting}[language=Python,linewidth=15.5cm,xleftmargin=.05\textwidth,xrightmargin=.05\textwidth,backgroundcolor=\color{yellow!10}]
data = np.load('Sampled_PROBES.npz')
n_T=3000; Fs=3000; dt=1/Fs
P2_U=data['P2_U'][0:n_T];  
P3_U=data['P3_U'][0:n_T]; 
# Prepare the time axis
t=np.linspace(0,dt*(n_T-1),n_T) 
# Define impulse response and 'input'
h=np.ones(3)/3; x=P2_U
# We padd to use the frequency domain:
h=np.concatenate([h,np.zeros(len(x)-3)]) 
# The transfer function will be
H=np.fft.fft(h)
# This is circular convolution
X=np.fft.fft(x)
# Add noise
N=np.random.normal(0, 0.5*np.std(x), len(x))
y=np.real(np.fft.ifft(X*H))+N
		\end{lstlisting}
	\end{centering}		
	
	We compute the power spectral densities $S_{XX}$ and $S_{YY}$ as follows:
	
	\begin{centering}
		\begin{lstlisting}[language=Python,linewidth=15.5cm,xleftmargin=.05\textwidth,xrightmargin=.05\textwidth,backgroundcolor=\color{yellow!10}]
N_P=300
f,Sxx_den = signal.welch(x, Fs, nperseg=N_P,scaling='density')
f, Syy_den = signal.welch(y, Fs, nperseg=N_P,scaling='density')
		\end{lstlisting}
	\end{centering}	
	
\hspace{2mm}	The reader should consult the documentation of the \textit{welch} function. This function performs mean removal by default, while the scaling option 'density' makes the output a density function (i.e. units are `V$^2$/Hz', with V the input units). The blocks for the averaging are here shown to be of size $300$ while the overlapping is set by default to $50\%$. The power spectral densities of the original and the manipulated signal is shown in Figure \ref{figEx3F}a). The spectra are quite similar at low frequencies and feature the same peak at $459 Hz$. However, they become increasingly different at higher frequencies.
	
\hspace{2mm}	We now compute the coherence function together with the transfer function of the moving average filter here implemented and plotting the results in the same graph:
	
	\begin{centering}
		\begin{lstlisting}[language=Python,linewidth=15.5cm,xleftmargin=.05\textwidth,xrightmargin=.05\textwidth,backgroundcolor=\color{yellow!10}]
# Prepare the transfer function of h
# curiosity: this is a nice tool for transf functions
wz, Hz = signal.freqz(h)
# Create the figure axis
fig, ax = plt.subplots(figsize=(6, 3))
# Compute the coherency
f, Cxy = signal.coherence(x, y, Fs, nperseg=N_P)
# Plot the results in decibels
plt.plot(f, 20*np.log10(Cxy),label='$C_{x,y}$')
# Plot the transfer function
plt.plot(wz/np.pi*Fs/2,20*np.log10(Hz), 'r',label='$|H(f)|$')
		\end{lstlisting}
	\end{centering}	
	
\hspace{2mm}	The result is in Figure \ref{figEx3F}b). It is clear that the coherency decreases following the same trend as the transfer function of the LTI system implemented: frequencies that are almost removed from the original signal leave the stage to the added random (incoherent) noise. The reader should repeat the exercise without adding noise.
	
	\begin{center}%
		\setcounter{subfigure}{0}%
		\begin{minipage}{.49\linewidth}
			\includegraphics[width=\linewidth]{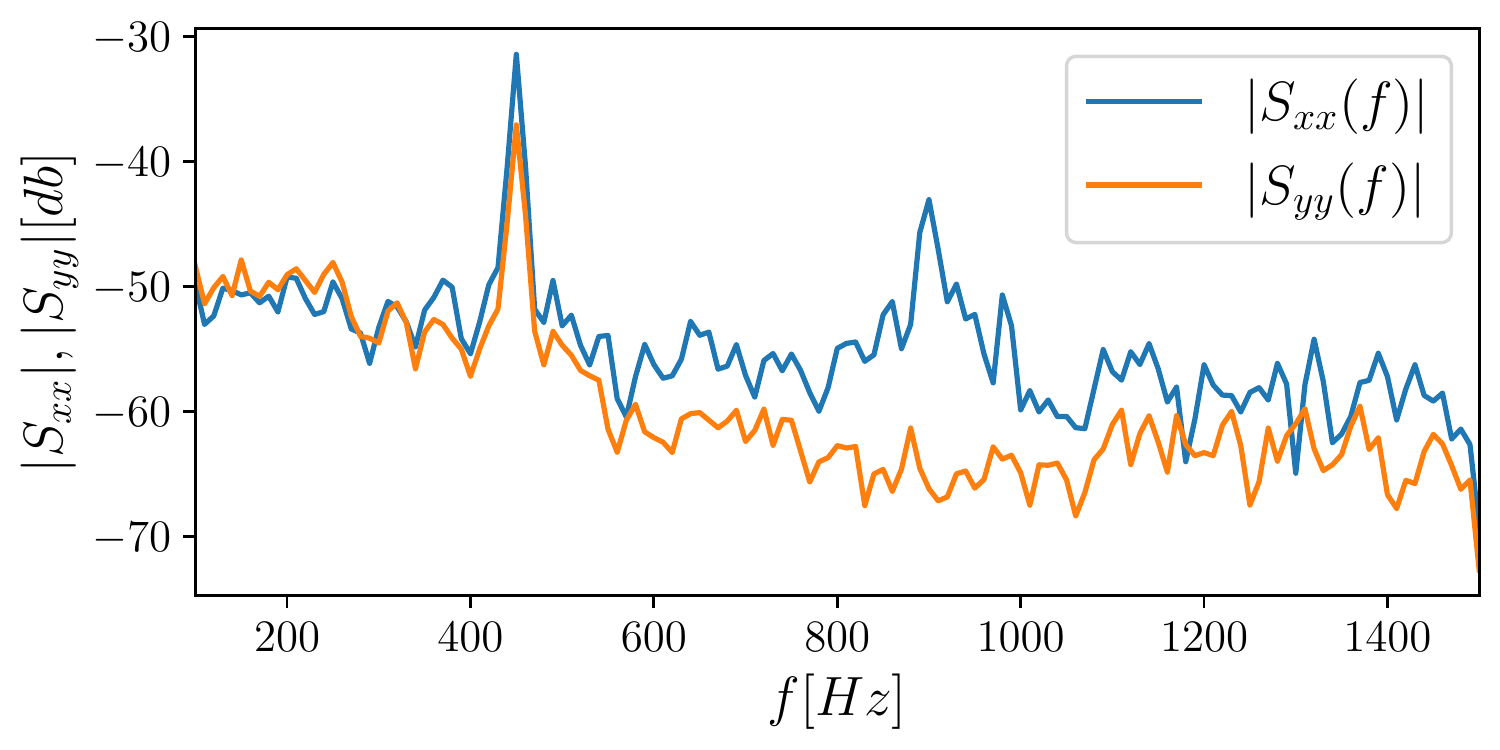}
			\label{Ex3aa)}%
			\vspace{-3mm}
			\center{a)}
		\end{minipage}
		\begin{minipage}{.49\linewidth}
			\includegraphics[width=\linewidth]{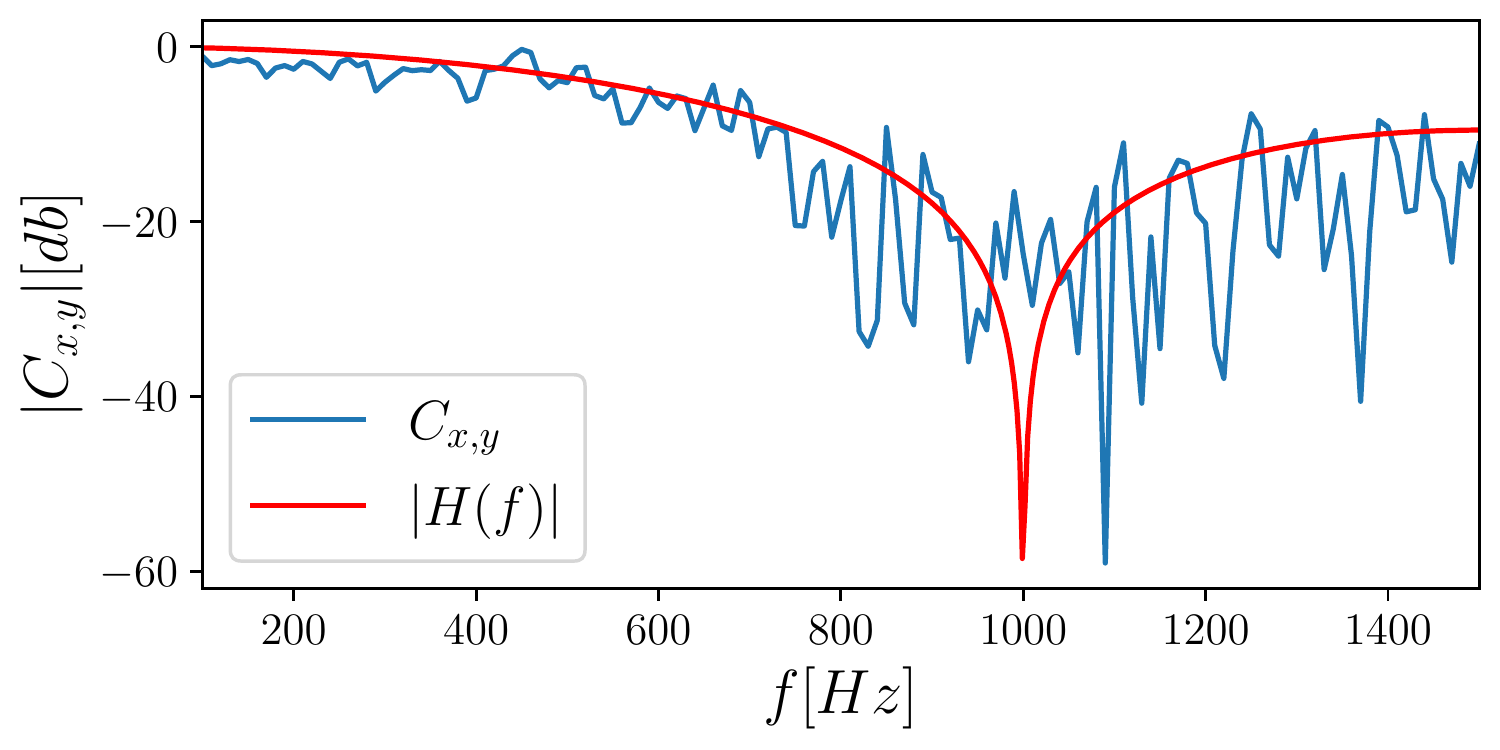}%
			\label{Ex3bb)}%
			\vspace{-3mm}
			\center{b)}
		\end{minipage}	
		\vspace{-0.05cm}
		\captionof{figure}{	Fig (a): Power spectral densities of the original signal (stream-wise velocity component in P1) together with its manipulated version (b):cross-coherency between the signal and its manipulated version, together with the magnitude of the transfer function used in the manipulation.}	
		\label{figEx3F}
	\end{center}

\hspace{2mm}	Finally, the cross coherency between P2 and P3 is shown in Figure \ref{fig11}. The overall level of coherence is low, except for the regions at $f_n\approx 459 Hz$. This corresponds to the frequency of the vortex shedding, characterized by \emph{coherent structures}. We look for these in the next section.

	\begin{center}%
		\setcounter{subfigure}{0}%
		\begin{minipage}{1\linewidth}
			\centering
			\includegraphics[width=0.6\linewidth]{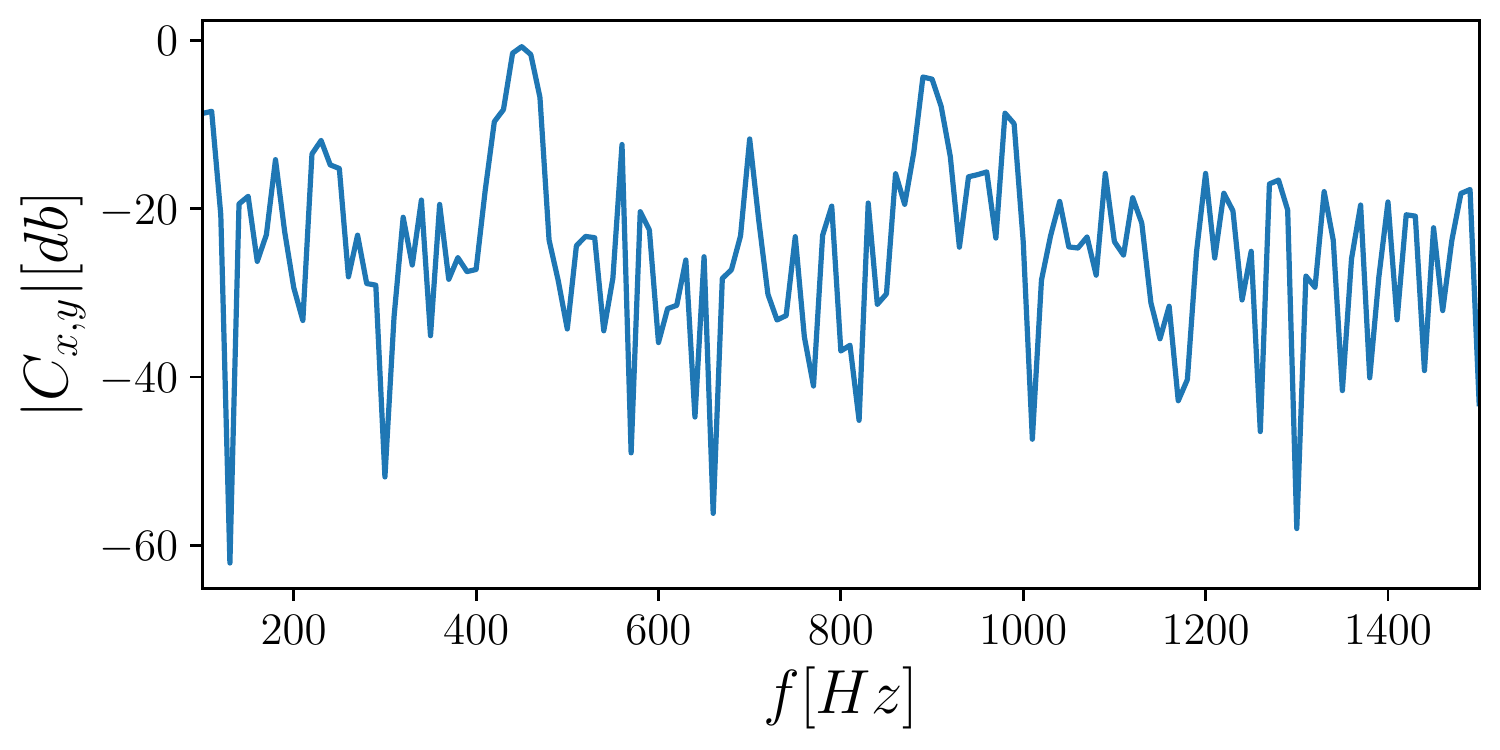}
			\label{fig11}%
			\captionof{figure}{Cross-coherency between the stream-wise velocity in P2 and P3. At approximately}
		\end{minipage}
		
	\end{center}

\end{tcolorbox}


\section{Classic Modal Analysis}\label{sec4}

In the last exercise, we observed a strong correlation between the signals in P2 and P3 at the frequency of the vortex shedding. In the previous section, we have considered time series arising by sampling a specific point (and we took care of including a subscript $n$ in all the defined quantities). 

In this section, we look for the statistics, the coherence and the frequency content of the flow field as a whole: the time series of interest is no longer a scalar, but a vector of dimension $n_s=n_x n_y$ for the components $u$ and $v$, and dimension $2 n_s$ if we take both components in the same `snapshot'. 

More specifically, let ${\vec{\bm{u}}}(\bm{x_i},t_k)=(\bm{u}(\bm{x_i},t),\bm{v}(\bm{x_i},t))$ denote the TR-PIV field over a uniform grid $\bm{x_i}\in\mathbb{R}^{n_x\times n_y}$, with $\bm{i}$ a matrix linear index, and a temporal discretization $t_k=\{(k-1)\Delta t\}^{k=n_t}_{k=1}$ with $\Delta t=1/f_s$ the time step and $f_s$ the sampling frequency. In what follows, bold fonts are used to indicate matrices.

The goal of modal analysis is to describe the data as a linear combination of elementary information referred to as \emph{modes}. Each mode has its own spatial structure and temporal evolution and can potentially describe essential features in the data. Each mode has a spatial structure $\bm{\vec{\phi}}_{k} (\bm{x}_i)$, a temporal structure $\psi_{r}(t_k)$, and an amplitude $\sigma_{r}$ that weights its importance. The sampled velocity field is thus expanded as 

\begin{equation}
	\label{EQ1}
	\vec{\bm{u}}(\bm{x_i},t_k)=\sum^{R}_{r=1}\,\vec{\bm{C}}_r (\bm{x}_i)\,\psi_{r} (t_k)=\sum^{R}_{r=1}\,\sigma_{r}\,\bm{\vec{\phi}}_{r} (\bm{x}_i)\,\psi_{r} (t_k)\,,
\end{equation} where the contribution of each mode is defined as $\sigma_r=||\vec{\bm{C}}_r (\bm{x}_i)||_2$ and the spatial structures are obtained via normalization $\bm{\vec{\phi}}_{r}=\vec{\bm{C}}_r (\bm{x}_i)/\sigma_r$. 

Assuming that all modes are (at least!) linearly independent, the number ($R$) of modes non-null amplitudes $\sigma_r$ defines the dimensionality of the data for a given basis. Truncating the summation to $\tilde{R}<R$ modes provides an approximation of order $\tilde{R}$ with respect to a specific decomposition.

These tools were initially developed for identifying coherent structures in turbulent flows \citep{Lumley1,Berkooz1993,Siro1} and to construct reduced-order models \citep{Holmes,Holmes1996,Benner2015} of fluid flows. Their range of application is nowadays much broader and include (among others) filtering \citep{Raiola2015}, image processing \citep{Mendez2017} and validation of numerical simulations \citep{Meyer}. By removing `irrelevant' modes, we use the decomposition as a filter, and we compress data. By focusing on the dynamics described by `a few' modes, we use decompositions as tools for constructing reduced-order models, which can significantly reduce the computational burdens in the simulations of large dynamical systems. A recent review of dimensionality reduction is proposed by \cite{Ahmed2021}.

In this section, we consider three decompositions in the form of eq.\eqref{EQ1}. These are the Discrete Fourier Transform (DFT) in subsection \ref{sec4p1}, the Proper Orthogonal Decomposition (POD) in subsection \ref{sec4p2} and the Dynamic Mode Decomposition (DMD) in subsection \ref{sec4p3}. These can be distinguished by the choice of the temporal structure $\psi_r(t_k)$ (see also \cite{MendezLS2}). We will not go into much implementation and algorithmic details (these are extensively treated in \cite{MendezLS2}) and we will use the {Python} package MODULO (see \cite{Ninni2020})

In what follows, we use the subscripts $\mathcal{F}$, $\mathcal{D}$, and $\mathcal{P}$ to denote, respectively, the modes linked to the DFT, the DMD and the POD.

\subsection[The DFT]{The DFT Modes}\label{sec4p1}

In the DFT, the temporal basis $\psi_{\mathcal{F}r}(t_k)$ is the Fourier basis, which we have already encountered in \eqref{Fourier_Pairs1}: that is $\psi_{\mathcal{F}r}(t_k)=1/\sqrt{n_t} \exp(2\pi f_r t_k)$ where $f_r=r \Delta f $, with $r=0,\dots n_t-1$, $\Delta f=f_s/n_t$ the frequency resolution and $f_s=1/\Delta t$ the sampling frequency. 

The only difference with respect to the previous section (and the classic implementations in Python or Matlab) is that we will now normalize the DFT by $1/\sqrt{n_t}$ and not $1/n_t$. This normalization makes the basis $\psi_{\mathcal{F}r}(t_k)$ \emph{orthonormal}: if we construct a vector $\psi_{\mathcal{F}r}[n]=1/\sqrt{n_t} \exp(2\pi f_r \,k\Delta)$, for any of the frequencies $f_r$ we will have $||\psi_r||_2=1$.

We now compute the DFT at \emph{every} location and arrange the results in the form of spatial modes. From a practical point of view, we can leverage the organization of the data into a snapshot matrix and compute the DFT at every row. Again, this will be very convenient and make the decomposition accessible in just one line of code.

Let us consider the DFT of the signal at one specific location $\mathbf{x}_i$:

\begin{equation}
	\label{Phi_F}
\mathcal{F}\{\vec{\bm{u}}(\bm{x_i},t_k)\} (\mathbf{x}_i,f_r)	= \frac{1}{\sqrt{n_t}}\sum^{n_t-1}_{k=0}\vec{\bm{u}}(\bm{x_i},t_k) e^{- 2\pi f_r \,t_k}\,= \vec{\bm{C}}_{\mathcal{F}r} (\bm{x}_i)\,.
\end{equation}

This operation can be interpreted in at least two ways. From a statistical point of view, this is a correlation (in time) between the temporal evolution of the signal at a certain position $\vec{\bm{u}}(\bm{x_i},t_k)$ and a certain basis element $\psi_{\mathcal{F}r} (t_k)$. From a linear algebra point of view, this is a projection (or `inner product') of the temporal evolution of the signal on to the basis element $\psi_{\mathcal{F}r} (t_k)$ (see \cite{MendezLS}).

The result of this operation is a set of coefficients $\vec{\bm{C}}_r (\bm{x}_i)$ (for the $u$ and the $v$ components) which is in essense a velociy field for every frequency $f_r$. This field represents the Fourier spectra of the time series of $u$ and the $v$ at that specific point $\mathbf{x}_i$ and pulsating at the single frequency $f_r$. For each of these fields, we can compute the norm (the space domain) as $\sigma_{\mathcal{F}r}=||\vec{\bm{C}}_{\mathcal{F}r}(\mathbf{x}_i)||_2$ and the normalized result will be denoted as $\bm{\vec{\phi}}_{\mathcal{F}r}=\vec{\bm{C}}_{\mathcal{F}r} (\bm{x}_i)/\sigma_{\mathcal{F}r}$. The set of $\bm{\vec{\phi}}_{\mathcal{F}r}$ gives the spatial structure (basis) of the DFT modes. 

The decomposition derived can be arranged as a matrix factorization of the snapshot matrix $\mathbf{D}=[\mathbf{d}_1,\mathbf{d}_2,\dots \mathbf{d}_{n_t}]\in \mathbb{R}^{n_s\times n_t}$, in which each column $\mathbf{d}_k$ contains the snapshot $\vec{\bm{u}}(\mathbf{x}_i,t_k)$ properly reshaped. For the DFT, this will be:

\begin{equation}
	\label{DFT_MATRIX}
	\mathbf{D}=\bm{\Phi}_{\mathcal{F}}\bm{\Sigma}_{\mathcal{F}}\mathbf{\Psi}_{\mathcal{F}}\,,
\end{equation} where $\bm{\Phi}_{\mathcal{F}}=[\phi_{\mathcal{F}1},\phi_{\mathcal{F}2},\dots \phi_{\mathcal{F}n_t}]\in \mathbb{R}^{n_s\times n_t}$ collects the spatial structures along its columns, $\bm{\Sigma}_{\mathcal{F}}$ is a diagonal matrix collecting the spectra $\sigma_{\mathcal{F}r}$ and $\bm{\Psi}_{\mathcal{F}}=[\psi_{\mathcal{F}1},\psi_{\mathcal{F}2},\dots \psi_{\mathcal{F}n_t}]\in \mathbb{R}^{n_t\times n_t}$ is the Fourier matrix collecting the temporal structures (note that this matrix is symmetric, i.e. $\bm{\Psi}_{\mathcal{F}}=\bm{\Psi}^T_{\mathcal{F}}$ ). Let us compute these for the provided dataset.

\begin{tcolorbox}[breakable, opacityframe=.1, title=Exercise 4: The DFT Modes of the TR-PIV data]
	
Consider the dataset in the first $t_k\leq 1$s. 
Using the \emph{fft} algorithm, compute the DFT modes. Plot the spectra $\sigma_{\mathcal{F}r}$ and the spatial structure of the modes with $f_r\approx 450 Hz$, namely the frequency of the vortex shedding.

\medskip
	\textbf{Solution}. We have the dataset already arranged in the form of snapshot matrices from the previous exercise. We load these and stack into a single snapshot matrix, to which we remove the time average:
	
		\begin{centering}
		\begin{lstlisting}[language=Python,linewidth=15.5cm,xleftmargin=.05\textwidth,xrightmargin=.05\textwidth,backgroundcolor=\color{yellow!10}]
# Load the snapshot matrices from Exercise 2
data = np.load('Snapshot_Matrices.npz')
Xg=data['Xg']; Yg=data['Yg']
D_U=data['D_U']; D_V=data['D_V']
# Assembly both components in one single matrix
D_M=np.concatenate([D_U,D_V],axis=0) # Reshape and assign
# We remove the time average (for plotting purposes)
D_MEAN=np.mean(D_M,1) # Temporal average (along the columns)
D=D_M-np.array([D_MEAN,]*n_t).transpose() # Mean Removed
		\end{lstlisting}
	\end{centering}

The computation of the DFT fields $\vec{\bm{C}}_{\mathcal{F}r}$ can be done in one line of code:

		\begin{centering}
	\begin{lstlisting}[language=Python,linewidth=15.5cm,xleftmargin=.05\textwidth,xrightmargin=.05\textwidth,backgroundcolor=\color{yellow!10}]
PHI_SIGMA = (np.fft.fft(D, n_t, 1)) / (n_t ** 0.5)
	\end{lstlisting}
\end{centering}

\hspace{2mm}Note that this is the product $\bm{\Phi}_{\mathcal{F}}\bm{\Sigma}_{\mathcal{F}}$ in \eqref{DFT_MATRIX} (hence the name). The remaining steps are (1) a normalization of all the columns, storing the results of the amplitudes in a vector:

\begin{centering}
		\begin{lstlisting}[language=Python,linewidth=15.5cm,xleftmargin=.05\textwidth,xrightmargin=.05\textwidth,backgroundcolor=\color{yellow!10}]
# Initialize the PHI_F MATRIX (spatial structures)
PHI_F = np.zeros((D.shape[0], n_t), dtype=complex)  
# Initialize the SIGMA_F vector (DFT Spectra)
SIGMA_F = np.zeros(n_t)  
# Normalization loop
for r in tqdm(range(0, n_t)): 
 SIGMA_F[r] = abs(np.vdot(PHI_SIGMA[:, r], PHI_SIGMA[:, r]))** 0.5
 PHI_F[:, r] = PHI_SIGMA[:, r] / SIGMA_F[r]
		\end{lstlisting}
	\end{centering}

and (2) the sorting of the modes in decreasing order of amplitude:

	\begin{centering}
		\begin{lstlisting}[language=Python,linewidth=15.5cm,xleftmargin=.05\textwidth,xrightmargin=.05\textwidth,backgroundcolor=\color{yellow!10}]
# find indices for sorting in decreasing order
Indices = np.flipud(np.argsort(SIGMA_F))  
# Sort all the sigmas
Sorted_Sigmas = SIGMA_F[Indices]  
# Sort all the frequencies accordingly.
Sorted_Freqs = Freqs[Indices]  
# Sorted Spatial Structures Matrix
Phi_F = PHI_F[:, Indices]  
# Sorted Amplitude Matrix
SIGMA_F = np.diag(Sorted_Sigmas)  
		\end{lstlisting}
	\end{centering}
	
where the vector `Freqs' is the one computed using scipy's function

\begin{centering}
	\begin{lstlisting}[language=Python,linewidth=15.5cm,xleftmargin=.05\textwidth,xrightmargin=.05\textwidth,backgroundcolor=\color{yellow!10}]
Freqs = np.fft.fftfreq(n_t) * Fs  # Compute the frequency bins
	\end{lstlisting}
\end{centering}

\hspace{2mm}This is essentially $f_r=r\Delta f$ defined in the previous section. The resulting spectra is plotted in Figure \ref{figEx41}a) versus the associated frequencies and in Figure \ref{figEx41}b) in terms of the mode index $r$. The first plot shows that, as expected, the leading modes are the ones at the frequency of the vortex shedding. The second plot shows that the `decay rate' of the amplitude is modest, and many modes are needed to approximate this dataset despite its strong oscillatory nature. We conclude that despite the periodicity in the data, the convergence of the DFT is not particularly strong.
\begin{center}%
	\setcounter{subfigure}{0}%
	
	\begin{minipage}{.49\linewidth}
		\includegraphics[width=\linewidth]{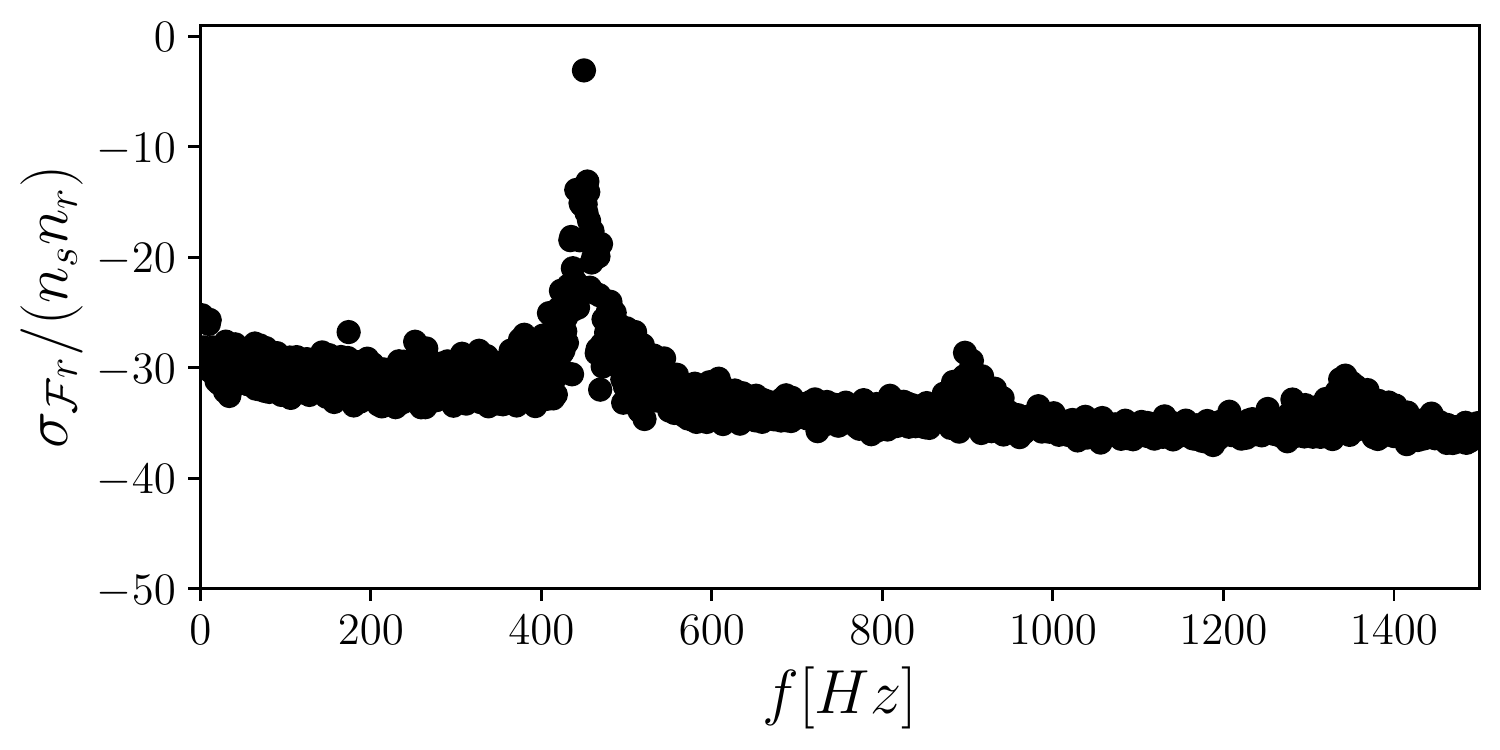}
		\label{Ex4a)}%
		\vspace{-3mm}
		\center{a)}
	\end{minipage}
	\begin{minipage}{.49\linewidth}
		\includegraphics[width=\linewidth]{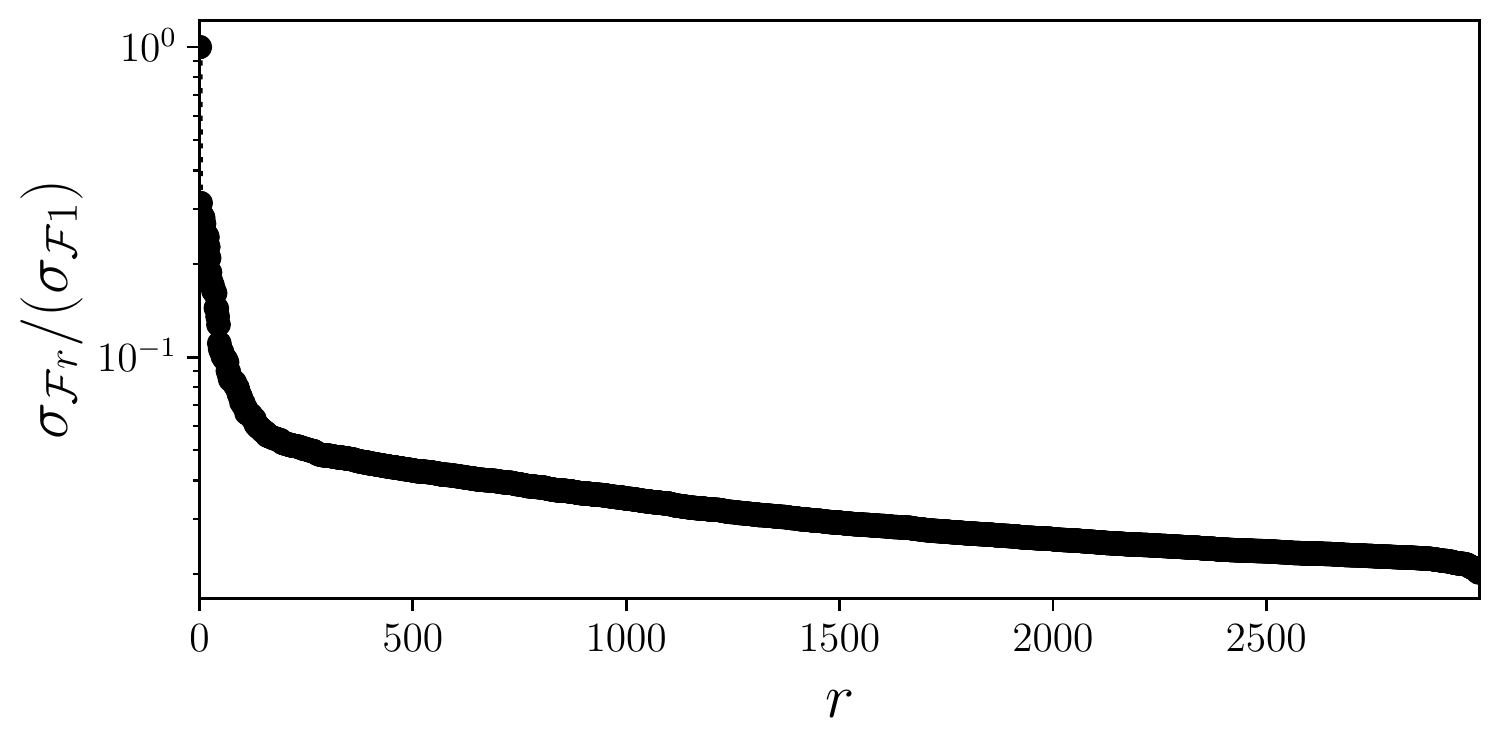}%
		\label{Ex4b)}%
		\vspace{-3mm}
		\center{b)}
	\end{minipage}	
	\vspace{-0.05cm}
	\captionof{figure}{	Fig (a): Amplitudes of the DFT modes $\sigma_{\mathcal{F}r}$ as a function of the associated frequency $f_r$. The leading modes are close to the vortex shedding frequency. Fig (b): amplitude of the DFT modes $\sigma_{\mathcal{F}r}/\sigma_{\mathcal{F}1}$, in decreasing order, as a function of the index $r$. }	
	\label{figEx41}
\end{center}

\hspace{2mm}The spatial structures can then be shown after a reshaping. The leading structure (see entry Sorted\_Freqs[0] has f\_0=450 Hz) and the reader should note that modes come in pairs (only the postive portion of the spectra is here shown, from $f_r=0$ to $f_r=fs/2$). The real and the imaginary part of the leading DFT modes can be reshaped as follows:

\begin{centering}
	\begin{lstlisting}[language=Python,linewidth=15.5cm,xleftmargin=.05\textwidth,xrightmargin=.05\textwidth,backgroundcolor=\color{yellow!10}]
r=0; 
# Take the real part of mode r
Phi_F_Mode_U_r=np.real(Phi_F[0:nxny,r]).reshape((n_x,n_y))
Phi_F_Mode_V_r=np.real(Phi_F[nxny::,r]).reshape((n_x,n_y))
# Take the imaginary part of mode r
Phi_F_Mode_U_i=np.imag(Phi_F[0:nxny,r]).reshape((n_x,n_y))
Phi_F_Mode_V_i=np.imag(Phi_F[nxny::,r]).reshape((n_x,n_y))
	\end{lstlisting}
\end{centering}

\hspace{2mm}These fields are shown in Figure \ref{figEx42}. Their superposition describes the travelling of vortices whose size is comparable to the cylinder. The two modes provide a simple kinematic approximation of the vortex shedding mechanism and explain why the signals in probe P2 and P3 have a strong level of coherency at $f_n\approx 450 $ Hz: both probes are located in regions where the first mode is `strongly present'. 

\hspace{2mm}While this exercise nicely showcased the strengths of the DFT; a simple variant can quickly showcase its limitations: the reader is invited to repeat the same exercise taking now all the $n_t=13000$ snapshots of the dataset. The convergence decreases significantly, and two dominant frequencies will now appear (linked to the vortex shedding in the two stationary conditions, as described in Sec.\ref{sec2}). These will have their associated modes, but the DFT will leave an important question open: \emph{when} are these modes active? 

\hspace{2mm}An extension of tools from time-frequency analysis (e.g. Wavelets, \cite{StefanoLS}) to the `modal' formalism could address these points but does not always improve the convergence and keeps the decomposition 'general', i.e. with bases not adapted to the dataset at hand. The POD introduced in the next session gives the best possible convergence and tailors its basis to the data. 

	\begin{center}%
		\setcounter{subfigure}{0}%
		
		\begin{minipage}{.49\linewidth}
			\includegraphics[width=\linewidth]{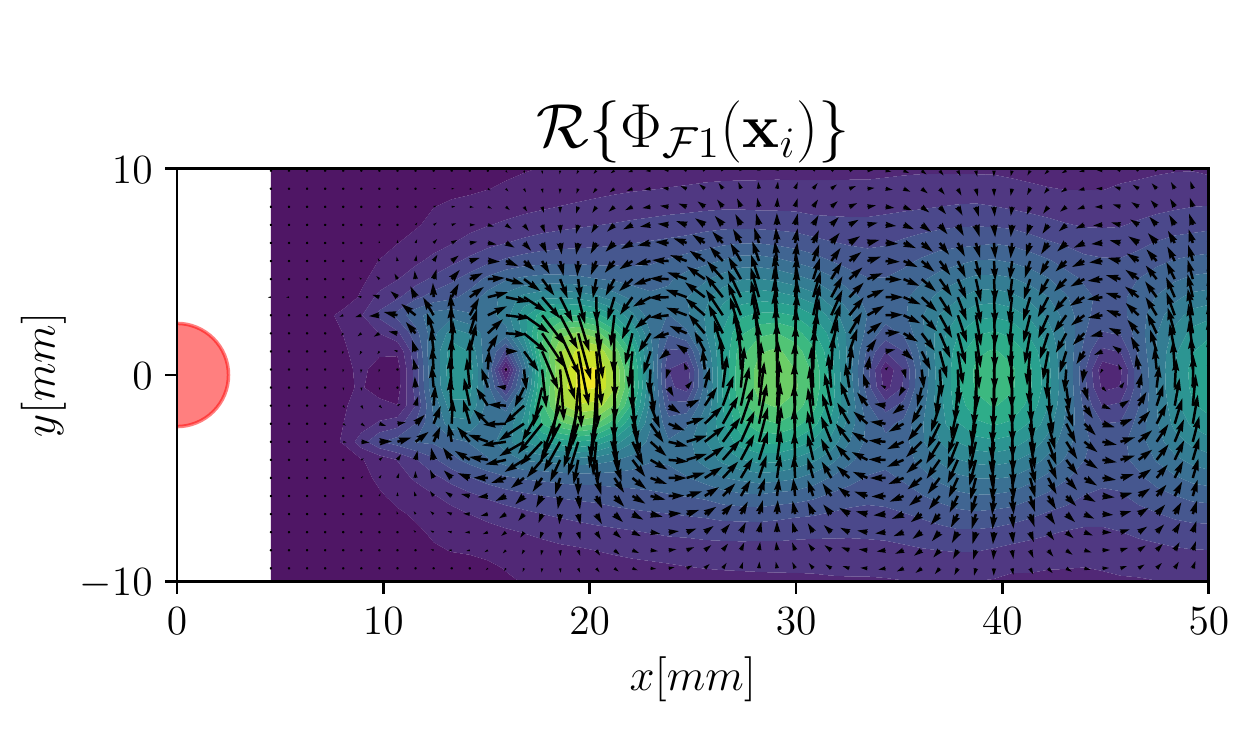}
			\label{Ex4aaa)}%
			\vspace{-3mm}
			\center{a)}
		\end{minipage}
		\begin{minipage}{.49\linewidth}
			\includegraphics[width=\linewidth]{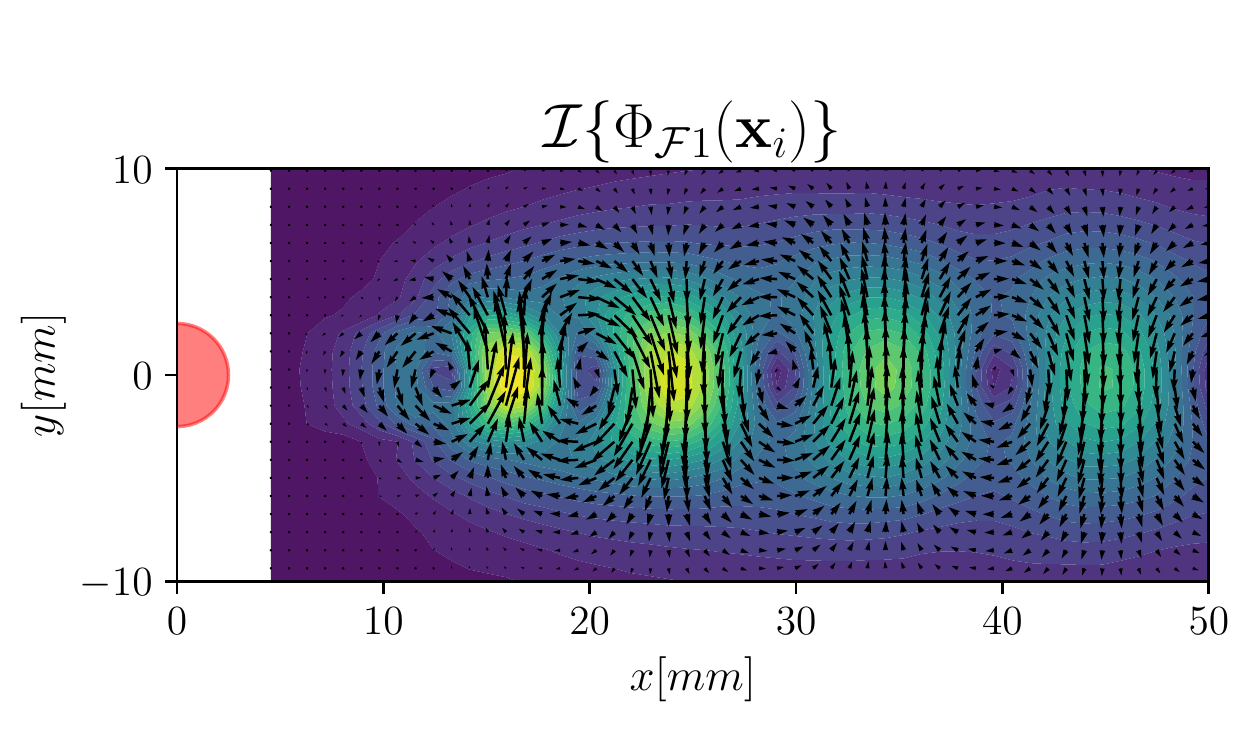}%
			\label{Ex4bbb)}%
			\vspace{-3mm}
			\center{b)}
		\end{minipage}	
		\vspace{-0.05cm}
		\captionof{figure}{	Real (a) and imaginary (b) parts of the spatial structure of the leading DFT mode, capturing the vortex shedding. }	
		\label{figEx42}
	\end{center}
\end{tcolorbox}


\subsection[The POD]{The Proper Orthogonal Decomposition}\label{sec4p2}

The previous exercise has highlighted the frequency localization capabilities of the DFT, but also its (generally) poor convergence. The DFT does not qualify as a `data-driven' decomposition because the basis employed is pre-defined: in the DFT, we use the Fourier basis regardless of the data at hand.

We now enter into data-driven decompositions and thus seek to tailor the basis to the data. The Proper Orthogonal Decomposition (POD) is the oldest technique and arguably the most robust and simple tool. The decomposition is known with different names in different fields, and the reader is referred to \cite{ScottLS} for a review.

The POD is the fundamental `energy-based' decomposition: it provides the modes with the highest amplitude, and such that the convergence of an approximation with increasing number of modes is optimal. Identifying the temporal structures satisfying this condition is a constrained optimization problem (see \cite{Holmes1996} or \cite{Bishop2006} in the framework of Principal Component Analysis, PCA) and we here give the solution without proofs: the POD's temporal structures $\psi_{\mathcal{P}r}$ must be eigenvectors of the temporal correlation matrix $\bm{K}[k,n]=\langle \vec{\bm{u}}(\bm{x_i},t_k),\vec{\bm{u}}(\bm{x_i},t_n)\rangle_S$, with $\langle \bullet\rangle_S$ the expectation operator in the space domain. Hence, we must have:

\begin{equation}
	\label{EIG}
	\bm{K}\,\psi_{\mathcal{P}r} (t_k)=\lambda_r\,\psi_{\mathcal{P}r} (t_k)\quad \forall r\,\in[1,R]\longrightarrow \bm{K}=\sum^{R}_{r=1} \lambda_r\,\psi_{\mathcal{P}r} (t_k)\psi^T_{\mathcal{P}r} (t_k)\,,
\end{equation} having considered each temporal structure as a column vector, and having recalled, in the last summation, that the correlation matrix is symmetric and positive definite. The $\psi_{\mathcal{P}r}$ are, therefore, orthonormal like the $\psi_{\mathcal{F}r}$ in the DFT. What gives the POD the special status of `proper' orthogonality, however, is that this specific choice of the temporal structure also makes the associated spacial structures $\phi_{\mathcal{P}r}$ orthonormal.

In addition to optimal convergence, this `proper orthogonality' brings two interesting by-products (see also \cite{Mendez2019,MendezLS2}). The first is that the amplitudes of the modes are $\sigma_{\mathcal{P}r}=\sqrt{{\lambda}_r}$ and thus the normalization step in the calculation of the spatial structures in \eqref{EQ1} is no longer necessary. Then, it is possible to compute the spatial structures via correlation:

\begin{equation}
	\label{Siro}
	\bm{\vec{\phi}}_{k} (\bm{x}_i)=\frac{1}{\sigma_r}\bigl \langle \vec{\bm{u}}(\bm{x_i},t_k), \psi_{\mathcal{P}r}(t_k) \bigr \rangle_T=\frac{1}{\sigma_r} \sum^{n_t}_{k=1} \vec{\bm{u}}(\bm{x_i},t_k)\,\psi_{\mathcal{P}r}(t_k)\,.
\end{equation}

The second by-product is that the spatial structures are eigenvectors of the spatial correlation matrix $\mathbf{C}[\bm{i},\bm{i}]=\langle \vec{\bm{u}}(\bm{x_i},t_k),\vec{\bm{u}}(\bm{x_j},t_k)\rangle_S$, with $\langle \bullet\rangle_S$ the inner product in the time domain. Their orthogonality allows also to write:

\begin{equation}
	\label{Classic}
	\psi_{\mathcal{P}r}(t_k)=\frac{1}{\sigma_r}\bigl \langle \vec{\bm{u}}(\bm{x_i},t_k), \bm{\vec{\phi}}_{k} (\bm{x}_i) \bigr \rangle_S=\frac{1}{\sigma_r} \sum^{n_s}_{i=1} \vec{\bm{u}}(\bm{x_i},t_k)\,\bm{\vec{\phi}}_{k} (\bm{x}_i)\,.
\end{equation}

Therefore, one might decide to work with the diagonalization of $\mathbf{K}$ or $\mathbf{C}$ depending on wheter one has $n_t\ll n_s$ or $n_s\ll 1$, and compute by correlation the remaining structures (either the $\phi_{\mathcal{P}r}$'s via \eqref{Siro} or the $\psi_{\mathcal{P}r}$ via \eqref{Classic}).

Once completed, the arranging the POD in the matrix formalism as in \eqref{DFT_MATRIX} becomes: 

\begin{equation}
	\label{POD_Matrix}
	\mathbf{D}=\bm{\Phi}_{\mathcal{P}}\bm{\Sigma}_{\mathcal{P}}\mathbf{\Psi}^T_{\mathcal{P}}\,.
\end{equation} 

For datasets uniformly spaced in time and space, this is a Singular Value Decomposition (SVD) of $\mathbf{D}$. We see this decomposition in action in the next exercise.

\begin{tcolorbox}[breakable, opacityframe=.1, title=Exercise 5: The POD Modes of the TR-PIV data]
	
	Compute the POD of the full dataset (i.e. $n_t=13200$, corresponding to $t_k$). Show the amplitude of the modes, their spatial structures and the frequency content of their temporal structures.

	\medskip
	\textbf{Solution}.  We should re-run the script from Exercise 2 and build the full dataset of $n_t=13200$ snapshot. Assuming that this is stored in a matrix D as before, the first step is to compute the temporal correlation matrix. This is simply $D^T D$. However, because the computational cost of this operation hits the limit of the author's laptop, we here use the memory saving feature of the MODULO package \citep{Ninni2020}, recently implemented in Python by \cite{Schena}. The following code loads the data and computes the snapshot matrix:
	
	\begin{centering}
		\begin{lstlisting}[language=Python,linewidth=15.5cm,xleftmargin=.05\textwidth,xrightmargin=.05\textwidth,backgroundcolor=\color{yellow!10}]
# Load the snapshot matrices from Exercise 2
data = np.load('Snapshot_Matrices.npz')
Xg=data['Xg']; Yg=data['Yg']
D_U=data['D_U']; D_V=data['D_V']
# Assembly both components in one single matrix
D_M=np.concatenate([D_U,D_V],axis=0) # Reshape and assign
# We remove the time average (for plotting purposes)
D_MEAN=np.mean(D_M,1) # Temporal average (along the columns)
D=D_M-np.array([D_MEAN,]*n_t).transpose() # Mean Removed
		\end{lstlisting}
	\end{centering}
	
\hspace{2mm}To use the MODULO package, we first create a MODULO object with some important information, then run the POD computation in one line: 
	
	\begin{centering}
		\begin{lstlisting}[language=Python,linewidth=15.5cm,xleftmargin=.05\textwidth,xrightmargin=.05\textwidth,backgroundcolor=\color{yellow!10}]
m = MODULO(data=D, 
           n_Modes = n_r,
           MEMORY_SAVING=True, 
           N_PARTITIONS=11)
# Prepare (partition) the dataset
D = m._data_processing() # Note that this will cancel D!
# Compute the mPOD
Phi_P, Psi_P, Sigma_P = m.compute_POD()
		\end{lstlisting}
	\end{centering}

\hspace{2mm}The first line in the script prepares the object declaring that the data to decompose is in the matrix D, that we look for the first 500 modes, and that the memory saving feature will be requested with $11$ partitions \citep{Ninni2020}. This means that the code will break the dataset into 11 chunks and store it in temporary files, releasing some memory. Then, during the computation of the correlation matrix, the code loads and use one piece at a time. This allows computing the temporal correlation matrix saving considerable computational resources. The partitioning is created as soon as the second line in the script is run. Finally, the last line contains the computation of the POD.

\hspace{2mm}The rest of the code is an exercise in reshaping and plotting data. Figure \ref{fig511} shows the plot of the POD mode amplitude from the diagonal of Sigma\_P. Interestingly, the amplitude of all modes tends to zero for $r>3$: the first three modes contains most of the information in the data. 

	 \begin{center}%
		\setcounter{subfigure}{0}%
		\begin{minipage}{1\linewidth}
			\centering
			\includegraphics[width=0.6\linewidth]{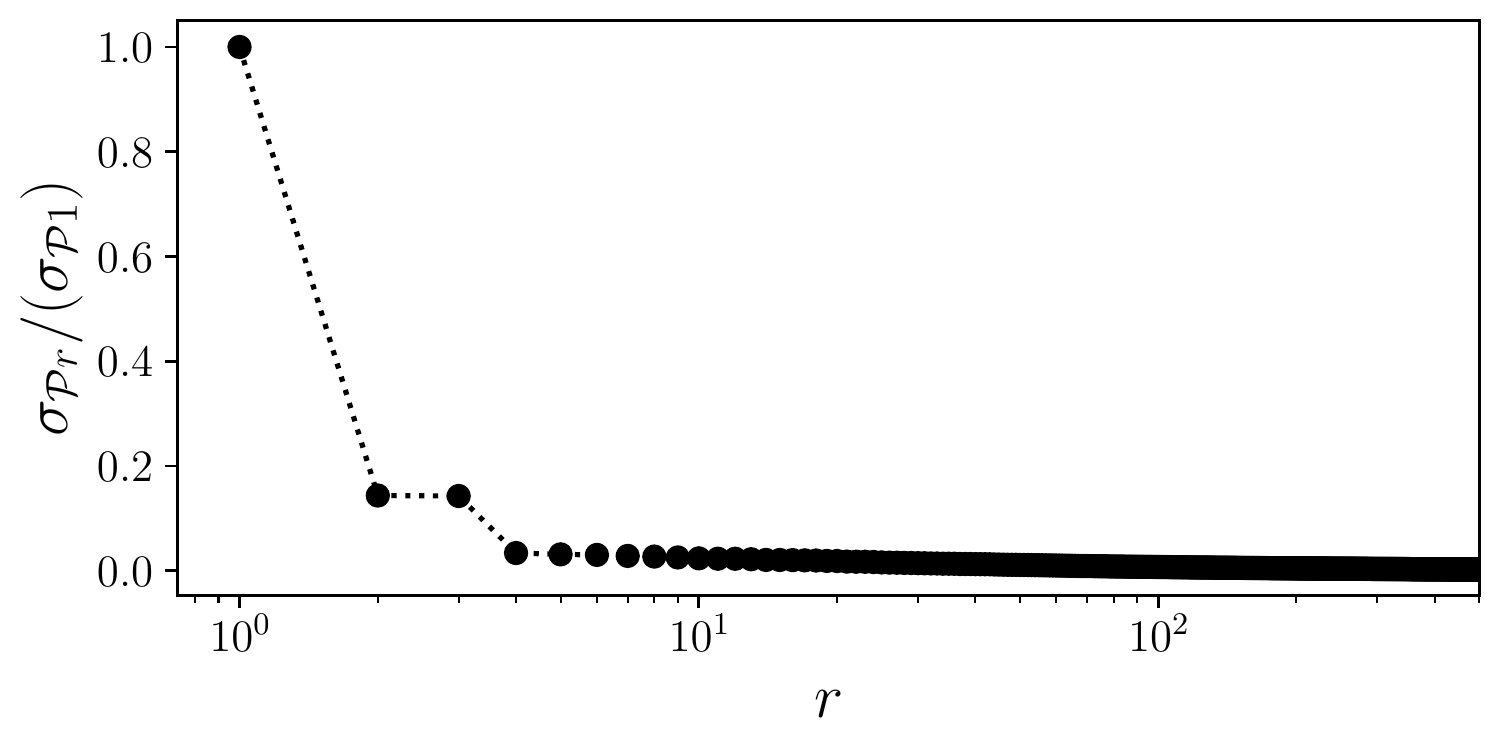}
			\label{fig511}%
			\captionof{figure}{	Amplitude decay of the POD Modes for the full dataset.}
		\end{minipage}
		
	\end{center}
	
	\hspace{2mm}As in the previous exercise, the spatial structures are first reshaped onto the PIV grid and then plotted as fields. Figure \ref{figEx55a} collects the spatial structures and the spectra of the first three leading modes. These modes are extensively discussed in \cite{Mendez2020}. Briefly, the first mode describes the large scale variation of the flow, while the other two represent the vortex shedding in its entire evolution. Because POD modes are real, travelling patterns can only be described by POD pairs whose temporal structures are in quadrature.
	
	\hspace{2mm}The compression achievable by the POD is interesting for many purposes. From a `dynamical system' perspective, the dataset can be seen as a discrete system in $\mathbb{R}^{n_s}$, with $n_s=4320$ spatial points ($2130$ per components). Predicting the flow field a time $t_{k+1}$ requires predicting $n_s=4320$ numbers. If one approximates the dataset with respect to the (reduced) POD basis:
	
\begin{equation}
	\label{POD_A}
	\tilde{\bm{u}}(\mathbf{x}_i,t_k)=\sigma_{\mathcal{P}1}\bm{\vec{\phi}}_{\mathcal{P}1} (\bm{x}_i) \psi_{\mathcal{P}2} (t_k) +\sigma_{\mathcal{P}2}\bm{\vec{\phi}}_{\mathcal{P}1} (\bm{x}_i) \psi_{\mathcal{P}2} (t_k) +\sigma_{\mathcal{P}3}\bm{\vec{\phi}}_{\mathcal{P}3} (\bm{x}_i) \psi_{\mathcal{P}3} (t_k) \,,
\end{equation} the flow field at time $t_{k+1}$ is prescribed by $3$ numbers, namely the value taken by the temporal structures $\psi_{\mathcal{P}} (t_k) $. The three temporal coefficients evolve according to a dynamical system in $\mathbb{R}^3$ and the spatial structures in \eqref{POD_A} map (linearly!) this small dimensional system to the original one, producing an error which can be shown to be equal to $\sigma_4$ in a $l_2$ sense \citep{ScottLS, MendezLS}.

	\begin{center}%
		\setcounter{subfigure}{0}%
		
		\begin{minipage}{.49\linewidth}
			\includegraphics[width=\linewidth]{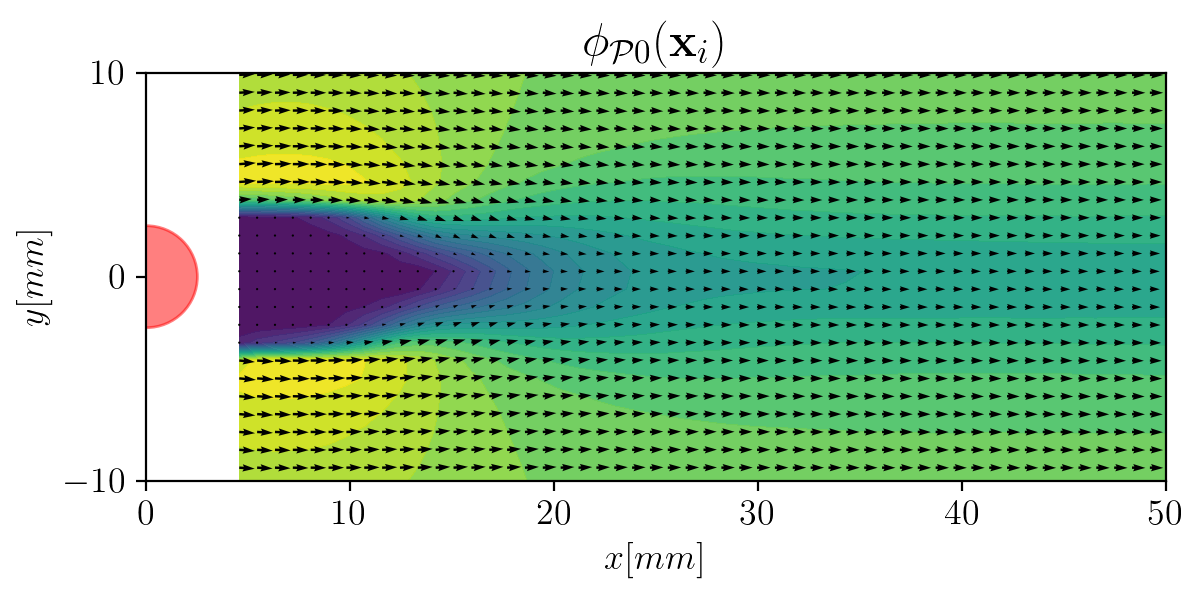}
			\label{Ex5aaaa)}%
			\vspace{-3mm}
			\center{a)}
		\end{minipage}
		\begin{minipage}{.49\linewidth}
			\includegraphics[width=\linewidth]{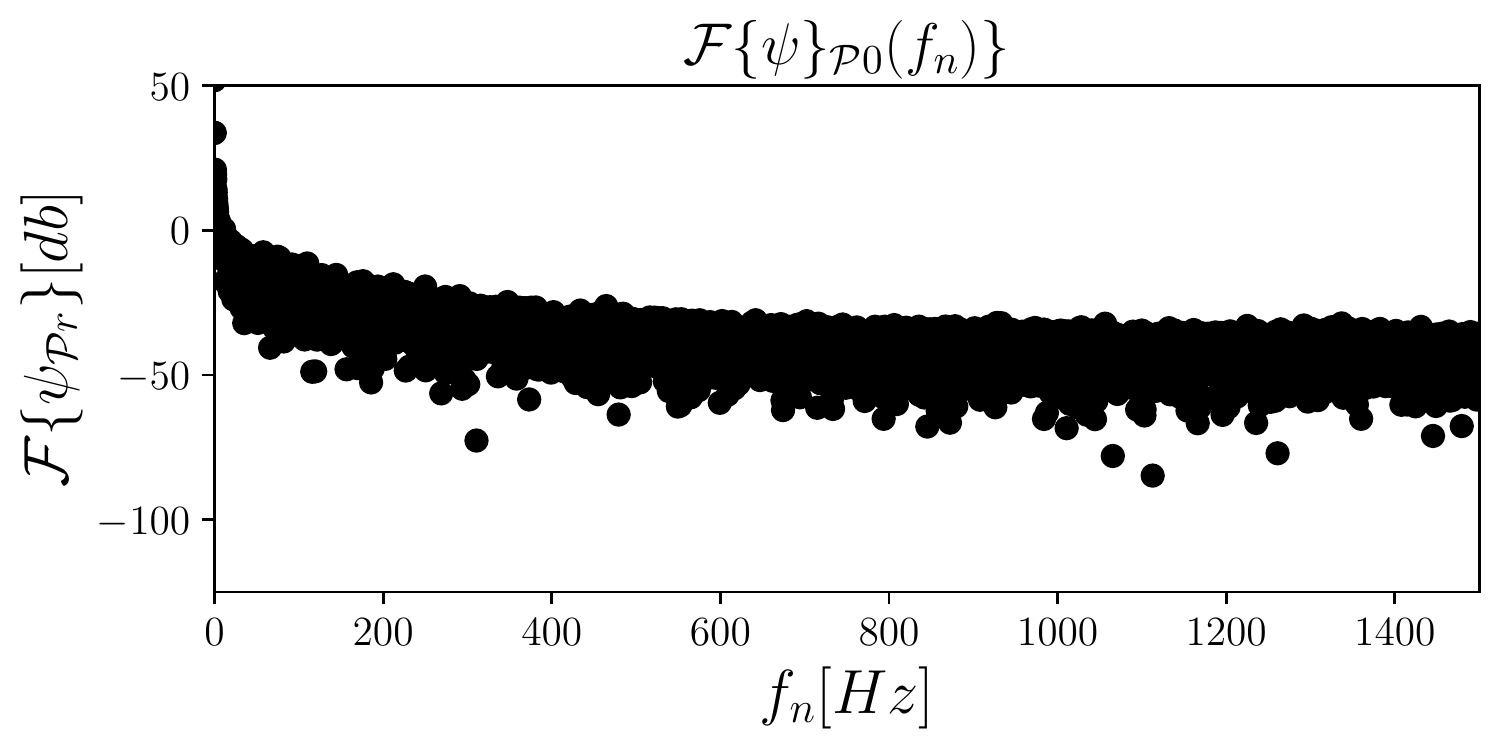}%
			\label{Ex5bbbb)}%
			\vspace{-3mm}
			\center{b)}
		\end{minipage}	
			\begin{minipage}{.49\linewidth}
		\includegraphics[width=\linewidth]{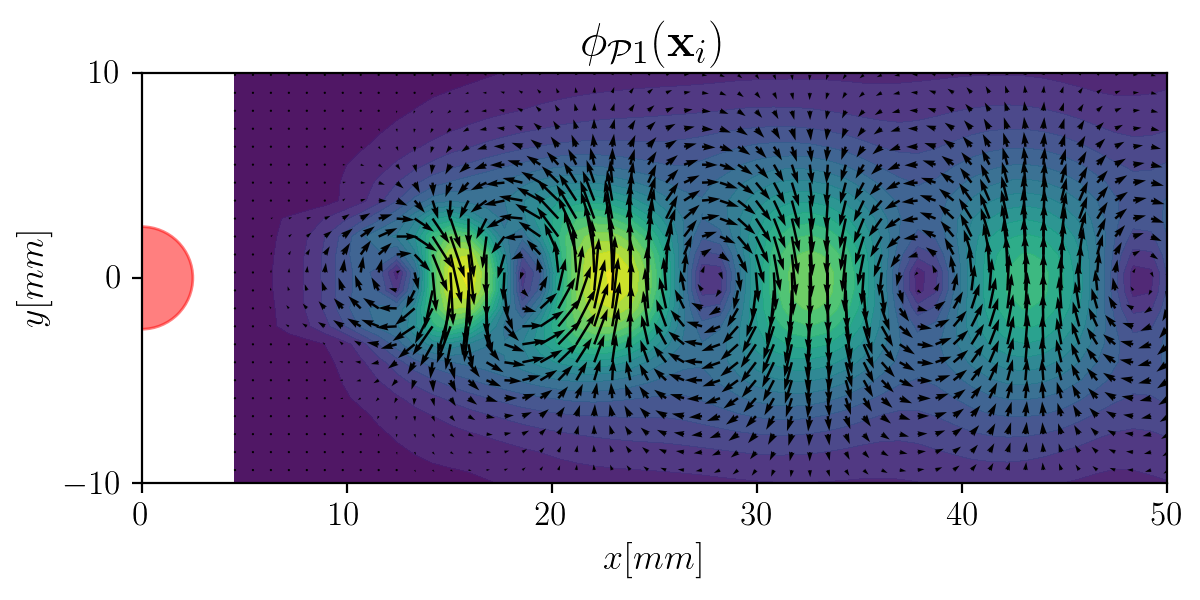}
		\label{Ex5aaa)}%
		\vspace{-3mm}
		\center{c)}
	\end{minipage}
	\begin{minipage}{.49\linewidth}
		\includegraphics[width=\linewidth]{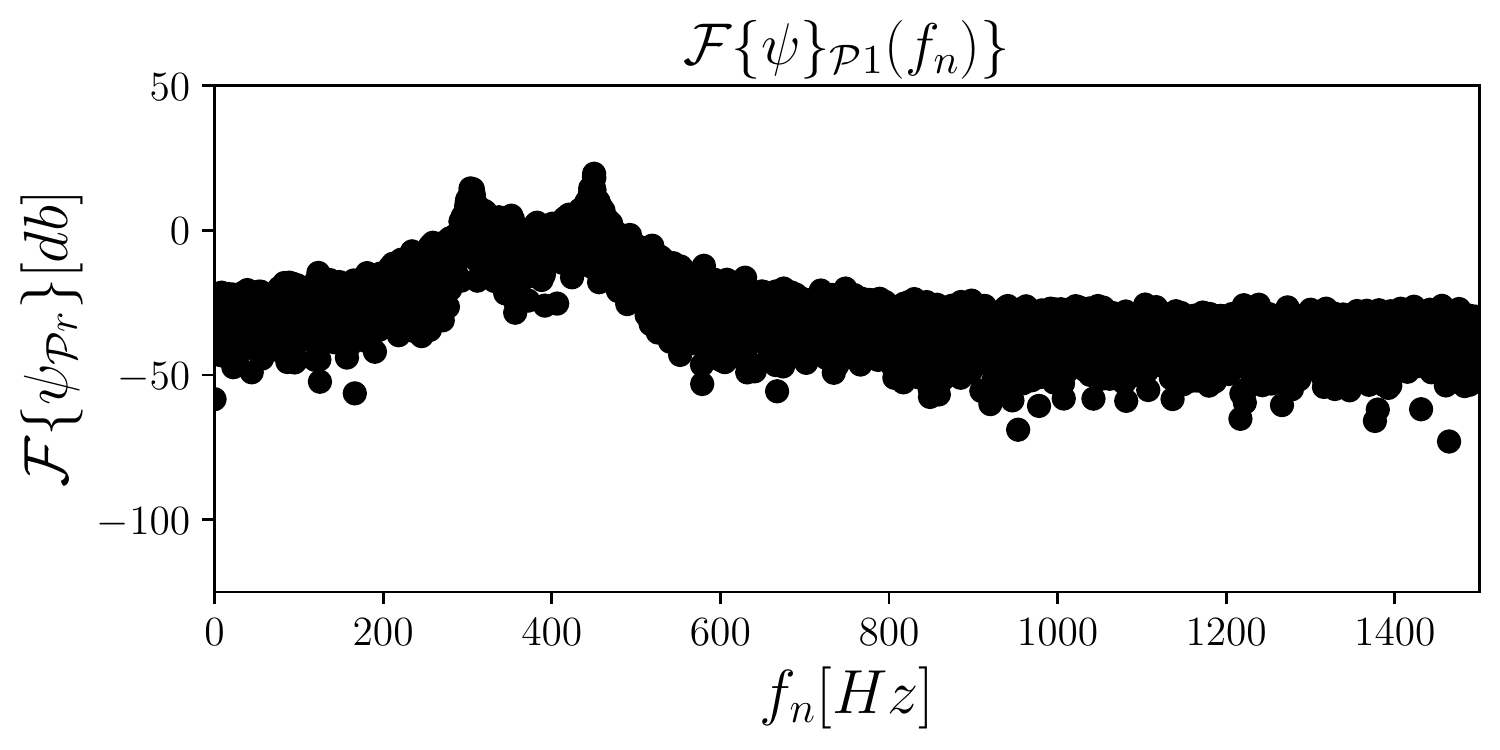}%
		\label{Ex5bbbbb)}%
		\vspace{-3mm}
		\center{d)}
	\end{minipage}	
		\begin{minipage}{.49\linewidth}
	\includegraphics[width=\linewidth]{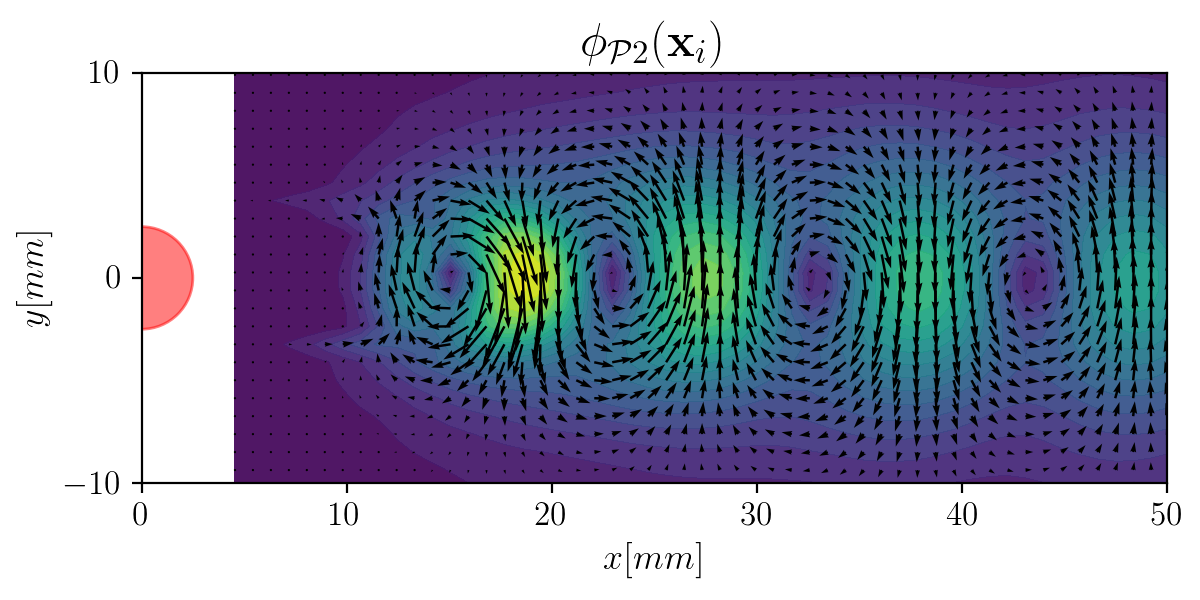}
	\label{Ex5aaaaa)}%
	\vspace{-3mm}
	\center{e)}
\end{minipage}
\begin{minipage}{.49\linewidth}
	\includegraphics[width=\linewidth]{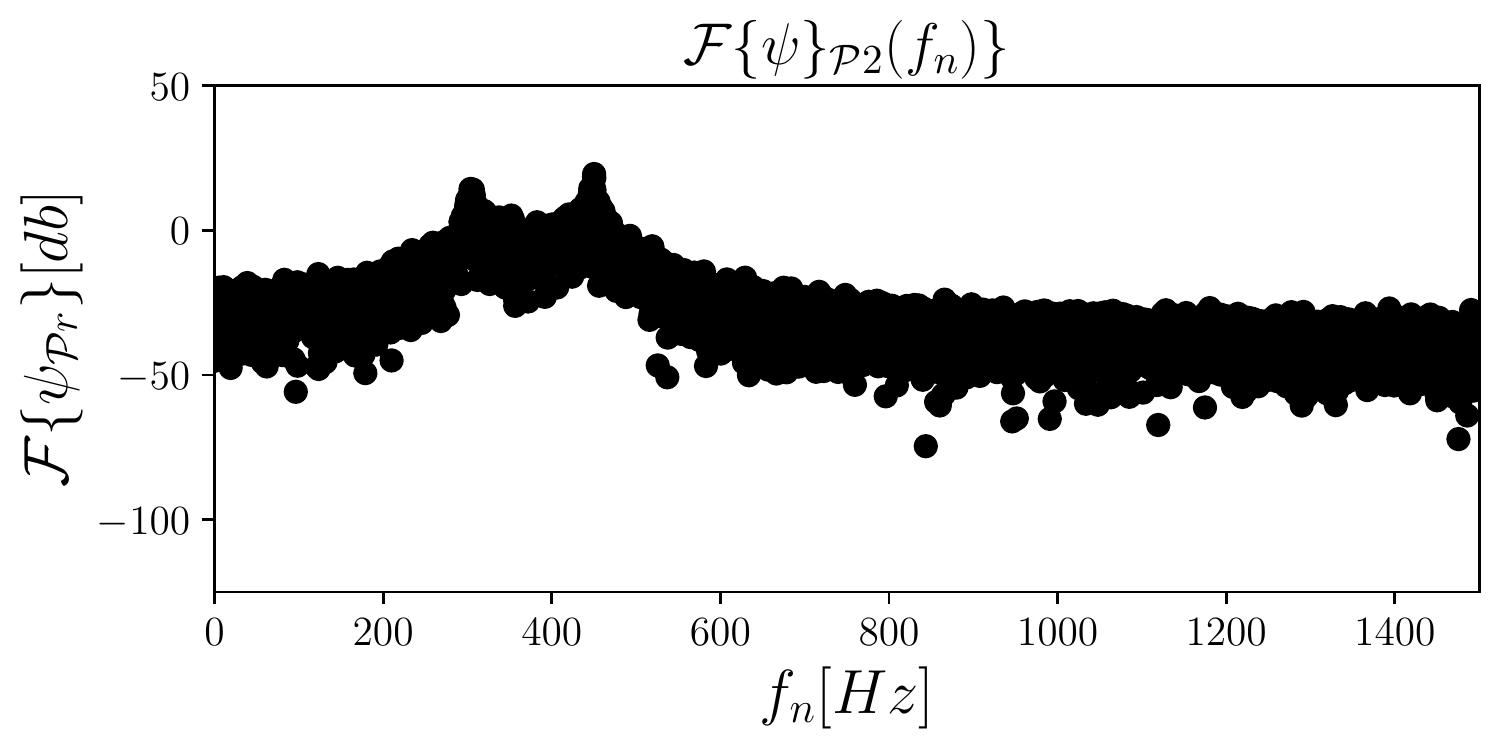}%
	\label{Ex5bbbbbb)}%
	\vspace{-3mm}
	\center{f)}
\end{minipage}	
		\vspace{-0.05cm}
		\captionof{figure}{	Spatial structures of the first three POD modes (a) and spectra of the associated temporal structures (b). }	
		\label{figEx55a}
	\end{center}

\hspace{2mm}Figure \ref{Orbit3D} shows the trajectory of the reduced dynamical system in $\mathbb{R}^3$, which is simply the plot of the three temporal structures. In stationary conditions, the modes $2$ and $3$ describe a circular orbit whose diameter is larger at larger velocities. The transient phase moves the system state from the large to the smaller orbits along the axis of the first mode. At all times, the system stays on a paraboloid. The existence of this paraboloid has been derived analytically at low Reynolds numbers ($Re=100$, see \cite{Noack_LS}); this exercise shows that this manifold continues to exists also at much higher $Re$ and during transient states.

 \begin{center}%
	\setcounter{subfigure}{0}%
	\begin{minipage}{1\linewidth}
		\centering
		\includegraphics[width=0.9\linewidth]{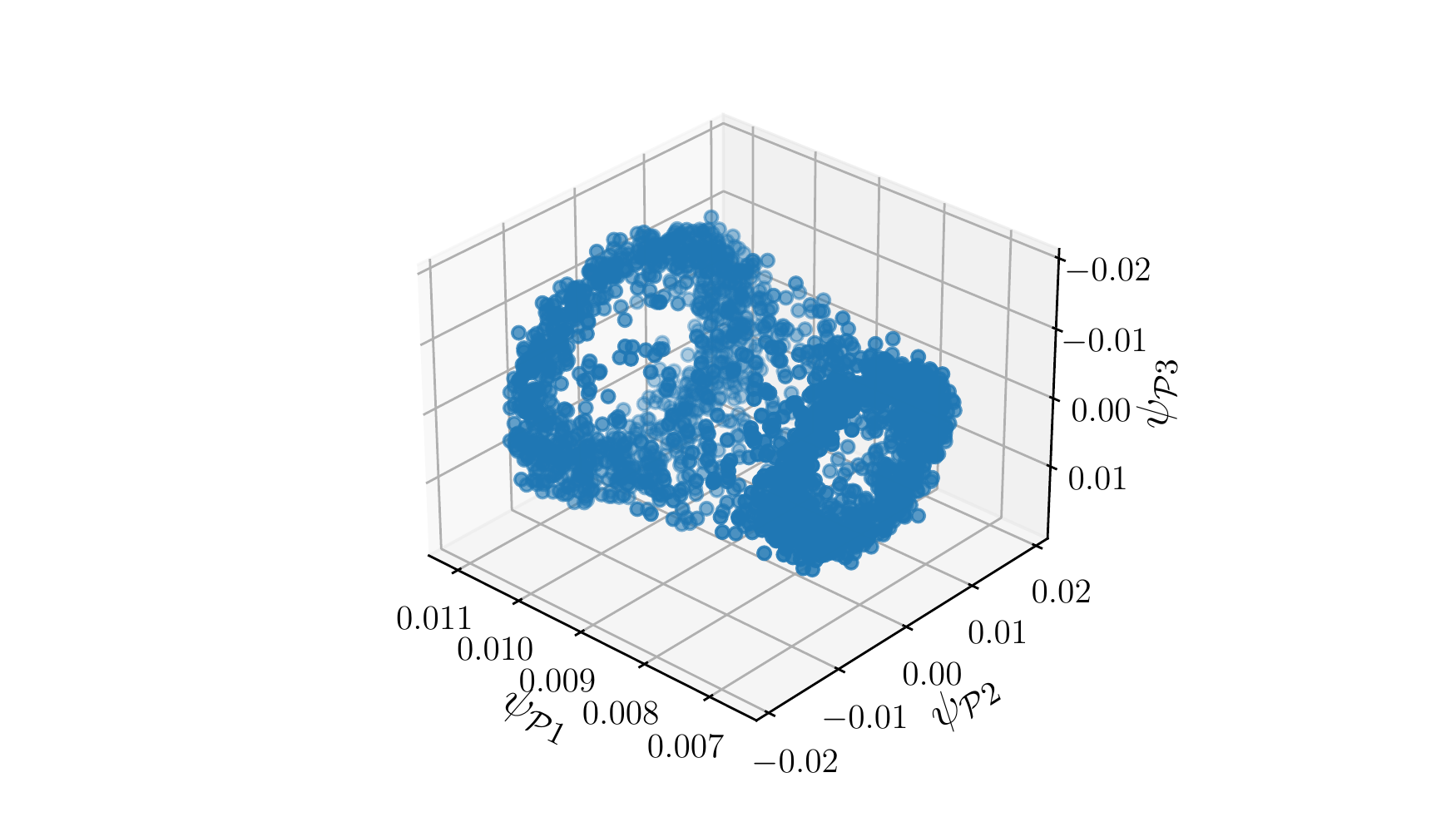}
		\label{Orbit3D}%
		\captionof{figure}{Evolution of the reduced order dynamical system in the POD basis.}
	\end{minipage}
	
\end{center}
\end{tcolorbox}

\subsection[The DMD]{The Dynamic Mode Decomposition}\label{sec4p3}
The DFT modes have one frequency each, but this frequency is constrained to be an integer multiple of the fundamental frequency $f_0=1/T$ with $T=n_t \Delta t$ the observation time. This potentially generates convergence issues due to \emph{windowing} problems. The POD modes have the best possible convergence, but its modes are of difficult interpretation: all the information is extremely compressed into a few modes. 

The Dynamic Mode Decomposition (DMD) solves potential windowing problems while putting the ``frequency-based'' formalism of the DFT into the ``data-driven'' or ``'tailored-basis'' formalism of the POD. The idea (introduced by \cite{Schmid}) is to fit a linear dynamical system onto a reduced dynamical system derived via POD. A different formulation was proposed by \cite{Rowley2} and the reader is referred to \cite{Chen2012} for an interesting discussion on the link between DMD and DFT.

The idea of fitting a linear dynamical system to the POD reduced-order model of the data has a long history, and it is interesting to note that modern DMD algorithms are remarkably similar to what was introduced in the late '80s as  Principal Oscillation Patterns (POP, \cite{Hasselmann1988,Storch1990}) or Linear Inverse Modeling (LIM, \cite{Penland1996,LIM1}) in climatology (see also \cite{Kutz2014,Mendez2020,Vlacav1}).

The basic algorithm described in this section is the one  implemented in MODULO and proposed by \cite{Kutz2014}, which is briefly described in this section. Let $\bm{D}_1=[d_1,\dots, d_{n_t-1}]\in \mathbb{R}^{n_t-1}$ and $\bm{D}_2=[d_2,\dots,d_{n_t}]\in \mathbb{R}^{n_t-1}$ be two portions of the snapshot matrix differing by a one shift in time. If a linear dynamical system 'moves' $\bm{D}_1$ to $\bm{D}_2$, then this system is characterized by a propagator matrix $\bm{P}\in{R}^{n_s\times n_s}$ such that $\bm{D}_2=\bm{P} \bm{D}_1$. The eigenvalues of this propagator governs the evolution of the system and its eigenvectors are the DMD modes we are looking for.

Solving for $\bm{P}$ is a canonical least square problem, which could be solved via the pseudo-inverse of $\bm{D}_1$. This can be computed via POD (i.e. SVD) of $\bm{D}_1$:

\begin{equation}
	\label{Prop}
	\bm{P}\approx \bm{D}_2 \mathbf{\Psi}_{\mathcal{P}}\bm{\Sigma}^{-1}_{\mathcal{P}}\bm{\Phi}^T_{\mathcal{P}}
\end{equation}

However, computing the eigendecomposition of $\bm{P}$ is a poorly conditioned and computationally prohibitive problem in most real applications. The solution is then to leverage on the compression achievable by POD. Projecting the whole problem onto a reduced POD basis $\tilde{\bm{\Phi}}_{\mathcal{P}}=[\phi_{\mathcal{P}1},\dots \phi_{\mathcal{P}n_r}]$, with $n_r\ll n_t$, the reduced system in $\mathbb{R}^{n_r\times n_r}$ becomes:

\begin{equation}
	\label{PROP}
	\bm{D}_2\approx \bm{P}\,\bm{D_1}\longrightarrow    \underbrace{\bm{\tilde{\Phi}}^{T}_{\mathcal{P}}\bm{D}_2}_{\tilde{\large{\bm{V}}_2}} \approx \underbrace{\bm{\tilde{\Phi}}^{T}_{\mathcal{P}} \bm{P} \bm{\tilde{\Phi}}_{\mathcal{P}}}_{\large{\tilde{\bm{S}}}} \,\underbrace{\bm{\tilde{\Phi}}_{\mathcal{P}}^{T}\,\bm{D}_1}_{\tilde{\large{\bm{V_1}}}} \longrightarrow \tilde{\bm{V_2}}\approx \tilde{\bm{S}} \,\tilde{\bm{V}_1}
\end{equation} where the $l_2$ approximation lies in the second step, i.e. in the assumption that $\bm{\tilde{\Phi}}_{\mathcal{P}}\,\bm{\tilde{\Phi}}^T_{\mathcal{P}}\approx \bm{I}$ with $\bm{I}$ the identity matrix of size $n_r\times n_r$.

In this reduced system, the reduced propagator is thus: 

\begin{equation}
	\bm{\tilde{S}}=\bm{\tilde{\Phi}}^{T}_{\mathcal{P}} \bm{P}\bm{\tilde{\Phi}}_{\mathcal{P}} = \bm{\tilde{\Phi}}^{T}_{\mathcal{P}} \bm{D}_2 \bm{\tilde{\Psi}}_P \bm{\tilde{\Sigma}}^{-1}_{\mathcal{P}}
\end{equation} having introduced \eqref{Prop}. The eigenvalue decomposition of this propagator $\bm{\tilde{S}}=\bm{Q} \bm{\Lambda} \bm{Q}^{-1}$ governs the evolution of the reduced system in \eqref{PROP}. 
Structures linked to eigenvalues that have $|\lambda_r|<1$ vanish at large times while those that have $|\lambda_r|>1$ explode. Structures linked to $|\lambda_r|=1$ are purely harmonic, like the DFT modes. However, their frequency is not constrained to be multiple of the fundamental tone. Finally, the (complex) spatial structure of the DMD can be computed by back-propagating in the original space using the POD basis: $\bm{\Phi}_{\mathcal{D}}=\bm{\tilde \Phi}_{\mathcal{P}}\,\bm{Q}$.

\begin{tcolorbox}[breakable, opacityframe=.1, title=Exercise 6: The DMD Modes of the TR-PIV data]
	
	As discussed in \cite{Mendez2020}, the DMD of the full dataset encounters serious convergence problems. After all, the trajectories in \ref{Orbit3D} cannot be approximated by a linear dynamical system. We thus focus again on the dataset in the first $t_k\leq 1s$ (that is $n_t=3000$). Compute the DMD spectra and show the spatial structure of the dominant mode.
		
	\medskip
	\textbf{Solution}.  Using MODULO, the call to the DMD function is a one-line command, provided that the snapshot matrix has been prepared as in the previous exercises. The DMD with n\_r=500 modes can be computed as follows:
	
	\begin{centering}
		\begin{lstlisting}[language=Python,linewidth=15.5cm,xleftmargin=.05\textwidth,xrightmargin=.05\textwidth,backgroundcolor=\color{yellow!10}]
#%% Perform the DMD with Modulo
# Initialize a 'MODULO Object'
m = MODULO(data=D,
MEMORY_SAVING=True,
n_Modes = n_r,
N_PARTITIONS=11)
# Prepare (partition) the dataset
D = m._data_processing() # Note that this will cancel D!
# Compute the POD
Phi_D, Lambda, freqs, a0s = m.compute_DMD()
		\end{lstlisting}
	\end{centering}
	
The output provides the spatial structures of the DMD (in Phi\_D), the eigenvalues of the reduced propagator (Lambda), the frequency (in Hz) linked to each eigenvalue and the initial amplitudes of all modes (a0's). We can now plot the DMD spectra in the complex plane:

	\begin{centering}
	\begin{lstlisting}[language=Python,linewidth=15.5cm,xleftmargin=.05\textwidth,xrightmargin=.05\textwidth,backgroundcolor=\color{yellow!10}]
fig, ax = plt.subplots(figsize=(4, 4)) # This creates the figure
plt.plot(np.imag(Lambda),np.real(Lambda),'ko')
circle=plt.Circle((0,0),1,color='b', fill=False)
ax.add_patch(circle)
ax.set_aspect('equal')
	\end{lstlisting}
\end{centering}

and in terms of frequencies:

	\begin{centering}
	\begin{lstlisting}[language=Python,linewidth=15.5cm,xleftmargin=.05\textwidth,xrightmargin=.05\textwidth,backgroundcolor=\color{yellow!10}]
fig, ax = plt.subplots(figsize=(6, 4)) # This creates the figure
plt.plot(freqs,np.abs(Lambda),'ko') 
	\end{lstlisting}
\end{centering}
The results are shown in Figure \ref{figEx6}a) and b) respectively.
We note that all the computed eigenvalues have absolute values smaller than one and will thus vanish in the long run. Two of these have $|\lambda|=0.9940742$ (see fig a): these are almost on the unitary circle in the complex domain and therefore are nearly harmonics. Interestingly, these correspond to $f_n=450.27 Hz$, i.e. the frequency of the vortex shedding (see Figure b). The spatial structures are complex and can be plotted as done for the Fourier modes. The reader finds the scripts in the provided Exercise\_6.py file. The spatial structures are nearly indistinguishable from Fourier modes.

In terms of convergence, the DMD \emph{does not converge}. The temporal structures of this decomposition are not even orthogonal, despite the provided case exhibiting strong periodicity. For a more extensive discussion on the convergence problems of the algorithm, the reader is referred to \cite{Mendez2019,Mendez2020,MendezLS2}.

\begin{center}%
	\setcounter{subfigure}{0}%
	
	\begin{minipage}{.49\linewidth}
		\includegraphics[width=0.8\linewidth]{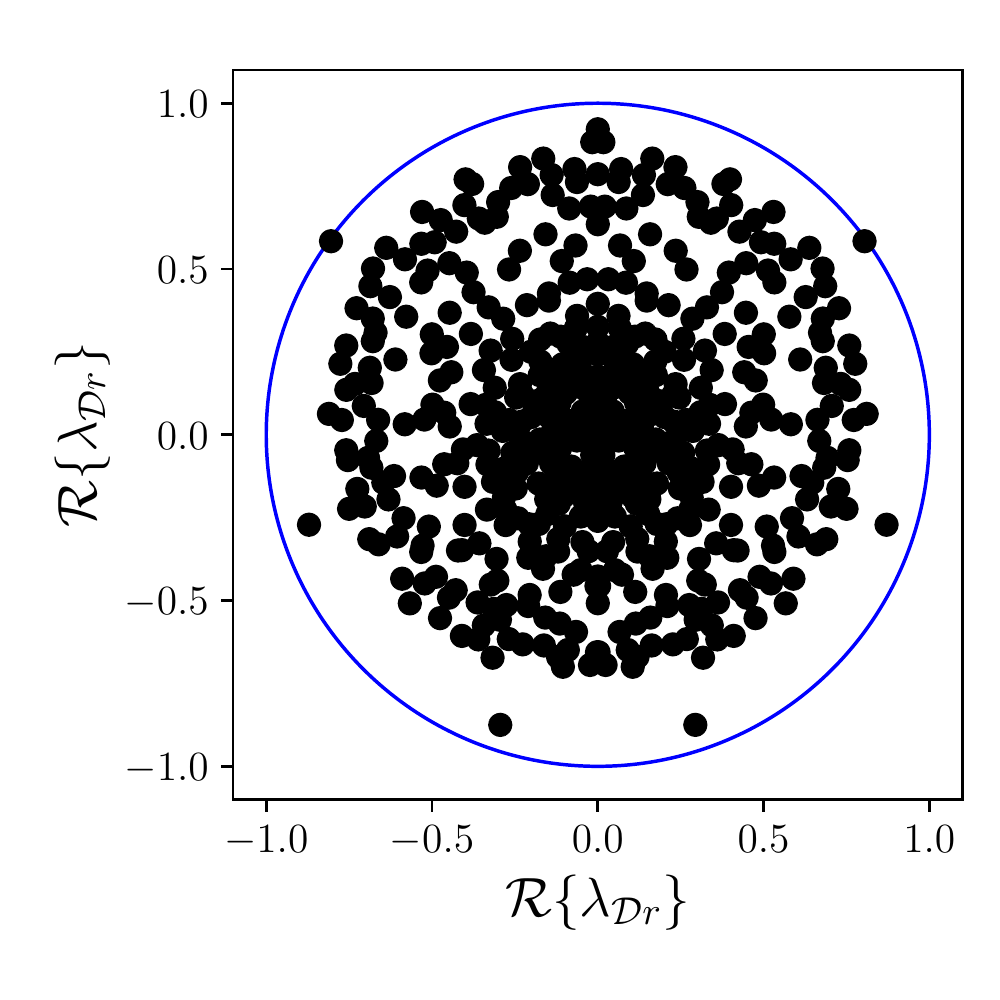}
		\label{Ex6a)}%
		\vspace{-3mm}
		\center{a)}
	\end{minipage}
	\begin{minipage}{.49\linewidth}
		\includegraphics[width=\linewidth]{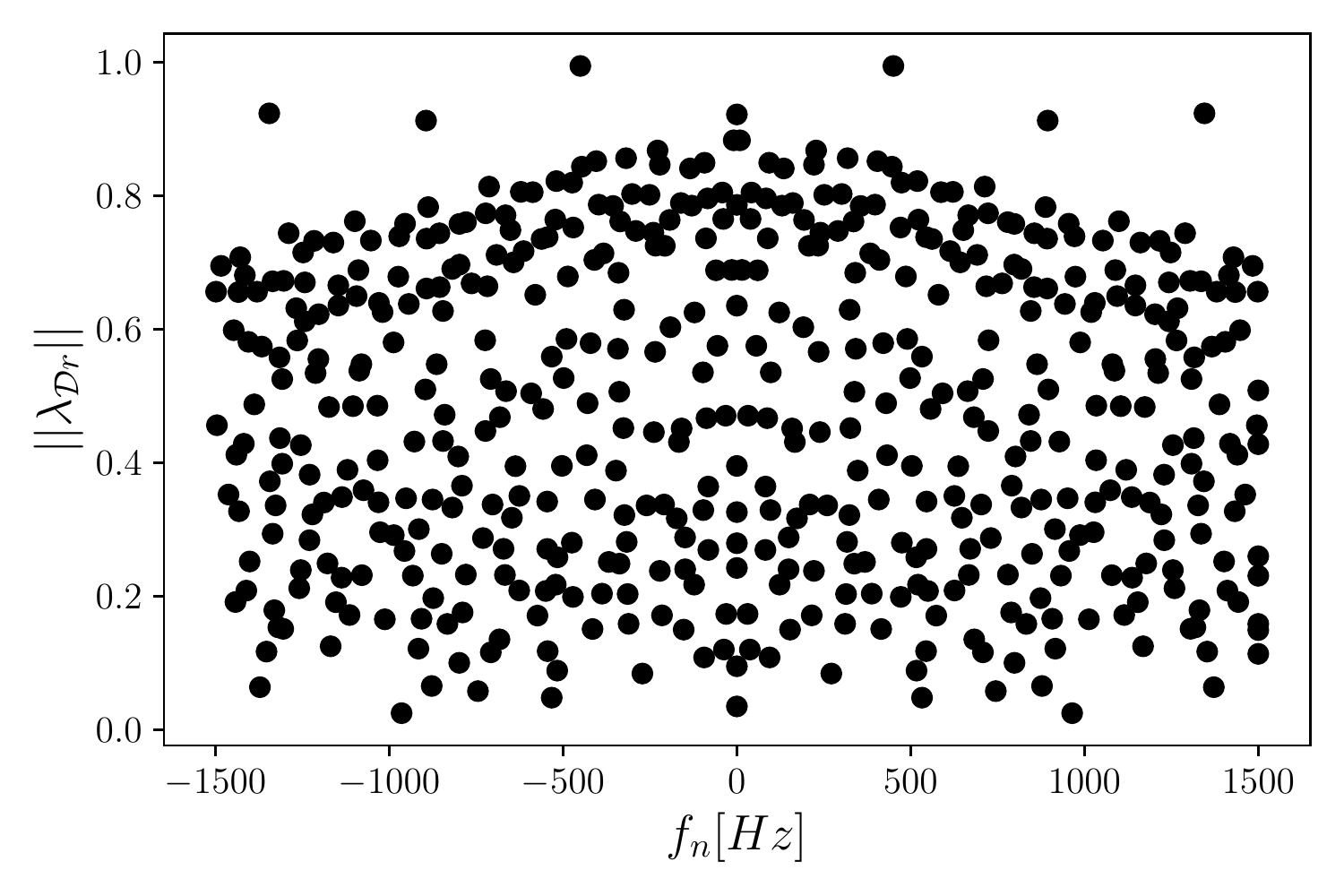}%
		\label{Ex6b)}%
		\vspace{-3mm}
		\center{b)}
	\end{minipage}	
	\vspace{-0.05cm}
	\captionof{figure}{	Fig (a): DMD spectra for the TR-PIV dataset in $t_k\leq 1$, plotted in the complex plane together with the unitary circle. Fig (b): absolute values of the DMD eigenvalues as a function of the corresponding frequency in Hertz. The modes with $|\lambda|\approx 1$ correspond to the vortex shedding. }	
	\label{figEx6}
\end{center}

\end{tcolorbox}

\section{Beyond Classic Decompositions}\label{sec5}

The previous sections gave a brief overview of the `classic' decompositions. We here briefly discuss two possible extensions.

\subsection[The mPOD]{The Multiscale Proper Orthogonal Decomposition}\label{sec5p1}

In the POD, the temporal structures $\psi_{\mathcal{P}}(t_k)$ are eigenvectors of the temporal correlation matrix. These have unconstrained frequency content derived under the energy optimality criterium. As a result (see exercise 4), this decomposition finds an effective representation of the data, but the spectra of each mode are broad. In the DMD, the temporal structures $\psi_{\mathcal{D}}(t_k)$ are eigenvectors of the reduced propagator $\tilde{\bm{S}}$ which best approximate the dataset as a linear dynamical system. These have one (complex) frequency, i.e. $\psi_{\mathcal{D}r}(t_k)=\exp(\mathrm{j}\omega_r t_k)$ with $\omega_r\in \mathbb{C}$. As a result (see exercise 5), this decomposition finds the `optimal' frequencies, but the resulting decomposition has convergence problems even in simple datasets.

The limits of these two `extreme approaches' (i.e. `energy optimality at all costs' or 'one-frequency per mode' at all costs) have motivated many hybrid methods (see \cite{Mendez2019,MendezLS2} for a brief literature review).

The Multiscale Proper Orthogonal Decomposition, first proposed in \cite{Mendez_ICNAM,Mendez_Journal_2} aims at finding a compromise between the energy optimality of the POD and the spectral purity of the DMD. The idea is to introduce the notion of optimality within a prescribed range of frequencies. As extensively illustrated in \cite{Mendez2019} and \cite{MendezLS2}, the mPOD is essentially a combination of POD and Multi-resolution Analysis. The MRA can be conveniently performed on the temporal correlation matrix.

The first step of the mPOD is the partitioning of the temporal correlation matrix $\mathbf{K}$ into the contribution of $M$ scales. This is done by a set of FIR filters with impulse response $\mathbf{h}_1, \mathbf{h}_2\dots \mathbf{h}_M$:  

\begin{equation}
	\label{K_Deco}
	\mathbf{K}= \sum^M_{m=1} \mathbf{K} \circledast \mathbf{h}_m = \mathbf{K}^{(1)}+\mathbf{K}^{(2)}+\dots \mathbf{K}^{(M)}\,,
\end{equation}  where $\circledast$ denotes convolution and $\mathbf{K}^{(m)}$ is correlation matrix for the scale $m$. The mPOD temporal structures are eigenfunctions of the different contributions, i.e. $\mathbf{K}^{(m)} \psi^{(m)}_r =\lambda^{(m)}_r \psi^{(m)}_r$. The filters in Eq.~\ref{K_Deco} are designed with two criteria:  they have separable impulse response and have non-overlapping band-pass regions in the frequency domain. Then, given the filter's transfer function

\begin{equation}
	\mathbf{H}_m(f_i,f_j)=\frac{1}{n^2_t}\sum_{l} \sum_{n} \mathbf{h}_m(t_l,t_n) e^{-2\pi (f_i t_l+f_n t_k)}\,,
\end{equation} the product $\mathbf{H}_m \mathbf{H}_n$ is identically zero for all $m\neq n$. 

The mPOD use Eq.~\ref{K_Deco} to approximate the correlation matrix using only filters with unitary transfer function along the diagonal, i.e. $\mathbf{H}_m (f_n,f_n)\neq0$. Then, it is possible to show that the eigenvectors of all the correlation matrices $\mathbf{K}^{(m)}$ are mutually orthogonal. Moreover, their frequency content is limited by the transfer function $\mathbf{H}_m$ of the filter identifying the corresponding scale: this means that structures from different scales have no common frequencies. Given $\widehat{\psi}^{(m)}_r(f_n)=\mathcal{F}\{\psi^{(m)}_r\}$ the Fourier transform of the r-th eigenvectors of $\mathbf{K}^{(m)}$, the product $\psi^{(m)}_r \psi^{(n)}_j$ is identically zero for all $m\neq n$ and $\widehat{\psi}^{(m)}_r(f_n)=0$ if $\mathbf{H}_m(f_n,f_n)=0$.

The temporal basis $\psi=[\psi^{(1)},\psi^{(2)}\dots \psi^{(M)}]$ collecting the eigenvectors of all scales is the mPOD basis. This is orthogonal by construction and spans $\mathbb{R}^{n_t}$: it thus allows for a lossless decomposition while having modes that are optimal within the prescribed range of frequencies.

The associated spatial structures are finally computed via projection and normalization, as done for the DFT in \eqref{Phi_F} or the POD\footnote{Note that the same kind of operation cannot be done with the DMD because the modes are generally not orthogonal.} in \eqref{Siro}:

\begin{equation}
	\label{Proj}
	\sigma_{\mathcal{M}r} \vec{\bm{\phi}}_{\mathcal{M}r} (\mathbf{x}_i)=\langle \vec{\bm{u}}(\mathbf{x}_i,t),\psi_r(t) \rangle_T= \sum^{n_t}_{k=1} \vec{\bm{u}}(\bm{x_i},t_k)\,\psi_{\mathcal{M}r}(t_k)\,,
\end{equation} with $\sigma_{\mathcal{M}r}=||\langle \vec{\bm{u}}(\mathbf{x}_i,t),\psi_r(t) \rangle_T||_2$. 

The software package MODULO provides an efficient implementation of the mPOD given a frequency splitting vector. We use it in the next exercise.

\begin{tcolorbox}[breakable, opacityframe=.1, title=Exercise 7: The mPOD Modes of the TR-PIV data]
	
	Compute the mPOD of the full dataset (i.e. $n_t=13200$, corresponding to $t_k$). Consider the following frequency band (in Hz): F\_V=np.array([10,290,320,430,470]).

	\hspace{2mm}In practice, we look for mPOD modes that have frequency content limited by the entries in F\_V. Modes that have frequency content in the range $10$-$290$ Hz should not have frequency content in the range $290$-$320$ Hz, etc. This decomposition has six scales: $0$-$10$, $10$-$290$ $\dots$ until $470$-$1500$ Hz. 
	
	\hspace{2mm}The goal is to identify the structures associated with the first stationary phase, hence a vortex shedding at $f_n=450$ Hz,  from the ones associated with the second stationary phase, shedding at $f_n=303$ Hz. Moreover, we are interested in the large scale evolution of the flow, which we consider describable with modes having frequency $f_n\leq 10$ Hz. This explains the choice of F\_V.

	\hspace{2mm}In this exercise, the reader computes the spatial structures, the temporal structures and the Fourier spectra of the modes linked to the two shedding frequencies.
		
	\medskip
	\textbf{Solution}.  Assuming that the data is arranged in a snapshot matrix like in the previous exercises, the mPOD with the prescribed parameters is computed as follows
	
	\begin{centering}
		\begin{lstlisting}[language=Python,linewidth=15.5cm,xleftmargin=.05\textwidth,xrightmargin=.05\textwidth,backgroundcolor=\color{yellow!10}]
m = MODULO(data=D, MEMORY_SAVING=True, n_Modes = n_r,
     N_PARTITIONS=11)
# Prepare (partition) the dataset
D = m._data_processing() # Note that this will cancel D!
# mPOD Settings
# Frequency splitting Vector
F_V=np.array([10,290,320,430,470])
# This vector collects the length of the filter kernels.
Nf=np.array([1201,801,801,801,801]); 
# Decide which scales to keep
Keep=np.array([1,1,0,1,0])
# Others
Ex=1203; boundaries = 'nearest'; MODE = 'reduced'
Phis, Psis, Sigmas = m.compute_mPOD(Nf, Ex, F_V, Keep,
         boundaries, MODE, 1/Fs)
		\end{lstlisting}
	\end{centering}
	
	\hspace{2mm}The first four lines are the same as for all decompositions in MODULO. Line 7 prescribes the frequency splitting vector proposed in the exercise. MODULO operates with FIR filter \citep{Hayes2011} and the vector N\_f collects the length of the associated impulse responses (which are computed by default using Hamming windows). Line 11 defines the `Keep' vector: if 1, the corresponding scale is kept; if 0, it is skipped. By setting some entries of this vector to 0, the mPOD also acts as a filter in the frequency domain.
	
	\hspace{2mm}The last options, in line 13, prescribe the maximum extension of the signals (Ex) during the filtering and the necessary boundary extension for the filtering (the convolutions in \eqref{K_Deco}): in this case, the option `nearest' pads the signals with the last or the first entry. Finally, the mode 'reduced' controls whether the basis to export is a reduced one or a complete one. The results are stored as a matrix factorization like the POD, and hence the rest of the exercise is a plotting exercise no different from Exercise 5. We thus skip the coding and show the results in Figure \ref{fig7aaa}. 
	
	\begin{center}%
		\setcounter{subfigure}{0}%
		
		\begin{minipage}{.49\linewidth}
			\includegraphics[width=\linewidth]{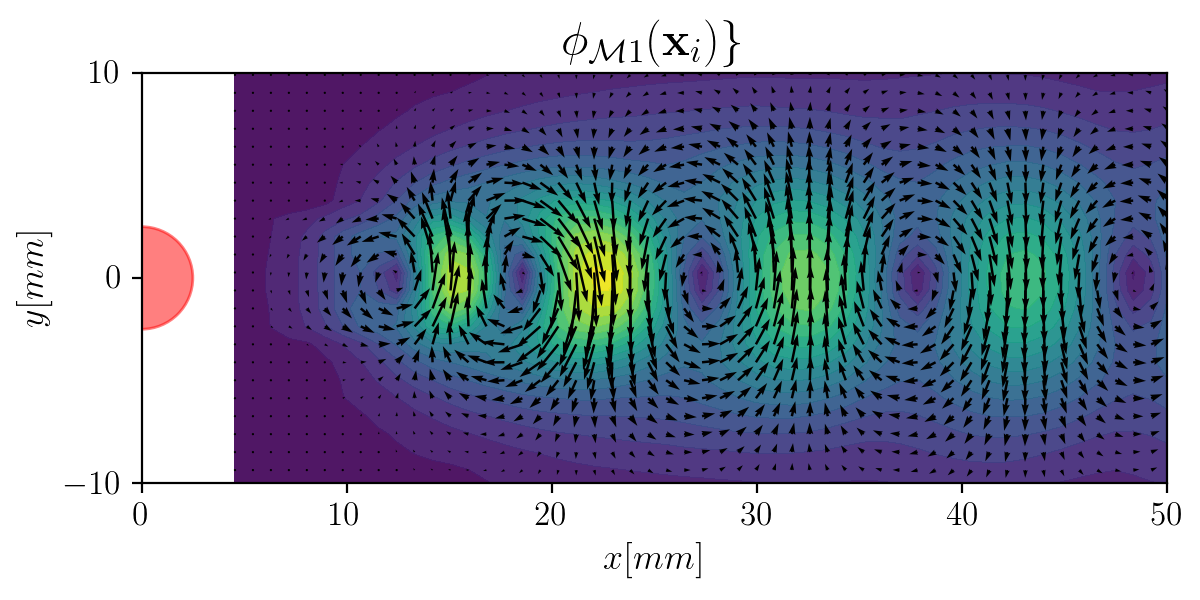}
			\label{Ex5aaaaaa)}%
			\vspace{-3mm}
			\center{a)}
		\end{minipage}
		\begin{minipage}{.49\linewidth}
			\includegraphics[width=\linewidth]{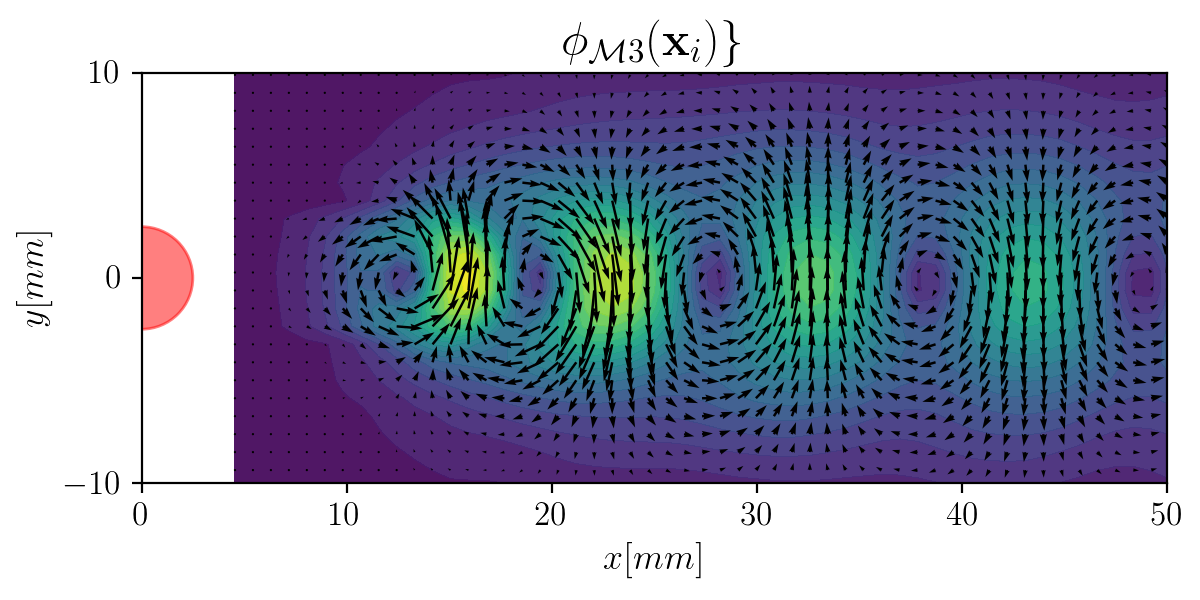}%
			\label{Ex5bbbbbbb)}%
			\vspace{-3mm}
			\center{b)}
		\end{minipage}	
		\begin{minipage}{.49\linewidth}
			\includegraphics[width=\linewidth]{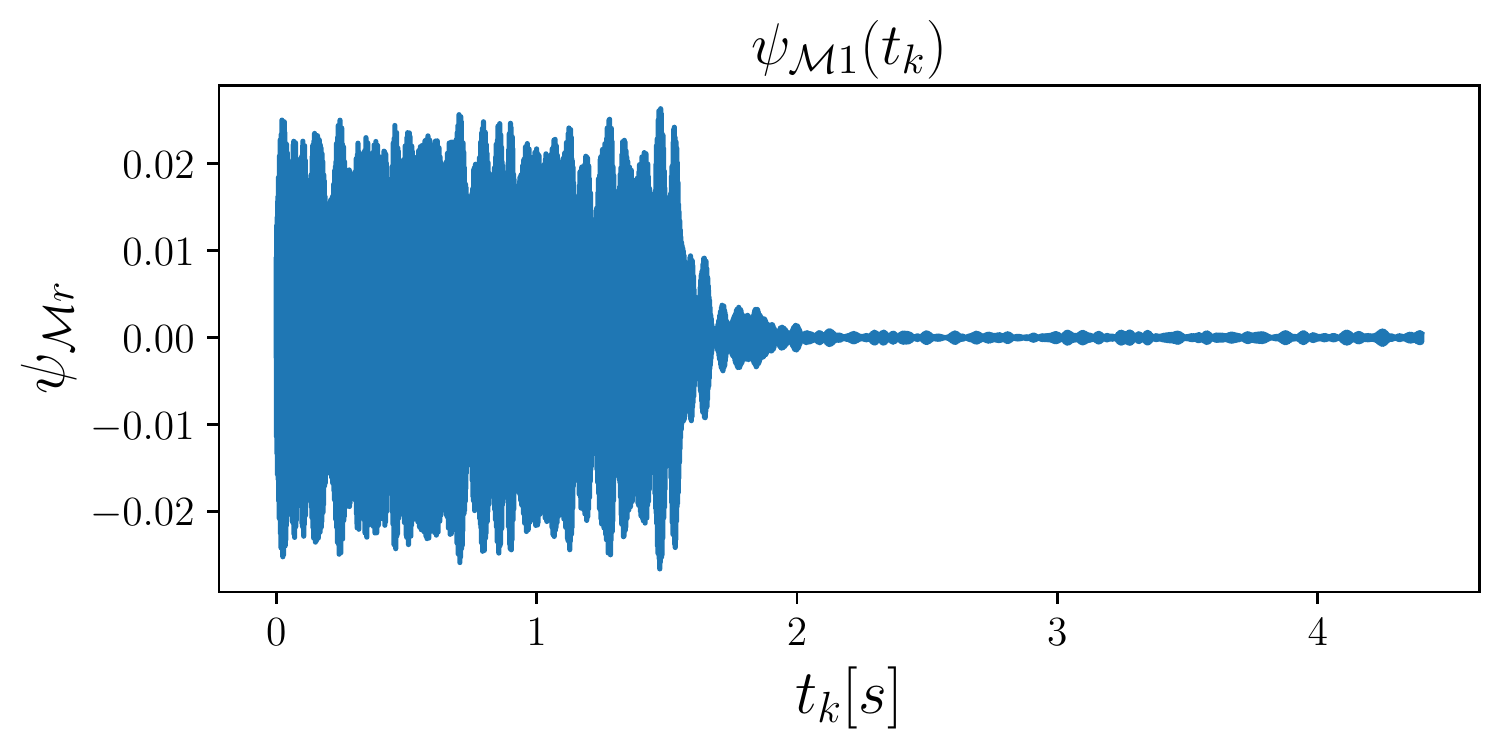}
			\label{Ex5aaaaaaaa)}%
			\vspace{-3mm}
			\center{c)}
		\end{minipage}
		\begin{minipage}{.49\linewidth}
			\includegraphics[width=\linewidth]{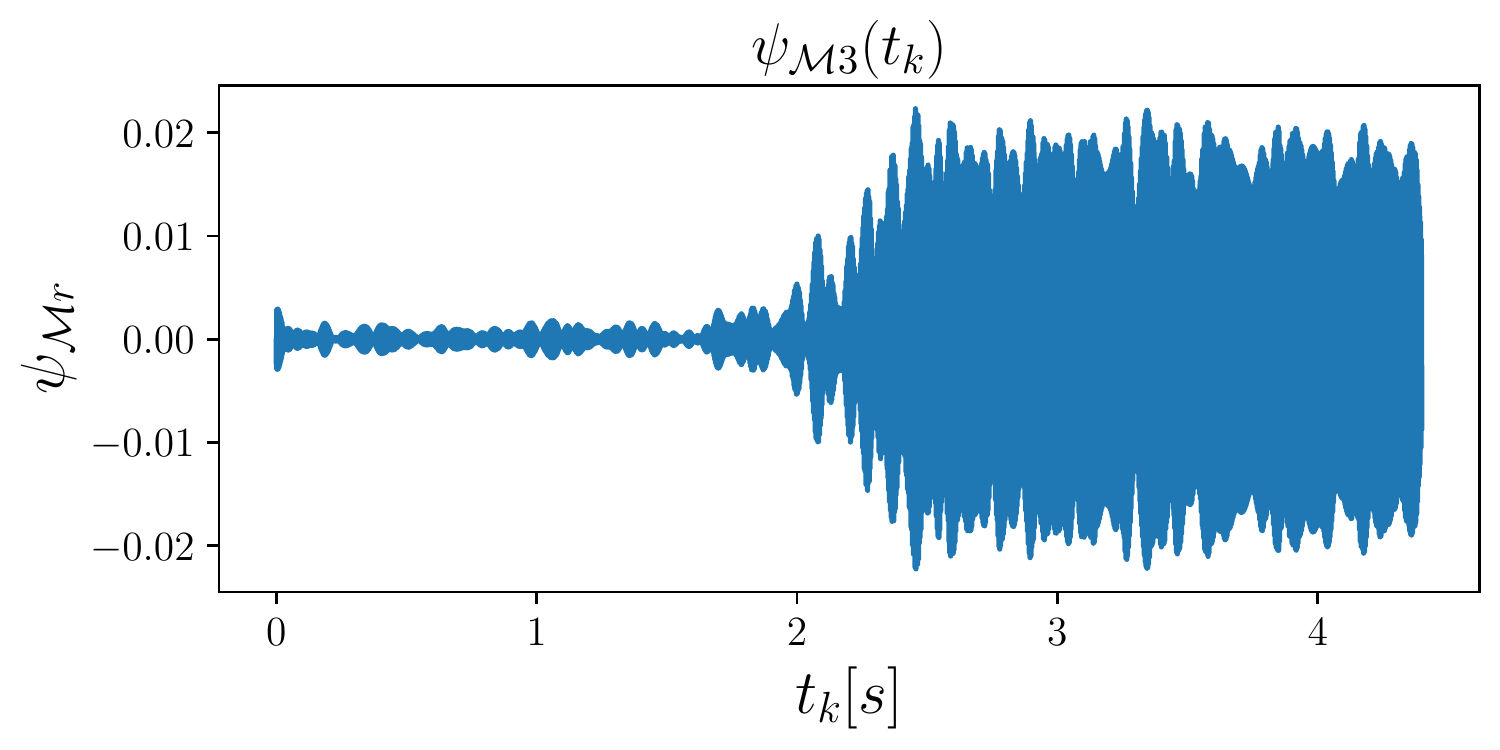}%
			\label{Ex5bbbbbbbbb)}%
			\vspace{-3mm}
			\center{d)}
		\end{minipage}	
		\begin{minipage}{.49\linewidth}
			\includegraphics[width=\linewidth]{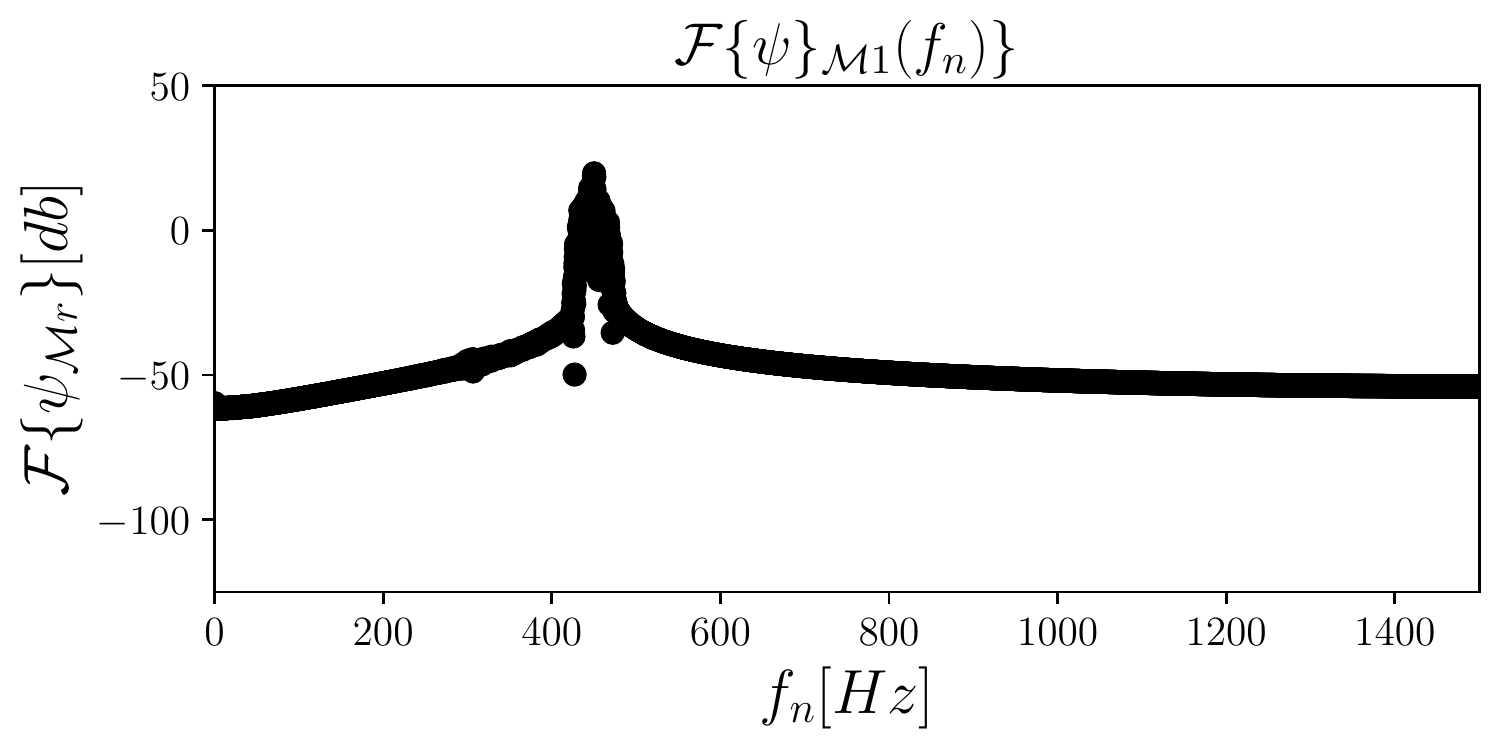}
			\label{Ex5aaaaaaa)}%
			\vspace{-3mm}
			\center{e)}
		\end{minipage}
		\begin{minipage}{.49\linewidth}
			\includegraphics[width=\linewidth]{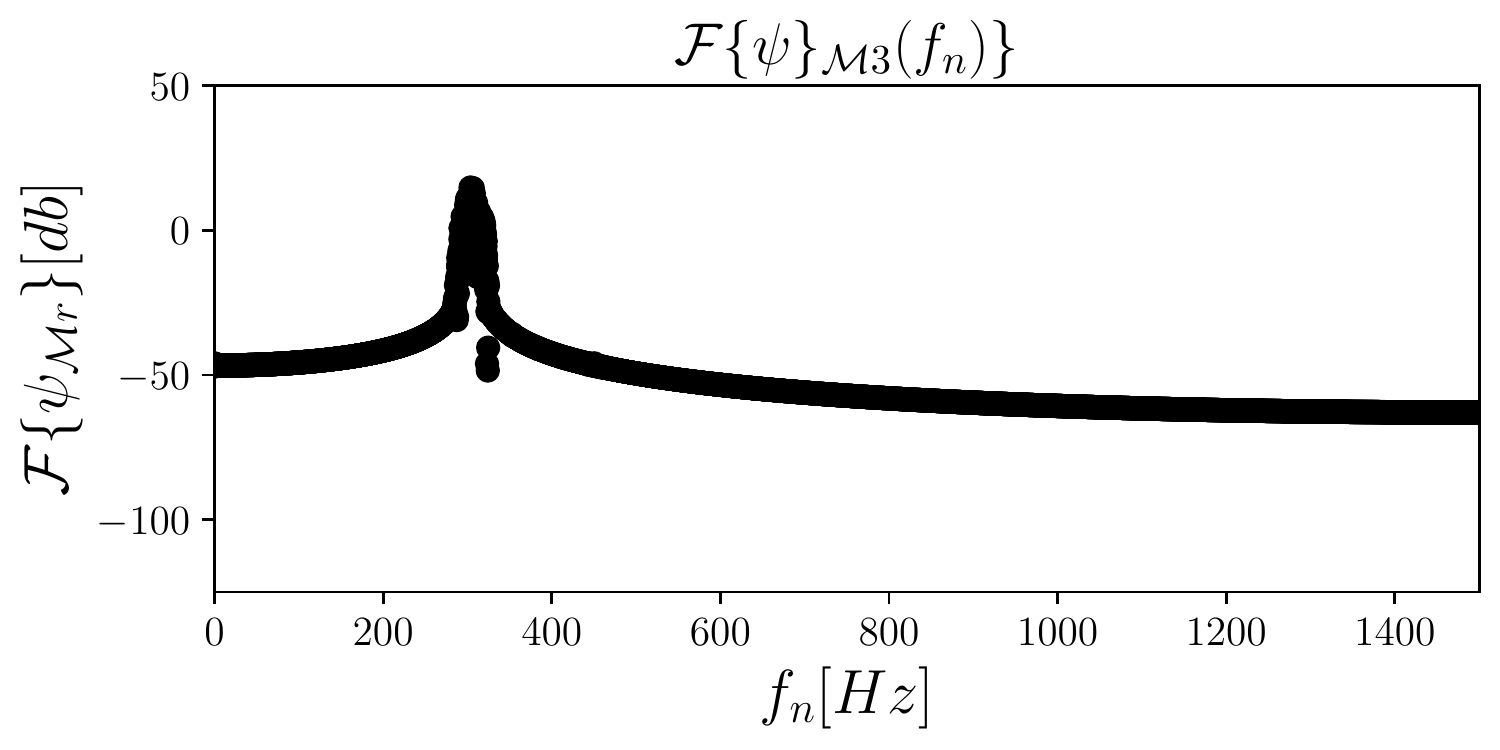}%
			\label{Ex5bbbbbbbb)}%
			\vspace{-3mm}
			\center{f)}
		\end{minipage}	
		\vspace{-0.05cm}
		\captionof{figure}{	Figure (a) and (b). Spatial structures of the two modes linked to the vortex shedding during the two stationary conditions. The temporal structures in (c) and (d) show that the modes are active within certain time intervals, corresponding to the stationary phases. The Fourier spectra of the temporal structures (e) and (f) show that these modes are also localized in the frequency domain. }	
		\label{fig7aaa}
	\end{center}
	
\hspace{2mm} As expected, the mPOD modes capture the modes for the shedding frequencies in the different time intervals. Interestingly, the mPOD inherits from its MRA architecture the ability to localize modes in the frequency domain (these have a relatively narrow frequency bandwidth) with the ability to localize modes in the time domain (these are active only within a specific time interval).

\end{tcolorbox}

\subsection{The Kernel PCA}\label{sec5p2}

In exercise 5, the reader experienced the remarkable compression capabilities of the POD. With only three modes, it is possible to build an approximation of a complex flow with an $l_2$ error of $\sigma_{\mathcal{P}4}/\sigma_{\mathcal{P}1}=0.1797$. This is the best possible result for \emph{any} linear decomposition.

What if we move to a \emph{nonlinear} framework? Nonlinear dimensionality reduction is a well-established subfield of machine learning \citep{Alpaydin2014,Bishop2006}, but its application to fluid mechanics is at its infancy (see \cite{Noack_LS,Ahmed2021,Mendez_OPT_LS}).
As for all the other sections, the field is too broad to be explored in this lecture, but a glance at the general ideas might be given on one page. Moreover, powerful libraries such as scikit-learn \citep{scikit-learn} make many nonlinear decompositions accessible in just one line of code. 

We consider the most popular nonlinear tool for dimensionality reduction: the Kernel Principal Component Analysis (kPCA) by \cite{Schoelkopf1997}. Instead of working with the snapshot matrix $\mathbf{D}$, we work with a \emph{kernelized} version of it: $\xi (\mathbf{D})\in\mathbb{R}^{n_F\times n_t}$. 

Here $\xi()$ is a kernel function which maps the snapshots $\vec{\bm{u}}(\mathbf{x}_i,t_k)\in \mathbb{R}^{n_s}$ into the so-called feature space $\xi(\vec{\bm{u}})(\mathbf{x}_i,t_k)\in \mathbb{R}^{n_F}$. While the spatial structures of the POD (PCA in machine learning terminology) $\phi_{\mathcal{P}}$ are eigenvectors of $\mathbf{D}\mathbf{D}^T$, the spatial structures of the kPCA (or kPOD, in a sense) $\phi_{\mathcal{K}}$ are eigenvectors of $\xi(\mathbf{D}) \xi(\mathbf{D})^T$. These are (orthonormal) nonlinear functions of the dataset and their evolution in time (the temporal structures $\psi_{\mathcal{K}}$) can be computed by projecting the data onto this basis.

If we focus on the forward problem, i.e. the compression problem, we are looking for the projection of the datasets onto the spatial structures $\phi_{\mathcal{K}}$. If the kernel functions are well-chosen, these projections can be computed without knowing the $\phi_{\mathcal{K}}$'s \citep{Bishop2006}!

The trick to perform this computation is known as \emph{kernel trick} \citep{Bishop2006,Schoelkopf1997} and goes as follows. Let us write the $\phi_{\mathcal{K} r}$ as linear combinations of features:

\begin{equation}
	\label{eq8}
	\phi_{\mathcal{K} r}=\sum^{n_p}_{i=1}a_{r}(t_k)\xi(\vec{\bm{u}})(\mathbf{x}_i,t_k)=\xi(\mathbf{D}) \mathbf{a}_r\,.,
\end{equation} where $\mathbf{a}_r\in\mathbb{R}^{n_t}$ is the vector collecting the coefficients of the combination. Introduce this ansatz into the eigenvalue problem and multiply by $\xi (\mathbf{D})^T$ on both sides:

\begin{equation}
	\label{eq9}
	\xi(\mathbf{D})^T(\xi(\mathbf{D})\xi(\mathbf{D})^T) \xi(\mathbf{D}) \mathbf{a}_r =\lambda_r \xi(\mathbf{D})^T \xi(\mathbf{D}) \mathbf{a}_r\rightarrow \mathbf{K}^2\mathbf{a}_r
	=\lambda_r \mathbf{K}\mathbf{a}_r\,.
\end{equation} where $\mathbf{K}=\xi(\mathbf{D})^T\xi(\mathbf{D})$ collects the inner products in the feature space. If the kernel function is carefully chosen \citep{Bishop2006}, the inner products defining these matrix can be computed via a \emph{kernel function}:

\begin{equation}
	\label{eq10}
	\mathbf{K}_{i,j}=\xi(\mathbf{d}_i)^T\xi(\mathbf{d}_j)=\kappa(\mathbf{d}_i,\mathbf{d}_j)\,,
\end{equation} where $\mathbf{d}_i$ and $\mathbf{d_j}$ are two columns of $\mathbf{D}$, i.e. two snapshots of the data.
If this matrix is invertible, equation \eqref{eq9} reduces to $\mathbf{K}\mathbf{a}_r
=\lambda_r\mathbf{a}_r$, i.e. the eigedecomposition of $\mathbf{K}$. Given the eigenvectors $\mathbf{a}_r$, the projection of a feature vector $\xi(\mathbf{x}_i)$ onto the principal component $\phi_{\mathcal{K}r}$ in the feature space is:

\begin{equation}
	\label{eq11}
	\sigma_{\mathcal{K}r} \psi_{\mathcal{K}r}^T=\phi^T_{\mathcal{K} r} \xi(\mathbf{D})=\mathbf{a}^T_r\xi(\mathbf{D})^T\xi(\mathbf{D})=\mathbf{a}^T_r\kappa(\mathbf{D},\mathbf{D})\,. 
\end{equation} 

Computing these temporal structures is the forward kPCA problem. Reconstructing the dataset from these (i.e. the inverse kPCA), on the other hand, require the computation of the $\phi_{\mathcal{K}}$ and this is a complex nonlinear optimization problem. The reader is referred to \cite{ikPA} for more information. In the next exercise, we close this section by showing how to compute the kPCA of the provided TR-PIV dataset.

\begin{tcolorbox}[breakable, opacityframe=.1, title=Exercise 8: The kPCA Manifold of the TR-PIV data]
	
	Compute the best 3 dimensional manifold approximating the full dataset using forward kPCA. Then, use this manifold to retreive back an approximation of the velocity field. What is the least square error of the approximation ?
	
	\medskip
	\textbf{Solution}. As usual, assuming that the data is arranged in a snapshot matrix as for the previous exercises, this exercise can be solved in one line of code. Scikit-learn provides a powerful function for kPCA:
	
	\begin{centering}
		\begin{lstlisting}[language=Python,linewidth=15.5cm,xleftmargin=.05\textwidth,xrightmargin=.05\textwidth,backgroundcolor=\color{yellow!10}]
D_s=D.T
from sklearn.decomposition import KernelPCA
kpca = KernelPCA(kernel="rbf", n_components=3,
fit_inverse_transform=True, 
alpha=1,gamma=None)
Psi_K = kpca.fit_transform(D_s)
		\end{lstlisting}
	\end{centering}
	
\hspace{2mm}The convention used by scikitlearn is that of having data realizations (in our case snapshots of the TR-PIV measurement) along the rows of the feature matrix; hence the transposition. The kPCA is here performed using Gaussian Radial Basis functions (see \cite{Bishop2006,Mendez_OPT_LS}) as kernel functions, with shape factor $\gamma$ taken as a default (i.e. $1/n_s$ in this case). The call to the function is performed in two steps: line 3 create the kpca object while line 5 runs the function `fit\_transform' within the object. The parameter $\alpha$ is a regularization parameter used to compute the inversion (as described in \cite{ikPA}). The plot of the 3D manifold derived by the kPCA is simply:

	\begin{centering}
	\begin{lstlisting}[language=Python,linewidth=15.5cm,xleftmargin=.05\textwidth,xrightmargin=.05\textwidth,backgroundcolor=\color{yellow!10}]
fig = plt.figure(figsize=(7,4))
ax = fig.add_subplot(111, projection='3d')
ax = Axes3D(fig, elev=-150, azim=48)
ax.dist = 13
ax.scatter(Psi_K[0:n_t:2,1],
-Psi_K[0:n_t:2,2],
Psi_K[0:n_t:2,0], 'ko')
	\end{lstlisting}
\end{centering}

\hspace{2mm}This is shown in Figure \ref{Orbit3D_kPCA}. The plot of these structures is made to have the same orientation as the one obtained via POD.

 \begin{center}%
	\setcounter{subfigure}{0}%
	\begin{minipage}{1\linewidth}
		\centering
		\includegraphics[width=0.9\linewidth]{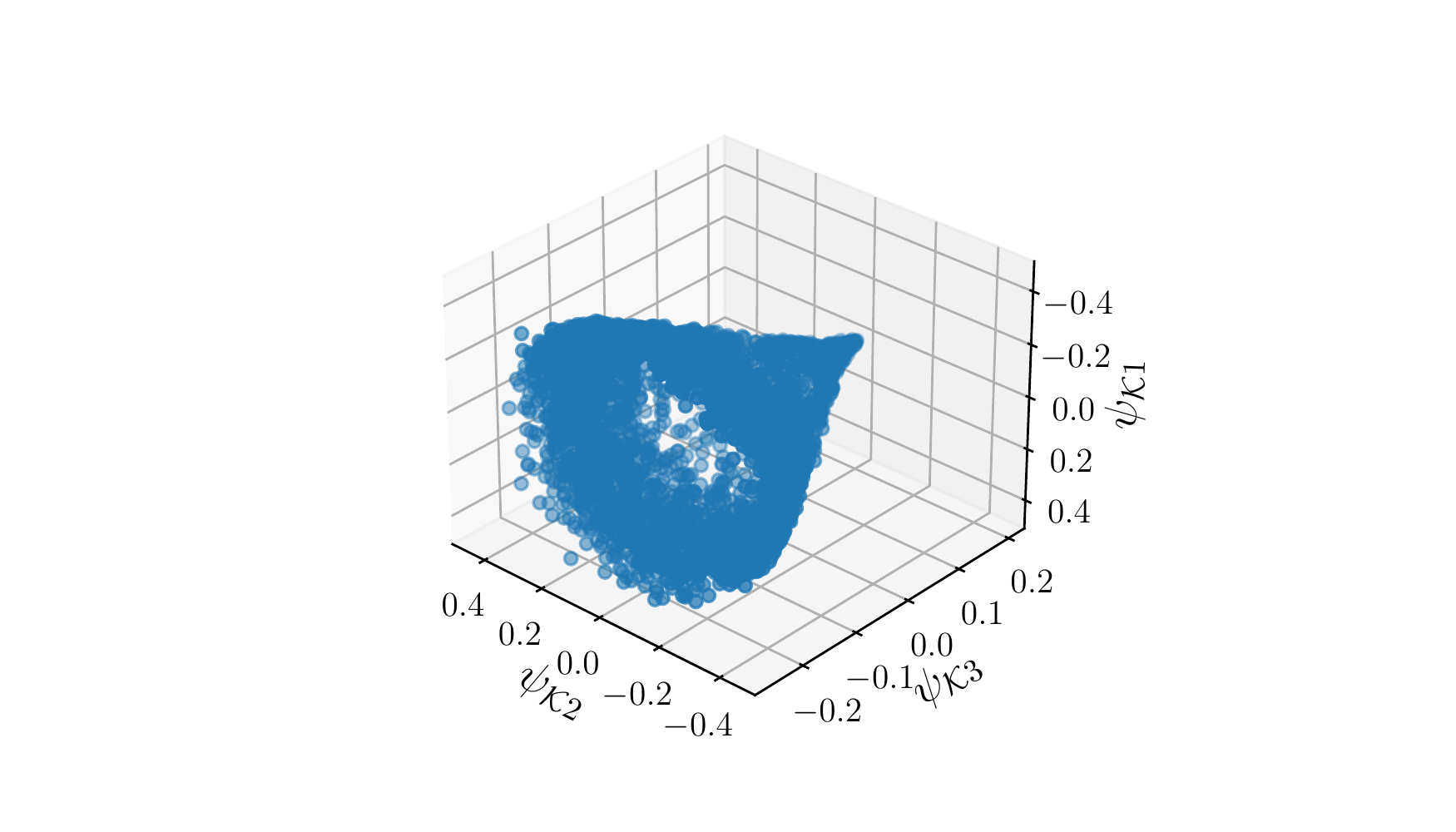}
		\label{Orbit3D_kPCA}%
		\captionof{figure}{Evolution of the reduced order dynamical system in the kPOD basis.}
	\end{minipage}

\end{center}

\hspace{2mm}It is thus possible to appreciate that the manifold is axisymmetric and that the system's orbits remain circular in both the stationary phases. However, the nonlinear mapping results in a different shape, with much narrower orbits at the highest velocity and much larger at the highest (see animations in the presentation). The transitory is also different. The most remarkable result is yet to come: the $l_2$ error obtained by mapping this manifold back to $\mathbb{R}^{3}$ can be computed as follows:

	\begin{centering}
	\begin{lstlisting}[language=Python,linewidth=15.5cm,xleftmargin=.05\textwidth,xrightmargin=.05\textwidth,backgroundcolor=\color{yellow!10}]
X_rec_kpca = kpca.inverse_transform(Psi_K )
Error_kpca=np.linalg.norm(D_s-X_rec_kpca)/np.linalg.norm(D_s)
print(Error_kpca)
	\end{lstlisting}
\end{centering}

\hspace{2mm}The reader should find the result in the console: 2.983e-5. This is practically zero and six orders of magnitudes smaller than the error produced by a POD approximation of the same dimension. Somehow, the manifold in Figure \ref{Orbit3D_kPCA} contains \emph{the same amount of information} available in the $4260$ dimensional space where the TR-PIV velocity fields live and evolve. Isn't this remarkable?

	\end{tcolorbox}

\section{Final Remarks}\label{sec6}

These notes offered a tour de force into signal processing, modal analysis and dimensionality reduction to process spatiotemporally resolved data like the one produced by modern image velocimetry. 

The lecture opened with statistical tools for time series analysis (Sec. \ref{sec3}), moved to classic modal analysis (Sec. \ref{sec4}) and explored modern extensions such as multiscale data-driven decompositions and nonlinear (kernel-based) methods (Sec. \ref{sec5}).

While the topics in Section \ref{sec3} are textbook material for students in fluid dynamics for at least $30$ years, the topics in Section \ref{sec4} have become standard tools in the researcher's toolbox more recently. It is thus not surprising that all statistical tools presented in Section \ref{sec3} have already been integrated into the theory and the modelling of fluid flows, while the same is not entirely true, yet, for the topics in Section \ref{sec4}. However, the trends are clear: today, these are popular tools to extract patterns, to construct simplified representations, or to filter experimental (and numerical) data. The reader is encouraged to dive into the provided exercises to gain computational proficiency on the subject. Finally, the topics in Section \ref{sec4} are at the forefront of fluid dynamics research and the ever-growing intersection between signal processing and machine learning. The field is very active, and the provided notes and exercises give a good starting point.


\bibliographystyle{apalike}
\bibliography{main}
\clearpage{\pagestyle{empty}\cleardoublepage}
\end{document}